\documentclass[sigconf]{aamas}

\usepackage{balance} 

\usepackage[utf8]{inputenc} 
\usepackage[T1]{fontenc}    
\usepackage{hyperref}       
\usepackage{url}            
\usepackage{booktabs}       
\usepackage{amsfonts}       
\usepackage{nicefrac}       
\usepackage{microtype}      

\usepackage{listings}
\definecolor{codegreen}{rgb}{0,0.6,0}
\definecolor{codegray}{rgb}{0.5,0.5,0.5}
\definecolor{codepurple}{rgb}{0.58,0,0.82}
\definecolor{backcolour}{rgb}{0.95,0.95,0.92}
\lstdefinestyle{mystyle}{
    backgroundcolor=\color{backcolour},   
    commentstyle=\color{codegreen},
    keywordstyle=\color{magenta},
    numberstyle=\tiny\color{codegray},
    stringstyle=\color{codepurple},
    basicstyle=\ttfamily\footnotesize,
    breakatwhitespace=false,         
    breaklines=true,                 
    captionpos=b,
    escapeinside={\%*}{*)},
    keepspaces=true,                 
    numbers=left,                    
    numbersep=1pt,                  
    showspaces=false,                
    showstringspaces=false,
    showtabs=false,                  
    tabsize=2
}
\usepackage{graphicx}
\usepackage{amsmath}
\usepackage{amsthm}
\usepackage{algorithm}
\usepackage{algorithmic}
\usepackage{amsmath,amsfonts,bm,bbm}

\def\eqref#1{equation~(\ref{#1})}
\def\Eqref#1{Equation~(\ref{#1})}

\def\Algref#1{Algorithm~\ref{#1}}

\def\1{\bm{1}}

\usepackage{cancel}

\usepackage{listings}
\usepackage{enumerate}

\DeclareMathAlphabet{\mathsfit}{\encodingdefault}{\sfdefault}{m}{sl}
\SetMathAlphabet{\mathsfit}{bold}{\encodingdefault}{\sfdefault}{bx}{n}

\newcommand{\R}{\mathbb{R}}

\DeclareMathOperator*{\argmax}{arg\,max}

\usepackage{amsthm}

\makeatletter
\newtheorem*{rep@theorem}{\rep@title}
\newcommand{\newreptheorem}[2]{%
\newenvironment{rep#1}[1]{%
 \def\rep@title{#2 \ref{##1}}%
 \begin{rep@theorem}}%
 {\end{rep@theorem}}}
\makeatother

\newtheorem{theorem}{Theorem}
\newreptheorem{theorem}{Theorem}

\newreptheorem{lemma}{Lemma}
\newtheorem{assumption}{Assumption}

\usepackage{epigraph}
\usepackage{wrapfig}

\usepackage{subcaption}

\usepackage[useregional]{datetime2}
\definecolor{green}{rgb}{0.0, 0.42, 0.24} 
\definecolor{orange}{rgb}{0.8, 0.33, 0.} 
\definecolor{blue}{rgb}{0.16, 0.32, 0.75} 
\definecolor{cobalt}{rgb}{0.0, 0.28, 0.67} 

\newcommand{\br}{\texttt{BR}}

\usepackage[toc,page]{appendix}
\addtocontents{toc}{\protect\setcounter{tocdepth}{-1}}

\setcopyright{ifaamas}
\acmConference[AAMAS '22]{Proc.\@ of the 21st International Conference
on Autonomous Agents and Multiagent Systems (AAMAS 2022)}{May 9--13, 2022}
{Online}{P.~Faliszewski, V.~Mascardi, C.~Pelachaud,
M.E.~Taylor (eds.)}
\copyrightyear{2022}
\acmYear{2022}
\acmDOI{}
\acmPrice{}
\acmISBN{}

\acmSubmissionID{278}

\title[ADIDAS]{Sample-based Approximation of Nash in\\Large Many-Player Games via Gradient Descent}

\author{Ian Gemp}
\affiliation{
  \institution{DeepMind}
  \city{London}
  \country{United Kingdom}}
\email{imgemp@deepmind.com}

\author{Rahul Savani}
\affiliation{
  \institution{University of Liverpool}
  \city{Liverpool}
  \country{United Kingdom}}
\email{rahul.savani@liverpool.ac.uk}

\author{Marc Lanctot}
\affiliation{
  \institution{DeepMind}
  \city{Edmonton}
  \country{Canada}}
\email{lanctot@deepmind.com}

\author{Yoram Bachrach}
\affiliation{
  \institution{DeepMind}
  \city{London}
  \country{United Kingdom}}
\email{yorambac@deepmind.com}

\author{Thomas Anthony}
\affiliation{
  \institution{DeepMind}
  \city{London}
  \country{United Kingdom}}
\email{twa@deepmind.com}

\author{Richard Everett}
\affiliation{
  \institution{DeepMind}
  \city{London}
  \country{United Kingdom}}
\email{reverett@deepmind.com}

\author{Andrea Tacchetti}
\affiliation{
  \institution{DeepMind}
  \city{London}
  \country{United Kingdom}}
\email{atacchet@deepmind.com}

\author{Tom Eccles}
\affiliation{
  \institution{DeepMind}
  \city{London}
  \country{United Kingdom}}
\email{eccles@deepmind.com}

\author{J\'anos~Kram\'ar}
\affiliation{
  \institution{DeepMind}
  \city{London}
  \country{United Kingdom}}
\email{janosk@deepmind.com}

\begin{abstract}
Nash equilibrium is a central concept in game theory. Several Nash solvers exist, yet none scale to normal-form games with many actions and many players, especially those with payoff tensors too big to be stored in memory. In this work, we propose an approach that iteratively improves an approximation to a Nash equilibrium through joint play. It accomplishes this by tracing a previously established homotopy that defines a continuum of equilibria for the game regularized with decaying levels of entropy. This continuum asymptotically approaches the \emph{limiting logit equilibrium}, proven by~\citeauthor{mckelvey1995quantal}~(\citeyear{mckelvey1995quantal}) to be unique in \emph{almost} all games, thereby partially circumventing the well-known equilibrium selection problem of many-player games. To encourage iterates to remain near this path, we efficiently minimize \emph{average deviation incentive} via stochastic gradient descent, intelligently sampling entries in the payoff tensor as needed. Monte Carlo estimates of the stochastic gradient from joint play are biased due to the appearance of a nonlinear max operator in the objective, so we introduce additional innovations to the algorithm to alleviate gradient bias. The descent process can also be viewed as repeatedly constructing and reacting to a polymatrix approximation to the game. In these ways, our proposed approach, \emph{average deviation incentive descent with adaptive sampling} (ADIDAS), is most similar to three classical approaches, namely homotopy-type, Lyapunov, and iterative polymatrix solvers. The lack of local convergence guarantees for biased gradient descent prevents guaranteed convergence to Nash, however, we demonstrate through extensive experiments the ability of this approach to approximate a unique Nash equilibrium in normal-form games with as many as seven players and twenty one actions (several billion outcomes) that are orders of magnitude larger than those possible with prior algorithms.
\end{abstract}


\newcommand{\BibTeX}{\rm B\kern-.05em{\sc i\kern-.025em b}\kern-.08em\TeX}

\begin{document}


\pagestyle{fancy}
\fancyhead{}


\maketitle 


\section{Introduction}
\label{intro}



Core concepts from game theory underpin many advances in multi-agent systems research. Among these, Nash equilibrium is particularly prevalent. Despite the difficulty of computing a Nash equilibrium~\citep{daskalakis2009complexity,chen2009settling}, a plethora of algorithms~\citep{lemke1964equilibrium,sandholm2005mixed,porter2008simple,govindan2003global,blum2006continuation} and suitable benchmarks~\citep{nudelman2004run} have been developed, however, none address large normal-form games with many actions and many players, especially those too big to be stored in memory.


In this work, we develop an algorithm for approximating a Nash equilibrium of a normal-form game with so many actions and players that only a small subset of the possible outcomes in the game can be accessed at a time. We refer the reader to~\citet{mckelvey1996computation} for a review of approaches for normal-form games. Several algorithms exactly compute a Nash equilibrium for small normal-form games and others efficiently approximate Nash equilibria for special game classes, however, practical algorithms for approximating Nash in large normal-form games with many players, e.g. 7, and many actions, e.g., 21, is lacking. Computational efficiency is of paramount importance for large games because a general normal-form game with $n$ players and $m$ actions contains $nm^n$ payoffs; simply enumerating all payoffs can be intractable and renders classical approaches ineligible. A common approach is to return the profile found by efficient no-regret algorithms that sample payoffs as needed~\citep{blum_mansour_2007} although~\citet{flokas2020no} recently proved that many from this family do not converge to mixed Nash equilibria in \emph{all} games, 2-player games included. 

While significant progress has been made for computing Nash in 2-player normal-form games which can be represented as a \emph{linear} complementarity problem, the many-player setting induces a \emph{nonlinear} complementarity problem, which is ``often hopelessly impractical to solve exactly''~(\cite{shoham2008multiagent}, p. 105).\footnote{While any n-player game can, in theory, be efficiently solved for approximate equilibria by reducing it to a two-player game, in practice this approach is not feasible for solving large games due to the blowups involved in the reductions. Details in Appx.~\ref{app:beyondtwo}.} The combination of high dimensionality ($m^n$ vs $m^2$ distinct outcomes) and nonlinearity (utilities are degree-$n$ polynomials in the strategies vs degree-$2$) makes many-player games much more complex.

This more general problem arises in cutting-edge multiagent research when learning~\citep{gray2020human} and evaluating~\citep{anthony2020learning} agents in Diplomacy, a complex 7-player board game. \citet{gray2020human} used no-regret learning to approximate a Nash equilibrium of subsampled games, however, this approach is brittle as we show later in Figure~\ref{fig:noregfail}. In \cite{anthony2020learning}, five Diplomacy bots were ranked according to their mass under an approximate Nash equilibrium. We extend that work to encourage convergence to a particular Nash and introduce sampling along with several technical contributions to scale evaluation to 21 Diplomacy bots, a >1000-fold increase in meta-game size.

Equilibrium computation has been an important component of AI in multi-agent systems~\citep{shoham2008multiagent}. It has been (and remains) a critical component of super-human AI in poker~\citep{Bowling15Poker,Moravcik17DeepStack,Brown17Libratus,brown2020combining}. 
As mentioned above, Nash computation also arises when strategically summarizing a larger domain by learning a lower dimensionality empirical game~\citep{wellman2006methods}; such an approach was used in the AlphaStar League, leading to an agent that beat humans in StarCraft~\citep{vinyals2019alphastar,vinyals2019grandmaster}.
Ultimately, this required solving for the Nash of a 2-player, 888-action game, which can take several seconds using state-of-the-art solvers on modern hardware. 
In contrast, solving an empirical game of Diplomacy, e.g., a 7-player 888-action game, would naively take longer than the current age of the universe.
This is well beyond the size of any game we inspect here, however, we approximate the Nash of games several orders of magnitude larger than previously possible, thus taking a step towards this ambitious goal.

\textbf{Our Contribution:} We introduce stochastic optimization into a classical \emph{homotopy} approach resulting in an algorithm that avoids the need to work with the full payoff tensor all at once and is, to our knowledge, the first algorithm generally capable of practically approximating a unique Nash equilibrium in large (billions of outcomes) many-player, many-action normal-form games. We demonstrate our algorithm on 2, 3, 4, 6, 7 and 10 player games (10 in Appx.~\ref{app:xtra_games}; others in \S\ref{exp}). We also perform various ablation studies of our algorithm (Appx.~\ref{app:ablations}), compare against several baselines including solvers from the popular \texttt{Gambit} library (more in Appx.~\ref{app:morealgs}), and examine a range of domains (more in Appx.~\ref{app:xtra_games}).


The paper is organized as follows. After formulating the Nash equilibrium problem for a general $n$-player normal-form game, we review previous work.
We discuss how we combine the insights of classical algorithms with ideas from stochastic optimization to develop our final algorithm, \emph{average deviation incentive descent with adaptive sampling}, or ADIDAS.
Finally, we compare our proposed algorithm against previous approaches on large games of interest from the literature: games such as Colonel Blotto~\citep{arad2012multi}, classical Nash benchmarks from the GAMUT library~\citep{nudelman2004run}, and games relevant to recent success on the $7$-player game Diplomacy~\citep{anthony2020learning,gray2020human}.

\section{Preliminaries}
In a finite $n$-player game in normal form, each player $i \in \{1,\ldots,n\}$ is given a strategy set $\mathcal{A}_i = \{a_{i1}, \ldots, a_{im_i}\}$ consisting of $m_i$ pure strategies. The pure strategies can be naturally indexed by non-negative integers, so we redefine $\mathcal{A}_i = \{0, \ldots, m_i - 1\}$ as an abuse of notation for convenience. Each player $i$ is also given a payoff or utility function, $u_i: \mathcal{A} \rightarrow \mathbb{R}$ where $\mathcal{A} = \prod_i \mathcal{A}_i$. In games where the cardinality of each player's strategy set is the same, we drop the subscript on $m_i$. Player $i$ may play a mixed strategy by sampling from a distribution over their pure strategies. Let player $i$'s mixed strategy be represented by a vector $x_i \in \Delta^{m_i-1}$ where $\Delta^{m_i-1}$ is the $(m_i-1)$-dimensional probability simplex embedded in $\mathbb{R}^{m_i}$. Each function $u_i$ is then extended to this domain so that $u_i(\boldsymbol{x}) = \sum_{\boldsymbol{a} \in \mathcal{A}} u_i(\boldsymbol{a}) \prod_{j} x_{ja_j}$ where $\boldsymbol{x} = (x_1, \ldots, x_n)$ and $a_j \in \mathcal{A}_j$ denotes player $j$'s component of the joint action $\boldsymbol{a} \in \mathcal{A}$. For convenience, let $x_{-i}$ denote all components of $\boldsymbol{x}$ belonging to players other than player $i$.

We say $\boldsymbol{x} \in \prod_i \Delta^{m_i-1}$ is a Nash equilibrium iff, for all $i \in \{1, \ldots, n\}$, $u_i(z_i, x_{-i}) \le u_i(\boldsymbol{x})$ for all $z_i \in \Delta^{m_i-1}$, i.e., no player has any incentive to unilaterally deviate from $\boldsymbol{x}$. Nash is most commonly relaxed with $\epsilon$-Nash, an additive approximation: $u_i(z_i, x_{-i}) \le u_i(\boldsymbol{x}) + \epsilon$ for all $z_i \in \Delta^{m_i-1}$. Later we explore the idea of regularizing utilities with a function $S^{\tau}_i$ (e.g., entropy) as follows:
\begin{align}
    u^{\tau}_i(\boldsymbol{x}) &= u_i(\boldsymbol{x}) + S^{\tau}_i(x_i, x_{-i}).
\end{align}

As an abuse of notation, let the atomic action $a_{i}$ also denote the $m_i$-dimensional ``one-hot" vector with all zeros aside from a $1$ at index $a_{i}$; its use should be clear from the context. And for convenience, denote by $H^i_{il} = \mathbb{E}_{x_{-il}}[u_i(a_i, a_l, x_{-il})]$ the Jacobian\footnote{See Appx.~\ref{appx:nfg_grad} for an example derivation of the gradient if this form is unfamiliar.} of player $i$'s utility with respect to $x_i$ and $x_l$; $x_{-il}$ denotes all strategies belonging to players other than $i$ and $l$ and $u_i(a_i, a_l, x_{-il})$ separates out $l$'s strategy $x_l$ from the rest of the players $x_{-i}$. We also introduce $\nabla^i_{x_i}$ as player $i$'s utility gradient. Note player $i$'s utility can now be written succinctly as $u_i(x_i, x_{-i}) = x_i^\top \nabla^i_{x_i} = x_i^\top H^i_{il} x_l$ for any $l$.

In a polymatrix game, interactions between players are limited to local, pairwise games, each of which is represented by matrices $H^i_{ij}$ and $H^j_{ij}$. This reduces the exponential $nm^n$ payoffs required to represent a general normal form game to a quadratic $n(n-1)m^2$, an efficiency we leverage later.



\subsection{Related work}

Several approaches exist for computing Nash equilibria of $n$-player normal form games\footnote{Note that Double-Oracle~\citep{mcmahan2003planning} and PSRO~\citep{lanctot2017unified} can be extended to n-player games, but require an n-player normal form meta-solver (Nash-solver) and so cannot be considered solvers in their own right. This work provides an approximate meta-solver.}. Simplicial Subdivision (SD)~\citep{van1987simplicial} searches for an equilibrium over discretized simplices; accuracy depends on the grid size which scales exponentially with the number of player actions. \citet{govindan2003global} propose a homotopy method (GW) that begins with the unique Nash distribution of a game whose payoff tensor has been perturbed by an arbitrary constant tensor. GW then scales back this perturbation while updating the Nash to that of the transformed game. GW is considered an extension of the classic Lemke-Howson algorithm (\citeyear{lemke1964equilibrium}) to $3$+ player games (see \S4.3, p. 107 of ~\citep{shoham2008multiagent}).
\begin{figure}
    \centering
    \includegraphics[width=0.5\textwidth]{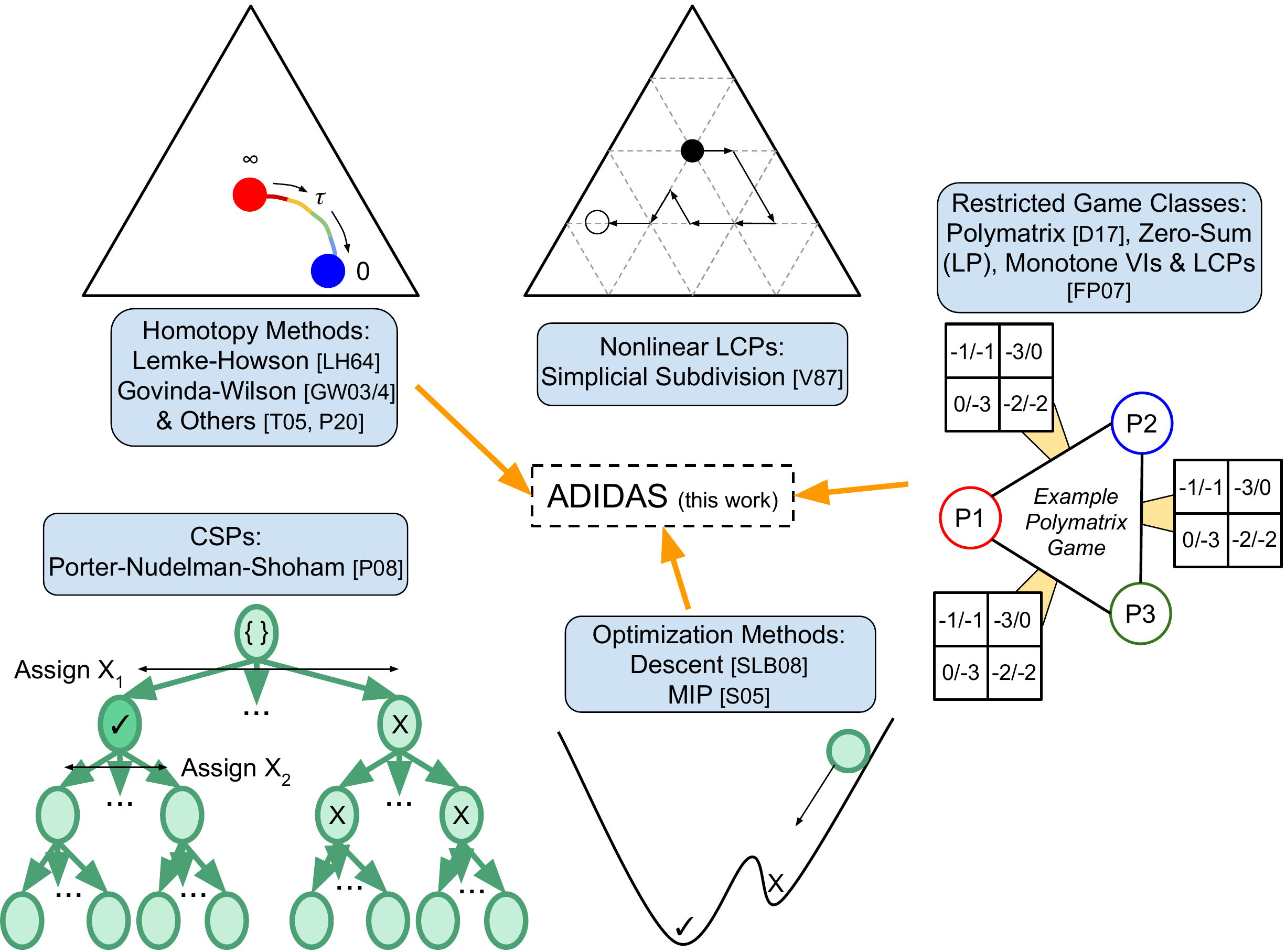}
    \caption{Algorithm Comparison and Overview.}
    \label{fig:algcomp}
\end{figure}
Another homotopy approach perturbs the payoffs with entropy bonuses, and evolves the Nash distribution along a continuum of quantal response equilibria (QREs) using a predictor-corrector method to integrate a differential equation~\citep{turocy2005dynamic} \textemdash we also aim to follow this same continuum. In a slightly different approach,~\citet{perolat2020poincar} propose an adaptive regularization scheme that repeatedly solves for the equilibrium of a transformed game. Simple search methods~\citep{porter2008simple} that approach Nash computation as a constraint satisfaction problem appear to scale better than GW and SD as measured on GAMUT benchmarks~\citep{nudelman2004run}. Lyapunov approaches minimize non-convex energy functions with the property that zero energy implies Nash~\citep{shoham2008multiagent}, however these approaches may suffer from convergence to local minima with positive energy. In some settings, such as polymatrix games with payoffs in $[0, 1]$, gradient descent on appropriate energy functions\footnote{\Eqref{popexp} but with $\max$ instead of $\sum$ over player regrets. Note that for symmetric games with symmetric equilibria, these are equivalent up to a multiplicative factor $n$.} guarantees a $(\frac{1}{2} + \delta)$-Nash in time polynomial in $\frac{1}{\delta}$~\citep{deligkas2017computing} and performs well in practice~\citep{deligkas2016empirical}.

\paragraph{Teaser} Our proposed algorithm consists of two key conceptual schemes. One lies at the crux of homotopy methods (see Figures~\ref{fig:algcomp} and~\ref{fig:homotopy}). We initialize the Nash approximation, $\boldsymbol{x}$, to the joint uniform distribution, the unique Nash of a game with infinite-temperature entropy regularization. The temperature is then annealed over time. To recover the Nash at each temperature, we minimize an appropriately adapted energy function via (biased) stochastic gradient descent. This minimization approach can be seen as simultaneously learning a suitable polymatrix decomposition of the game similarly to~\citet{govindan2004computing} but from batches of stochastic play, i.e., we compute Monte Carlo estimates of the payoffs in the bimatrix game between every pair of players by observing the outcomes of the players' joint actions (sampled from $\boldsymbol{x}$ after each update) rather than computing payoffs as exact expectations.
\nocite{facchinei2007finite}
\nocite{sandholm2005mixed}
\nocite{blum2006continuation}
\section{Deviation Incentive \& Warm-Up}


We propose minimizing the energy function in~\eqref{popexp} below, \emph{average deviation incentive} (ADI), to approximate a Nash equilibrium of a large, entropy-regularized normal form game. This loss measures, on average, how much a single agent can exploit the rest of the population by deviating from a joint strategy $\boldsymbol{x}$. For sake of exposition, we drop the normalizing constant from the denominator (number of players, $n$), and consider the sum instead of the average. This quantity functions as a \emph{loss} that can be minimized over $\mathcal{X} = \prod_i \Delta^{m_i-1}$ to find a Nash distribution. Note that when ADI is zero, $\boldsymbol{x}$ is a Nash. Also, if $\sum_k$ is replaced by $\max_k$, this loss measures the $\epsilon$ of an $\epsilon$-Nash, and therefore,~\eqref{popexp} is an upper bound on this $\epsilon$.  Lastly, note that, in general, this loss function is non-convex and so convergence to local, suboptimal minima is theoretically possible if naively minimizing via first order methods like gradient descent \textemdash we explain in \S\ref{annealing} how we circumvent this pitfall via temperature annealing. Let $\br_k = \br(x_{-k}) = \argmax_{z_k \in \Delta^{m_k-1}} u^{\tau}_k(z_k, x_{-k})$ be player $k$'s best response to all other players' current strategies where $u^\tau_k$ is player $k$'s utility regularized by entropy with temperature $\tau$ and formally define
\begin{align}
    \mathcal{L}^{\tau}_{adi}(\boldsymbol{x}) &= \sum_k \overbrace{u^{\tau}_k(\br_k, x_{-k}) - u^{\tau}_k(x_k,x_{-k})}^{\text{incentive to deviate to $\br_k$ vs $\boldsymbol{x}_k$}}. \label{popexp}
\end{align}

If $\tau=0$, we drop the superscript and use $\mathcal{L}_{adi}$. The Nash equilibrium of the game regularized with Shannon entropy is called a \emph{quantal response equilibrium}, QRE($\tau$) (see p. 152-154, 343 of~\cite{fudenberg1998theory}).

Average deviation incentive has been interpreted as a pseudo-distance from Nash in prior work, where it is referred to as \texttt{NashConv} \citep{lanctot2017unified}. We prefer average deviation incentive because it more precisely describes the function and allows room for exploring alternative losses in future work. The objective can be decomposed into terms that depend on $x_k$ (second term) and $x_{-k}$ (both terms). Minimizing the second term w.r.t. $x_k$ seeks strategies with high utility, while minimizing both terms w.r.t. $x_{-k}$ seeks strategies that cannot be exploited by player $k$. In reducing $\mathcal{L}_{adi}$, each player $k$ seeks a strategy that not only increases their payoff but also removes others' temptation to exploit them.

A related algorithm is Exploitability Descent (ED)~\citep{Lockhart19ED}.
Rather than minimizing $\mathcal{L}_{adi}$, each player independently maximizes their utility assuming the other players play their best responses. In the two-player normal-form setting, ED is equivalent to extragradient~\citep{korpelevich1976extragradient}~(see Appx.~\ref{app:ed_connection}). However, ED is only guaranteed to converge to Nash in two-player, zero-sum games. We include a comparison against ED as well as Fictitious-play, another popular multiagent algorithm, in Appx.~\ref{app:xtra_comparisons}. We also relate $\mathcal{L}_{adi}$ to Consensus optimization~\citep{mescheder2017numerics} in Appx.~\ref{app:consensus_connection}.



\subsection{Warm-Up}
\label{annealing}
\citet{mckelvey1995quantal} proved the existence of a continuum of QREs starting at the uniform distribution (infinite temperature) and ending at what they called the \emph{limiting logit equilibrium} (LLE). Furthermore, they showed this path is unique for \emph{almost all games}, partially circumventing the equilibrium selection problem. We encourage the reader to look ahead at Figure~\ref{fig:homotopy} for a visual of the homotopy that may prove helpful for the ensuing discussions.

In this work, we assume we are given one of these common games with a unique path (no branching points) so that the LLE is well defined (\textbf{Assumption~\ref{nobranches}}). Furthermore, we assume there exist no ``turning points'' in the temperature $\tau$ along the continuum (\textbf{Assumption~\ref{noturning}}). \citet{turocy2005dynamic} explains that even in generic games, temperature might have to be temporarily increased in order to remain on the path (principal branch) to the LLE. However, \citeauthor{turocy2005dynamic} also proves there exists a $\tau^*$ such that no turning points exist with $\tau > \tau^*$ suggesting that as long as we remain near the principal branch after $\tau^*$, we can expect to proceed to the LLE.

We follow the principal path by alternating between annealing the temperature and re-solving for the Nash at that temperature by minimizing $\mathcal{L}^{\tau}_{adi}$. We present a basic version of our approach that converges to the limiting logit equilibrium assuming access to exact gradients in Algorithm~\ref{alg_warmup} (proof in Appx.~\ref{app:conv}). We substitute $\lambda=\frac{1}{\tau}$ and initialize $\lambda=0$ in order to begin at infinite temperature. The proof of this simple warm-up algorithm relies on the detailed examination of the continuum of QREs proposed in~\cite{mckelvey1995quantal} and further analyzed in~\cite{turocy2005dynamic}. Theorem~\ref{conv_basic} presented below is essentially a succinct repetition of one of their known results (Assumptions~\ref{sensitivity} and~\ref{boa} below are expanded on in Appx.~\ref{app:conv}). In subsequent sections, we relax the exact gradient assumption and assume gradients are estimated from stochasic play (i.e., each agent samples an action from their side of the current approximation to the Nash).

\begin{algorithm}[H]
\begin{algorithmic}[1]
    \STATE Given: Total anneal steps $T_{\lambda}$, total optimizer iterations $T^*$, and anneal step size $\Delta \lambda$.
    \STATE $\lambda = 0$
    \STATE $\boldsymbol{x} \leftarrow \{ \frac{1}{m_i} \boldsymbol{1} \,\, \forall \,\, i\}$
    \FOR{$t_{\lambda} = 1: T_{\lambda}$}
        \STATE $\lambda \leftarrow \lambda + \Delta \lambda$ \label{line:anneal}
        \STATE $\boldsymbol{x} \leftarrow \texttt{OPT}(\texttt{loss} = \mathcal{L}^{\tau=\lambda^{-1}}_{adi}, \boldsymbol{x}_{init} = \boldsymbol{x}, iters = T^*)$ \label{line:descend}
    \ENDFOR
    \STATE return $\boldsymbol{x}$
\end{algorithmic}
\caption{Warm-up: Anneal \& Descend}
\label{alg_warmup}
\end{algorithm}

\begin{theorem}
\label{conv_basic}
Make assumptions~\ref{nobranches} and~\ref{noturning}. Also, assume the QREs along the homotopy path have bounded sensitivity to $\lambda$ given by a parameter $\sigma$ (Assumption~\ref{sensitivity}), and basins of attraction with radii lower bounded by $r$ (Assumption~\ref{boa}). Let the step size $\Delta \lambda \le \sigma (r - \epsilon)$ with tolerance $\epsilon$. And let $T^*$ be the supremum over all $T$ such that Assumption~\ref{boa} is satisfied for any inverse temperature $\lambda \ge \Delta \lambda$. Then, assuming gradient descent for \texttt{OPT}, \Algref{alg_warmup} converges to the limiting logit equilibrium $\boldsymbol{x}^*_{\lambda=\infty} = \boldsymbol{x}^*_{\tau=0}$ in the limit as $T_{\lambda} \rightarrow \infty$.
\end{theorem}

\subsection{Evaluating $\mathcal{L}^{\tau}_{adi}$ with Joint Play}
\label{bias_intuition}


In the warm up, we assumed we could compute exact gradients which required access to the entire payoff tensor. However, we want to solve very large games where enumerating the payoff tensor is prohibitively expensive. Therefore, we are particularly interested in minimizing $\mathcal{L}^{\tau}_{adi}$ when only given access to samples of joint play, $\boldsymbol{a} \sim \prod_i x_i$. The best response operator, $\br$, is nonlinear and hence can introduce bias if applied to random samples. For example, consider the game given in Table~\ref{tab:biased_game} and assume $x_2 = [0.5, 0.5]^\top$.

\begin{table}[!ht]
    \centering
    \begin{tabular}{c|c|c}
        $u_1$ & $a_{21}$ & $a_{22}$ \\ \hline
        $a_{11}$ & 0 & 0 \\
        $a_{12}$ & 1 & -2 \\
        $a_{13}$ & -2 & 1
    \end{tabular}
    \hspace{1.0cm}
    \begin{tabular}{c|c|c}
        $u_2$ & $a_{21}$ & $a_{22}$ \\ \hline
        $a_{11}$ & 0 & 0 \\
        $a_{12}$ & 0 & 0 \\
        $a_{13}$ & 0 & 0
    \end{tabular}
    \caption{A 2-player game with biased stochastic $\br$'s.}
    \label{tab:biased_game}
\end{table}
%
%
Consider computing (row) player $1$'s best response to a single action sampled from (column) player $2$'s strategy $x_2$. Either $a_{21}$ or $a_{22}$ will be sampled with equal probability, which results in a best response of either $a_{12}$ or $a_{13}$ respectively. However, the true expected utilities for each of player $1$'s actions given player $2$'s strategy are $[0, -0.5, -0.5]$ for which the best response is the first index, $a_{11}$. The best response operator completely filters out information on the utility of the true best response $a_{11}$. Intuitively, a \emph{soft} best response operator, demonstrated in equations (\ref{zero_temp})-(\ref{goldilocks}), that allows some utility information for each of the actions to pass through could alleviate the problem:
\begin{align}
    \mathbb{E}[\br{}^{\tau\rightarrow 0}] &= [0.00, 0.50, 0.50] \label{zero_temp}
    \\ \mathbb{E}[\br{}^{\tau=1}] &\approx [0.26, 0.37, 0.37]
    \\ \mathbb{E}[\br{}^{\tau=10}] &\approx [\mathbf{0.42}, 0.29, 0.29]. \label{goldilocks}
\end{align}
By adding an entropy regularizer to the utilities, $\tau \mathcal{H}(x_i)$, we induce a soft-$\br$. Therefore, the homotopy approach has the added benefit of partially alleviating gradient bias for moderate temperatures. Further empirical analysis of bias can be found in Appx.~\ref{appx:bias}.




\section{ADIDAS}

In the previous section, we laid out the conceptual approach we take and identified bias as a potential issue to scaling up computation with Monte Carlo approximation. Here, we inspect the details of our approach, introduce further modifications to reduce the issue of bias, and present our resulting algorithm ADIDAS. Finally, we discuss the advantages of our approach for scaling to large games.

\subsection{Deviation Incentive Gradient}
\label{ped_grads}



Regularizing the utilities with weighted Shannon entropy, $u^{\tau}_k(\boldsymbol{x}) = u_k(\boldsymbol{x}) + S^\tau_k(x_k, x_{-k})$, where $S^\tau_k(x_k, x_{-k}) = -\tau \sum_{a_k} x_{ka_k} \ln(x_{ka_k})$, leads to the following average deviation incentive gradient derived in Appx.~\ref{gen_pop_exp_grad} where $\br_j = \texttt{softmax}(\nabla^j_{x_j}/\tau)$ and $\texttt{diag}(v)$ creates a diagonal matrix with $v$ on the diagonal:
\begin{align}
    &\nabla_{x_i} \mathcal{L}^{\tau}_{adi}(\boldsymbol{x}) = -\overbrace{(\nabla^i_{x_i} - \tau (\ln(x_i) + 1))}^{\text{policy gradient}} \nonumber
    \\ &+ \sum_{j \ne i} \Big[ J_{x_i}(\br_j)^\top (\nabla^j_{x_j} - \tau (\ln(\br_j) + 1)) + H^j_{ij} (\br_j - x_j) \Big] \label{qre_grad}
    \\ &\text{with } J_{x_i}(\br_j) = \frac{1}{\tau} (\texttt{diag}(\br_j) - \br_j \br_j^\top) H^j_{ji}. \label{qre_jac}
\end{align}
%
In the limit, $\nabla_{x_i} \mathcal{L}^{\tau}_{adi}(\boldsymbol{x}) \stackrel{\tau\rightarrow 0^+}{=} -\nabla^i_{x_i} + \sum_{j \ne i} H^j_{ij} (\br_j - x_j)$. The first term is recognized as player $i$'s payoff or \emph{policy} gradient. The second term is a correction that accounts for the other players' incentives to exploit player $i$ through a strategy deviation.
%
Each $H^j_{ij}$ approximates player $j$'s payoffs in the bimatrix game between players $i$ and $j$. Recall from the preliminaries that in a polymatrix game, these matrices capture the game exactly. We also explore an adaptive Tsallis entropy in Appx.~\ref{gen_pop_exp_grad}.

\subsection{Amortized Estimates with Historical Play}
\label{amortized_estimates}


Section~\ref{bias_intuition} discusses the bias that can be introduced when best responding to sampled joint play and how the annealing process of the homotopy method helps alleviate it by softening the $\br$ operator with entropy regularization. To reduce the bias further, we could evaluate more samples from $\boldsymbol{x}$, however, this increases the required computation. Alternatively, assuming strategies have changed minimally over the last few updates (i.e., $\boldsymbol{x}^{(t-2)} \approx \boldsymbol{x}^{(t-1)} \approx \boldsymbol{x}^{(t)}$), we can instead reuse historical play to improve estimates. We accomplish this by introducing an auxiliary variable $y_i$ that computes an exponentially averaged estimate of each player $i$'s payoff gradient $\nabla^i_{x_i}$ throughout the descent similarly to~\citet{sutton2008convergent}.
%
We also use $y_i$ to compute an estimate of ADI, $\hat{\mathcal{L}}^{\tau}_{adi}$, as follows:
\begin{align}
    \hat{\mathcal{L}}^{\tau}_{adi}(\boldsymbol{x}, \boldsymbol{y}) &= \sum_k y_k^\top (\hat{\br}_k - x_k) +  S^\tau_k(\hat{\br}_k, x_{-k}) - S^\tau_k(x_k, x_{-k}) \label{amortized_exp_reg}
\end{align}
where $\hat{\br}_k = \argmax_{z_k \in \Delta^{m_k-1}} y_k^\top z_k + S^\tau_k(z_k, x_{-k})$ is computed with $y_k$ instead of $\nabla^k_{x_k}$.
Likewise, replace all $\nabla^k_{x_k}$ with $y_k$ and $\br_k$ with $\hat{\br}_k$ in equations (\ref{qre_grad}) and (\ref{qre_jac}) when computing the gradient:

\subsection{Putting It All Together}
\label{convergence}
\begin{algorithm}[H]
\begin{algorithmic}[1]
    \STATE Given: Strategy learning rate $\eta_x$, auxiliary learning rate $\eta_y$, initial temperature $\tau$ ($=100$), ADI threshold $\epsilon$, total iterations $T$, simulator $\mathcal{G}_i$ that returns player $i$'s payoff given a joint action.
    \STATE $\boldsymbol{x} \leftarrow \{ \frac{1}{m_i} \boldsymbol{1} \,\, \forall \,\, i\}$
    \STATE $\boldsymbol{y} \leftarrow \{ \boldsymbol{0} \,\, \forall \,\, i \}$
    \FOR{$t = 1: T$}
        \STATE $a_i \sim x_i \,\, \forall \,\, i$ \label{start}
        \FOR{$i \in \{1, \ldots, n\}$}
            \FOR{$j \ne i \in \{1, \ldots, n\}$}
                \STATE $H^i_{ij}[r,c] \leftarrow \mathcal{G}_i(r,c,a_{-ij}) \,\, \forall \,\, r \in \mathcal{A}_i, c \in \mathcal{A}_j$
            \ENDFOR
        \ENDFOR \label{end}
        \STATE $\nabla^i_{x_i} = H^i_{ij} x_j$ for any $x_j$ (or average the result over all $j$)
        \STATE $y_i \leftarrow y_i - \max(\frac{1}{t}, \eta_y) (\nabla^i_{x_i} - y_i)$
        \STATE $x_i \leftarrow x_i - \eta_x \nabla_{x_i} \hat{\mathcal{L}}^{\tau}_{adi}(\boldsymbol{x}, \boldsymbol{y})$ (def. in~\S\ref{amortized_estimates} and code in Appx.~\ref{app:code})
        \IF{$\hat{\mathcal{L}}^{\tau}_{adi}(\boldsymbol{x}, \boldsymbol{y}) < \epsilon$ (def. in~\eqref{amortized_exp_reg})}
            \STATE $\tau \leftarrow \frac{\tau}{2}$ \label{anneal}
        \ENDIF
    \ENDFOR
    \STATE return $\boldsymbol{x}$
\end{algorithmic}
\caption{ADIDAS}
\label{alg_saped}
\end{algorithm}

\begin{figure}
    \centering
    \begin{subfigure}[b]{.49\textwidth}
        \centering
        \includegraphics[width=0.5\textwidth]{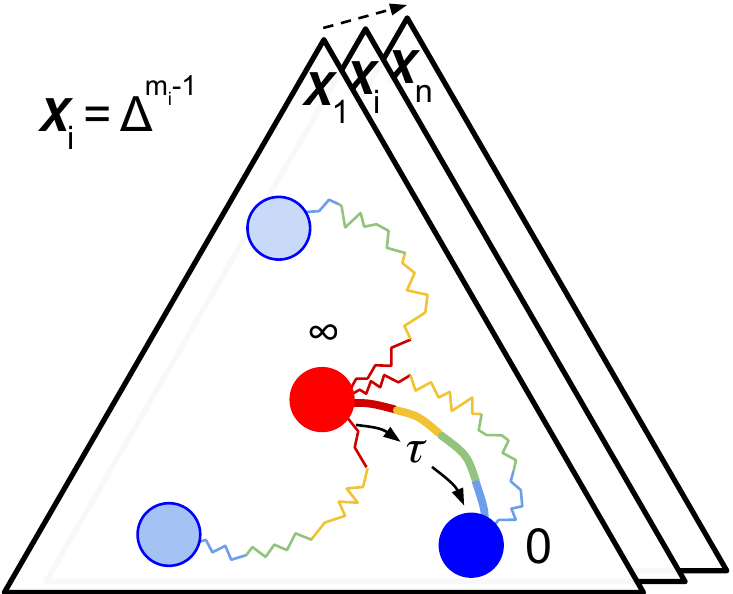}
        \caption{ADIDAS pathologies \label{fig:manynasherror}}
    \end{subfigure}
    \begin{subfigure}[b]{.49\textwidth}
        \centering
        \includegraphics[width=0.55\textwidth]{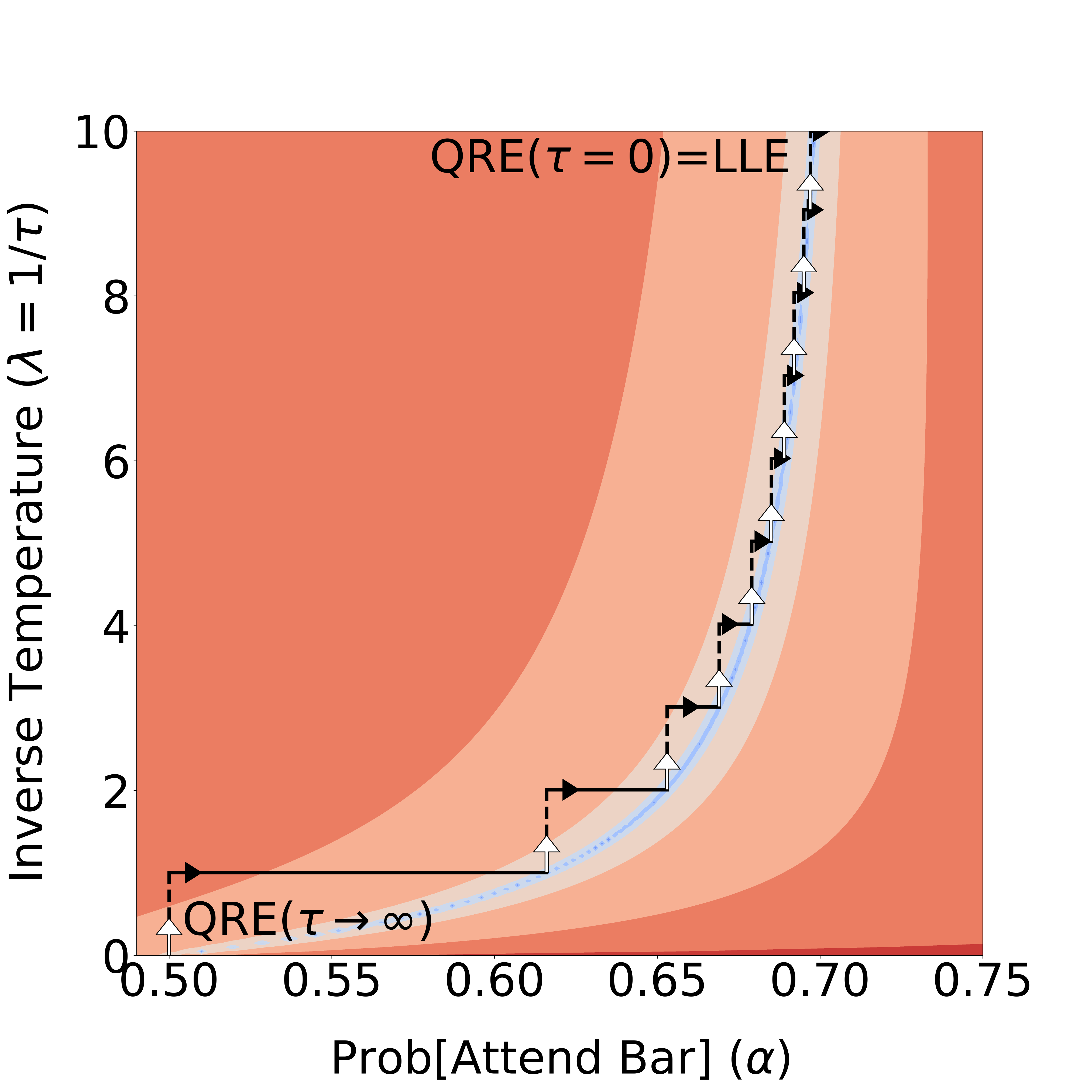}
        \caption{$10$-player, 2-action El Farol homotopy \label{fig:el_farol_homotopy}}
    \end{subfigure}
    \caption{(\subref{fig:manynasherror}) In the presence of multiple equilibria, ADIDAS may fail to follow the path to the uniquely defined Nash due to gradient noise, gradient bias, and a coarse annealing schedule. If these issues are severe, they can cause the algorithm to get stuck at a local optimum of $\mathcal{L}^{\tau}_{adi}$\textemdash see Figure~\ref{fig:dipmedexp_ate} in \S\ref{large_scale}. (\subref{fig:el_farol_homotopy}) Such concerns are minimal for the El Farol Bar stage game by~\citet{arthur1994complexity}. The solid black curves represent (biased) descent trajectories while the dashed segments indicate the temperature is being annealed.}
    \label{fig:homotopy}
\end{figure}

Algorithm~\ref{alg_saped}, ADIDAS, is our final algorithm.
ADIDAS attempts to approximate the unique continuum of quantal response equilibria by way of a quasi-stationary process\textemdash see Figure~\ref{fig:homotopy}. Whenever the algorithm finds a joint strategy $\boldsymbol{x}$ exhibiting $\hat{\mathcal{L}}^{\tau}_{adi}$ below a threshold $\epsilon$ for the game regularized with temperature $\tau$, the temperature is exponentially reduced (line~\ref{anneal} of ADIDAS) as suggested in~\citep{turocy2005dynamic}. Incorporating stochastic optimization into the process enables scaling the classical homotopy approach to extremely large games (large payoff tensors). At the same time, the homotopy approach selects a unique limiting equilibrium and, symbiotically, helps alleviate gradient bias, further amortized by the reuse of historical play.

\paragraph{Limitations:} As mentioned earlier, gradient bias precludes a rigorous convergence proof of ADIDAS. However, recent work showed that gradient estimators that are biased, but consistent worked well empirically~\citep{chen2018fastgcn} and follow-up analysis suggests consistency may be an important property~\citep{chen2018stochastic}. Bias is also being explored in the more complex Riemannian optimization setting where it has been proven that the amount of bias in the gradient shifts the stationary point by a proportional amount~\citep{durmus2020convergence}. Note that ADIDAS gradients are also consistent in the limit of infinite samples of joint play, and we also find that biased stochastic gradient descent maintains an adequate level of performance for the purpose of our experiments.

No-regret algorithms scale, but have been proven not to converge to Nash~\citep{flokas2020no} and classical solvers~\citep{mckelvey2014gambit} converge to Nash, but do not scale. ADIDAS suffers from gradient bias, an issue that may be further mitigated by future research. In this sense, ADIDAS is one of the few, if only, algorithms that can practically approximate Nash in many-player, many-action normal-form games.

\begin{table}[ht!]
    \centering
    \begin{tabular}{c|c|c|c}
        Alg Family & Classical & No-Regret & This Work \\ \hline \hline
        Convergence to Nash & Yes & No & Yes$^\dag$ \\ \hline
        Payoffs Queried & $nm^n$ & $Tnm^{\ddag}$ & $T(nm)^2$
    \end{tabular}
    \caption{Comparison of solvers. $^\dag$See \emph{Limitations} in \S\ref{convergence} and Appx.~\ref{app:conv:sto}. $^\ddag$Reduce to $T$ at the expense of higher variance.}
    \label{tab:nealg_comp}
\end{table}

\subsection{Complexity and Savings}
\label{scale}


A normal form game may also be represented with a tensor $U$ in which each entry $U[i, a_1, \ldots, a_n]$ specifies the payoff for player $i$ under the joint action $(a_1, \ldots, a_n)$. In order to demonstrate the computational savings of our approach, we evaluate the ratio of the number of entries in $U$ to the number of entries queried (in the sense of~\citep{babichenko2016query,fearnley2015learning,fearnley2016finding}) for computing a single gradient, $\nabla \mathcal{L}^{\tau}_{adi}$. This ratio represents the number of steps that a gradient method can take before it is possible to compute $\mathcal{L}^{\tau}_{adi}$ exactly in expectation.

Without further assumptions on the game, the number of entries in a general payoff tensor is $nm^n$. In contrast, computing the stochastic deviation incentive gradient requires computing $H_{ij}^j$ for all $i, j$ requiring less than $(nm)^2$ entries\footnote{Recall $\nabla^i_{x_i}$ can be computed with $\nabla^i_{x_i} = H^i_{ij} x_j$ for any $x_j$.}. The resulting ratio is $\frac{1}{n} m^{n-2}$.
%
%
For a $7$-player, $21$-action game, this implies at least $580,000$ descent updates can be used by stochastic gradient descent.

If the game is symmetric and we desire a symmetric Nash, the payoff tensor can be represented more concisely with $\frac{(m + n - 1)!}{n! (m - 1)!}$ entries (number of multisets of cardinality $n$ with elements taken from a finite set of cardinality $m$). The number of entries required for a stochastic gradient is less than $m^2$. Again, for a $7$-player $21$-action game, this implies at least $2,000$ update steps. Although there are fewer unique entries in a symmetric game, we are not aware of libraries that allow sparse storage of or efficient arithmetic on such permutation-invariant tensors. ADIDAS can exploit this symmetry.




\section{Experiments}
\label{exp}

We test the performance of ADIDAS empirically on very large games. We begin by considering Colonel Blotto, a deceptively complex challenge domain still under intense research~\citep{behnezhad2017faster,boix2020multiplayer}, implemented in OpenSpiel~\citep{LanctotEtAl2019OpenSpiel}. For reference, both the 3 and 4-player variants we consider are an order of magnitude ($>20\times$) larger than the largest games explored in~\citep{porter2008simple}. We find that no-regret approaches as well as existing methods from Gambit~\citep{mckelvey2014gambit} begin to fail at this scale, whereas ADIDAS performs consistently well. At the same time, we empirically validate our design choice regarding amortizing gradient estimates (\S\ref{amortized_estimates}). Finally, we end with our most challenging experiment, the approximation of a unique Nash of a 7-player, 21-action (> billion outcome) Diplomacy meta-game.

We use the following notation to indicate variants of the algorithms compared in Table~\ref{tab:competingalgs}. A $y$ superscript prefix, e.g., $^y$\texttt{QRE}, indicates the estimates of payoff gradients are amortized using historical play; its absence indicates that a fresh estimate is used instead. $\bar{x}_t$ indicates that the average deviation incentive reported is for the average of $\boldsymbol{x}^{(t)}$ over learning. A subscript of ${\infty}$ indicates best responses are computed with respect to the true expected payoff gradient (infinite samples). A superscript $auto$ indicates the temperature $\tau$ is annealed according to line~\ref{anneal} of~\Algref{alg_saped}. An $s$ in parentheses indicates lines~\ref{start}-\ref{end} of ADIDAS are repeated $s$ times, and the resulting $H^i_{ij}$'s are averaged for a more accurate estimate. Each game is solved on 1 CPU, except Diplomacy (see Appx.~\ref{app:runtime}).

\begin{table}[ht!]
    \begin{tabular}{l|l}
        \texttt{FTRL} & Simultaneous Gradient Ascent \\
        \texttt{RM} & Regret-Matching~\citep{blackwell1956analog} \\
        \texttt{ATE} & ADIDAS with Tsallis (Appx.~\ref{ate_exp}) \\
        \texttt{QRE} & ADIDAS with Shannon
    \end{tabular}
    \caption{Algorithms}
    \label{tab:competingalgs}
\end{table}
\vspace{-3em}
\begin{table}[ht!]
    \begin{tabular}{r|l}
        $\eta_x$ & $10^{-5}, 10^{-4}, 10^{-3}, 10^{-2}, 10^{-1}$ \\
        $\eta_x^{-1} \cdot \eta_y$ & $1, 10, 100$ \\
        $\tau$ & $0.0, 0.01, 0.05, 0.10$ \\
        $\Pi_{\Delta}(\nabla \mathcal{L}_{adi})$ & Boolean \\
        Bregman-$\psi(\boldsymbol{x})$ & $\{\frac{1}{2}||\boldsymbol{x}||^2, -\mathcal{H}(\boldsymbol{x})\}$ \\
        $\epsilon$ & $0.01, 0.05$
    \end{tabular}
    \caption{Hyperparameter Sweeps}
    \label{tab:hypsweeps}
\end{table}


Sweeps are conducted over whether to project gradients onto the simplex ($\Pi_{\Delta}(\nabla \mathcal{L}_{adi})$), whether to use a Euclidean projection or entropic mirror descent~\citep{beck2003mirror} to constrain iterates to the simplex, and over learning rates. Averages over $10$ runs of the best hyperparameters are then presented\footnote{Best hyperparameters are used because we expect running ADIDAS with multiple hyperparameter settings in parallel to be a pragmatic approach to approximating Nash.} except for Diplomacy for which we present all settings attempted (more in Appx.~\ref{app:more_dip}). Performance is measured by $\mathcal{L}_{adi}$, a.k.a. \texttt{NashConv}~\citep{lanctot2017unified}. For symmetric games, we enforce searching for a symmetric equilibrium (see Appx.~\ref{app:sym}).

\begin{figure}[!ht]
    \centering
    \begin{subfigure}[b]{.4\textwidth}
    \centering
    \includegraphics[width=0.8\textwidth]{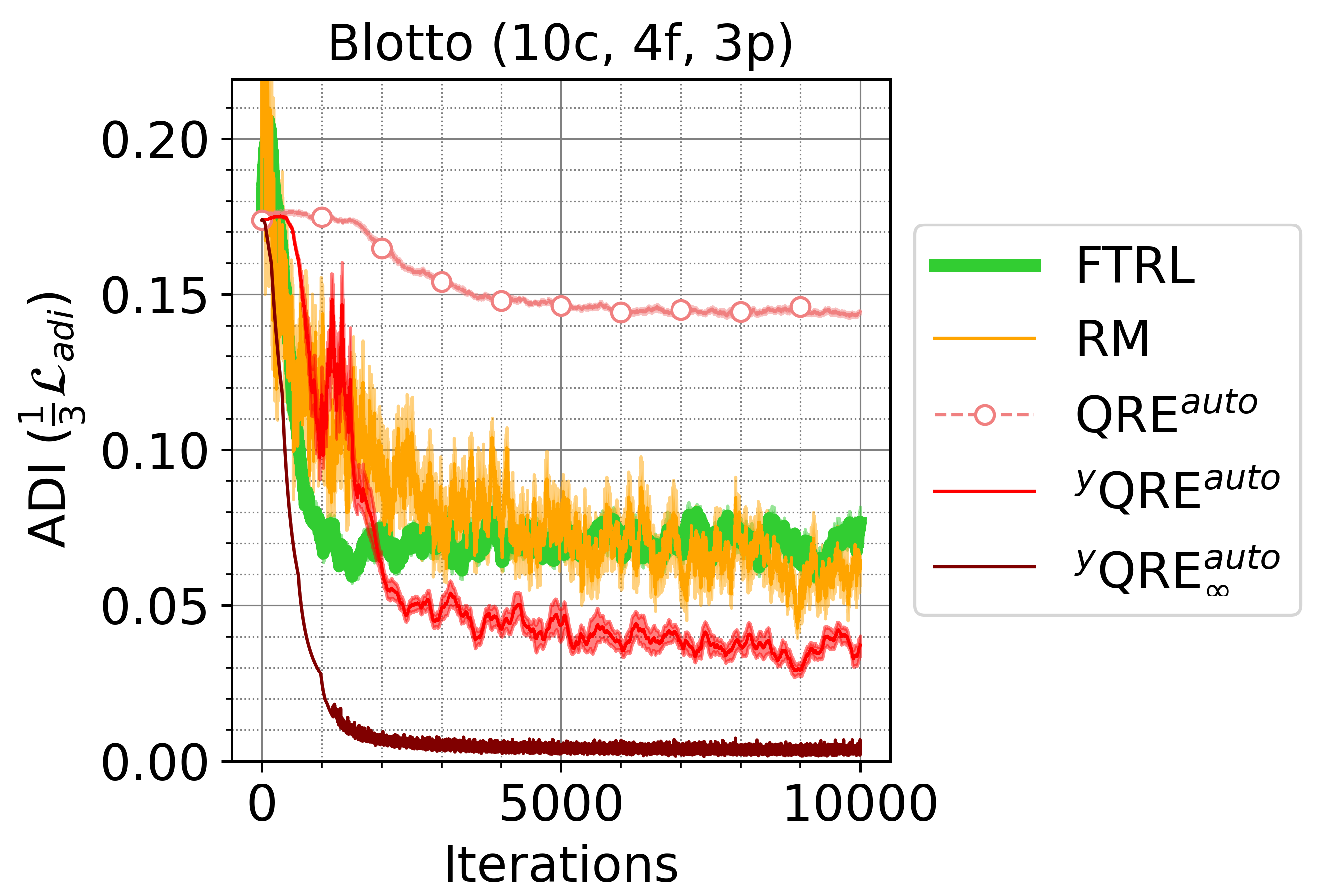}
    \caption{3-player, 286-action Blotto \label{fig:blotto3track_qre}}
    \end{subfigure}
    \begin{subfigure}[b]{.4\textwidth}
    \centering
    \includegraphics[width=0.8\textwidth]{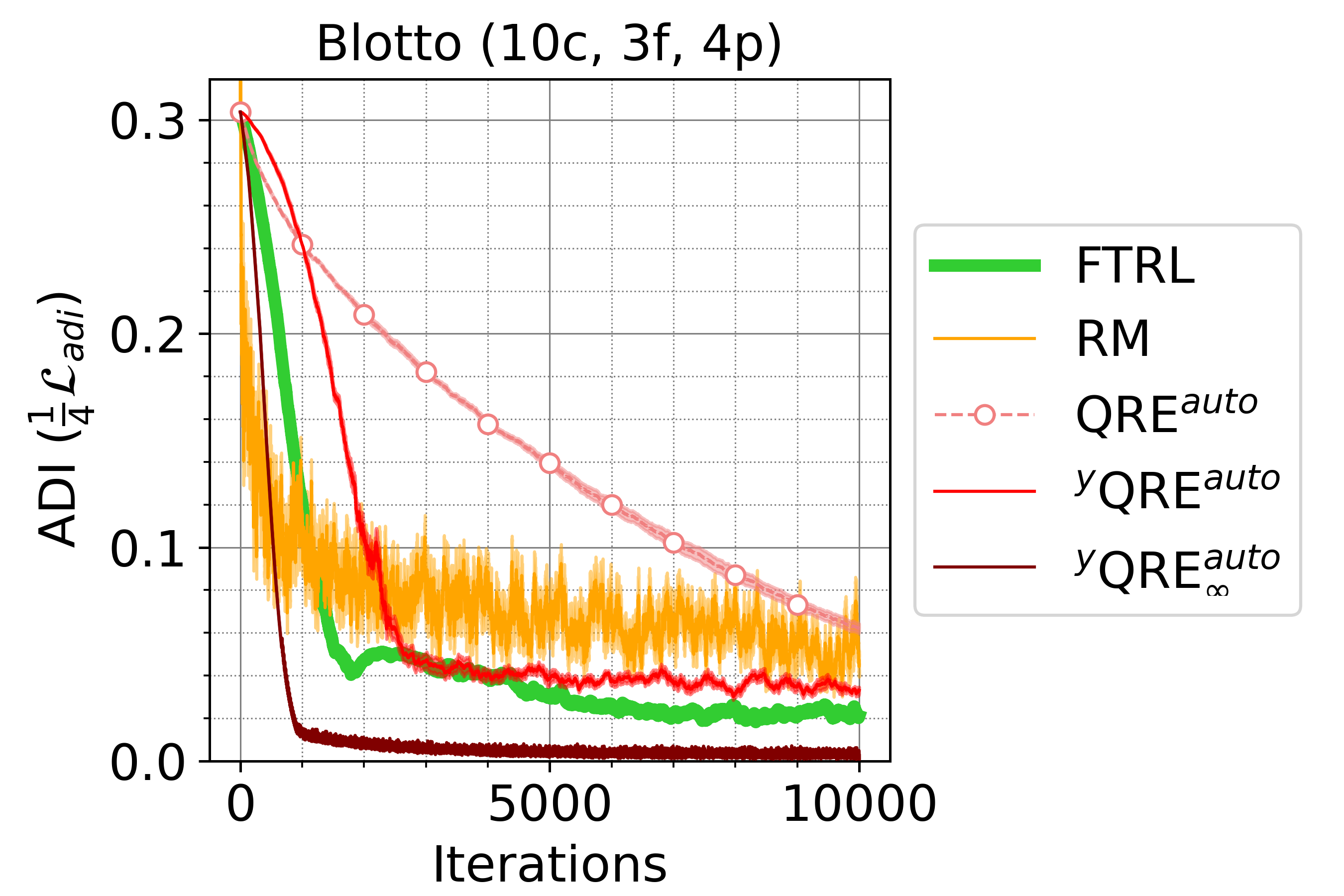}
    \caption{4-player, 66-action Blotto \label{fig:blotto4track_qre}}
    \end{subfigure}
    \vspace{-5pt}
    \caption{Amortizing estimates of joint play using $\boldsymbol{y}$ can reduce gradient bias, further improving performance (e.g., compare \texttt{QRE}$^{auto}$ to $^y$\texttt{QRE}$^{auto}$ in (\subref{fig:blotto3track_qre}) or (\subref{fig:blotto4track_qre})).}
    \label{fig:blotto_tracking_qre}
\end{figure}

For sake of exposition, we do not present all baselines in all plots, however, we include the full suite of comparisons in the appendix. Our experiments demonstrate that without any additional prior information on the game, ADIDAS is the only practical approach for approximating a Nash equilibrium over many-players and many-actions. We argue this by systematically ruling out other approaches on a range of domains. For example, in Figure~\ref{fig:blotto_tracking_qre}, \texttt{RM} reduces ADI adequately in Blotto. We do not present \texttt{RM} with improvements in Figure~\ref{fig:blotto_tracking_qre} such as using exact expectations, \texttt{RM}$_{\infty}$, or averaging its iterates, \texttt{RM}$(\bar{x}_t)$, because we show that both these fail to save \texttt{RM} on the GAMUT game in Figure~\ref{fig:noregfail}. In other words, we do not present baselines that are unnecessary for logically supporting the claim above. Code is available at \href{https://github.com/deepmind/open_spiel}{github.com/deepmind/open\_spiel}~\cite{LanctotEtAl2019OpenSpiel}.

\subsection{Med-Scale re. \S\ref{scale}}
\label{med_scale}
Govindan-Wilson is considered a state-of-the-art Nash solver, but it does not scale well to large games. For example, on a symmetric, $4$-player Blotto game with $66$ actions ($10$ coins, $3$ fields), GW, as implemented in Gambit, is estimated to take 53,000 hours\footnote{Public correspondence with primary \texttt{gambit} developer [\href{https://github.com/gambitproject/gambit/issues/261\#issuecomment-660894391}{link}].}. Of the solvers implemented in Gambit, none finds a symmetric Nash equilibrium within an hour\footnote{\texttt{gambit-enumpoly} returns several non-symmetric, pure Nash equilibria. Solvers listed in Appx.~\ref{app:gambit_solvers}. Symmetric equilibria are necessary for ranking in symmetric meta-games.}. Of those, \texttt{gambit-logit}~\citep{turocy2005dynamic} is expected to scale most gracefully. Experiments from the original paper are run on maximum $5$-player games ($2$-actions per player) and $20$-action games ($2$-players), so the $4$-player, $66$-action game is well outside the original design scope. Attempting to run $\texttt{gambit-logit}$ anyways with a temperature $\tau=1$ returns an approximate Nash with $\mathcal{L}_{adi}=0.066$ after $101$ minutes. In contrast, Figure~\ref{fig:blotto4track_qre} shows ADIDAS achieves a lower ADI in $\approx 3$ minutes.


\begin{figure}[!ht]
    \centering
    \begin{subfigure}[b]{.45\textwidth}
    \centering
    \includegraphics[width=0.8\textwidth]{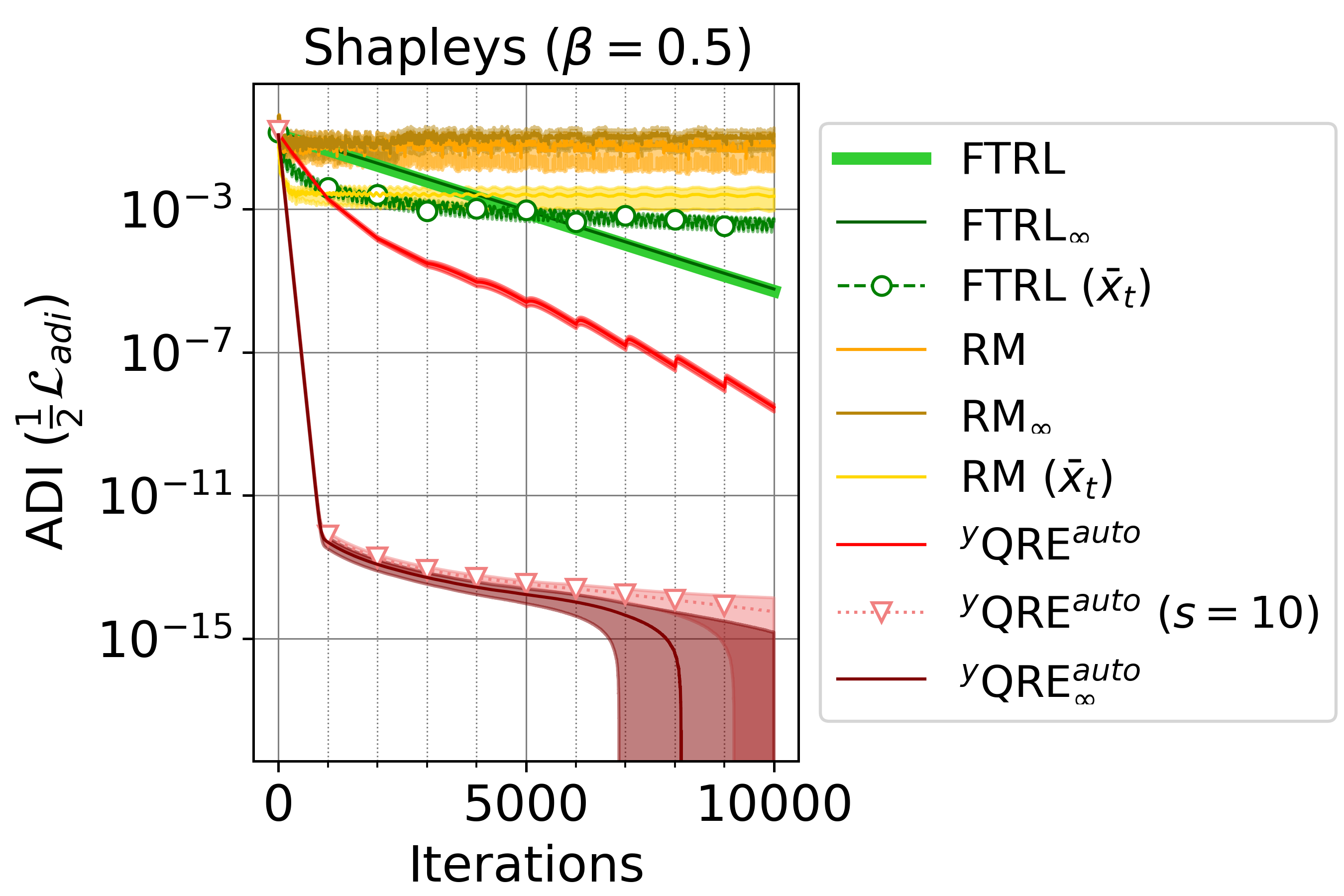}
    \caption{2-player, 3-action modified Shapley's \label{fig:modshapley_qre}}
    \end{subfigure}
    \begin{subfigure}[b]{.45\textwidth}
    \centering
    \includegraphics[width=0.8\textwidth]{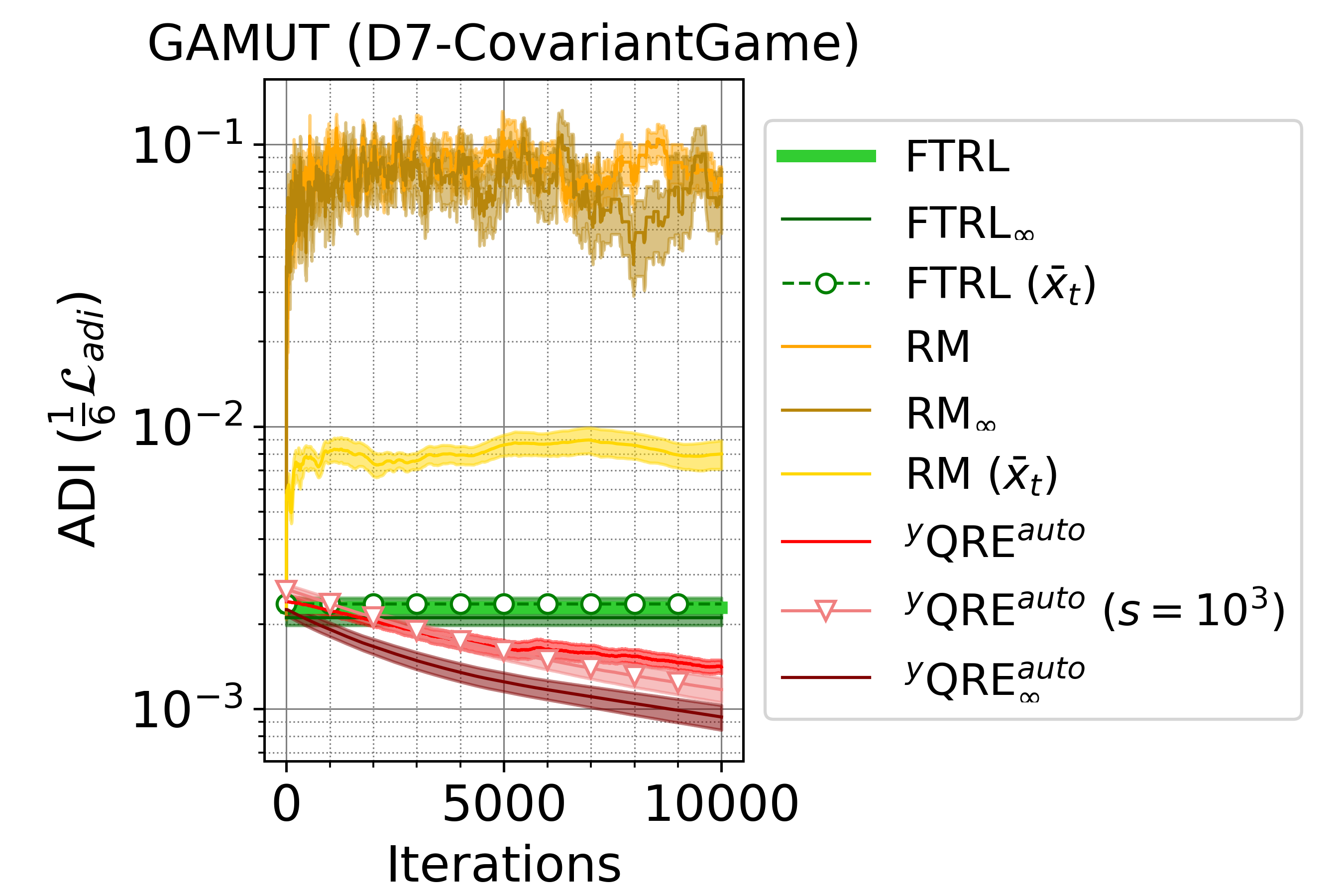}
    \caption{6-player, 5-action GAMUT-D7 \label{fig:gamutd7_qre}}
    \end{subfigure}
    \vspace{-5pt}
    \caption{ADIDAS reduces $\mathcal{L}_{adi}$ in both these nonsymmetric games. In contrast, regret matching stalls or diverges in game~(\subref{fig:modshapley_qre}) and diverges in game~(\subref{fig:gamutd7_qre}). \texttt{FTRL} makes progress in game~(\subref{fig:modshapley_qre}) but stalls in game~(\subref{fig:gamutd7_qre}). In game~(\subref{fig:modshapley_qre}), created by~\protect\citet{ostrovski2013payoff}, to better test performance, $\boldsymbol{x}$ is initialized randomly rather than with the uniform distribution because the Nash is at uniform.}
    \label{fig:noregfail}
\end{figure}

\paragraph{Auxiliary $y$ re. \S\ref{amortized_estimates}} The introduction of auxiliary variables $y_i$ are supported by the results in Figure~\ref{fig:blotto_tracking_qre}\textemdash $^y$\texttt{QRE}$^{auto}$ significantly improves performance over \texttt{QRE}$^{auto}$ and with low algorithmic cost.

\paragraph{No-regret, No-convergence re.~\S\ref{convergence}} In Figure~\ref{fig:blotto_tracking_qre}, \texttt{FTRL} and \texttt{RM} achieve low ADI quickly in some cases. \texttt{FTRL} has recently been proven not to converge to Nash, and this is suggested to be true of no-regret algorithms in general~\citep{flokas2020no,mertikopoulos2018cycles}. Before proceeding, we demonstrate empirically in Figure~\ref{fig:noregfail} that \texttt{FTRL} and \texttt{RM} fail on games where ADIDAS significantly reduces ADI. Note that GAMUT (D7) was highlighted as a particularly challenging problem for Nash solvers in~\citep{porter2008simple}.

\begin{figure}[!ht]
    \centering
    \begin{subfigure}[b]{.45\textwidth}
    \centering
    \includegraphics[width=0.75\textwidth]{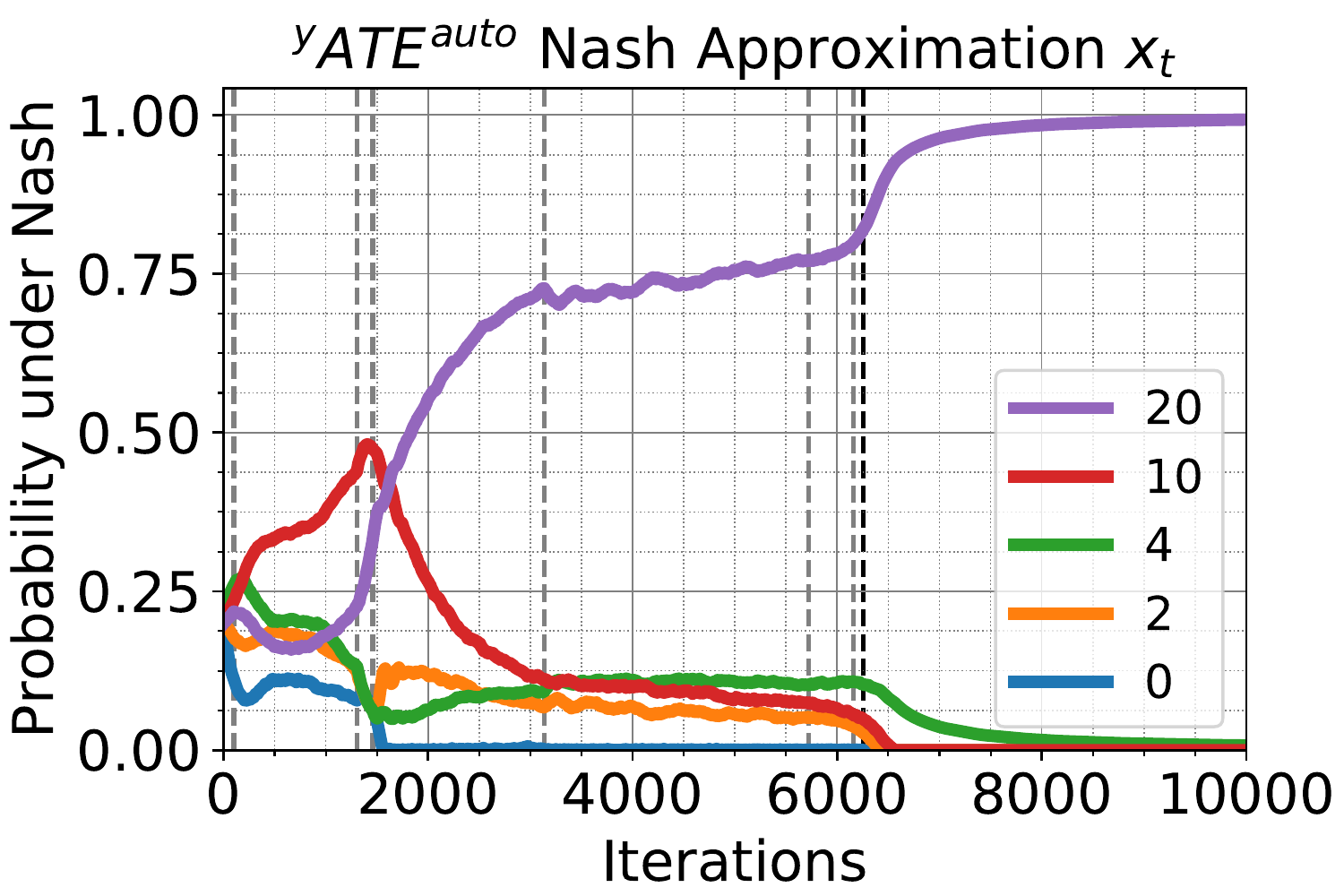}
    \caption{7-player, 5-action symmetric Nash ($x_t$) \label{fig:dipmednash_ate}}
    \end{subfigure}
    \begin{subfigure}[b]{.45\textwidth}
    \centering
    \includegraphics[width=0.75\textwidth]{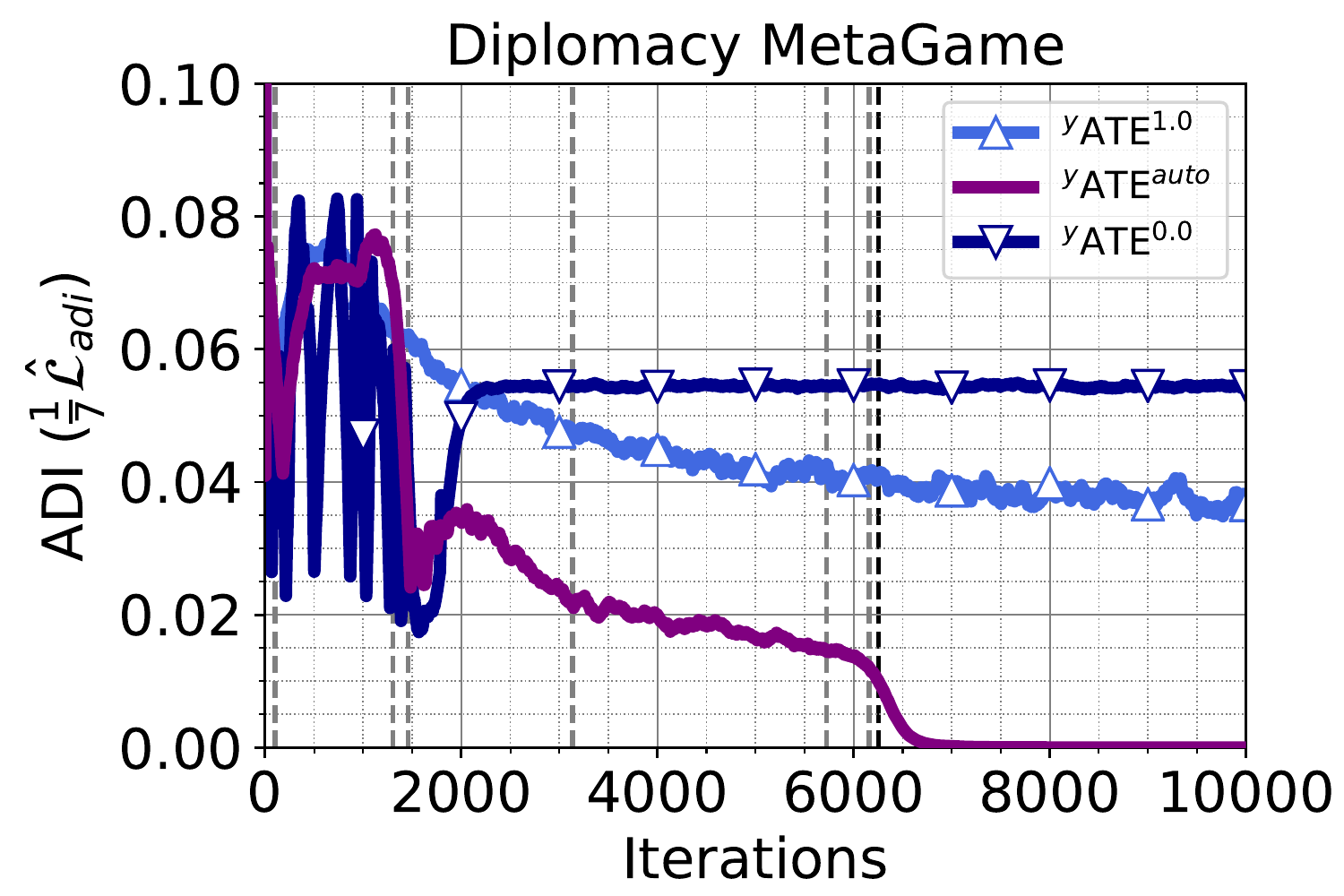}
    \caption{ADI Estimate \label{fig:dipmedexp_ate}}
    \end{subfigure}
    \vspace{-5pt}
    \caption{(\subref{fig:dipmednash_ate}) Evolution of the symmetric Nash approximation returned by ADIDAS for the $7$-player Diplomacy meta-game that considers a subset $\{0, 2, 4, 10, 20\}$ of the available $21$ bots; (\subref{fig:dipmedexp_ate}) ADI estimated from auxiliary variable $\boldsymbol{y}_t$. Black vertical lines indicate the temperature $\tau$ was annealed.}
    \label{fig:diplomacy_med_ate}
\end{figure}

\subsection{Large-Scale}
\label{large_scale}
Figure~\ref{fig:diplomacy_med_ate} demonstrates an empirical game theoretic analysis~\citep{wellman2006methods,jordan2007empirical,wah2016empirical} of a large symmetric $7$-player Diplomacy meta-game where each player elects $1$ of $5$ trained bots to play on their behalf. Each bot represents a snapshot taken from an RL training run on Diplomacy~\citep{anthony2020learning}. In this case, the expected value of each entry in the payoff tensor represents a winrate. Each entry can only be estimated by simulating game play, and the result of each game is a Bernoulli random variable (ruling out deterministic approaches, e.g., \texttt{gambit}). To estimate winrate within $0.01$ (ADI within $0.02$) of the true estimate with probability $95\%$, a Chebyshev bound implies more than $223$ samples are needed. The symmetric payoff tensor contains $330$ unique entries, requiring over $73$ thousand games in total. ADIDAS achieves near zero ADI in less than $7$ thousand iterations with $50$ samples of joint play per iteration ($\approx 5 \times$ the size of the tensor).
%
%
%
%
\paragraph{Continuum of QREs approaching LLE}
The purpose of this work is to approximate a unique Nash (the LLE) which ADIDAS is designed to do, however, the approach ADIDAS takes of attempting to track the continuum of QREs (or the continuum defined by the Tsallis entropy) allows returning these intermediate QRE strategies which may be of interest. Access to these intermediate approximations can be useful when a game playing program cannot wait for ADIDAS's final output to play a strategy, for example, in online play. Interestingly, human play appears to track the continuum of QREs in some cases where the human must both learn about the game (rules, payoffs, etc.) whilst also evolving their strategy~\citep{mckelvey1995quantal}.
\balance
Notice in Figure~\ref{fig:diplomacy_med_ate} that the trajectory of the Nash approximation is not monotonic; for example, see the kink around $2000$ iterations where bots $10$ and $20$ swap rank. The continuum of QRE's from $\tau=\infty$ to $\tau=0$ is known to be complex providing further reason to carefully estimate ADI and its gradients.

\paragraph{Convergence to a Local Optimum}
\balance
One can also see from Figure~\ref{fig:dipmedexp_ate} that $^y\texttt{ATE}^{0.0}$ has converged to a suboptimal local minimum in the energy landscape. This is likely due to the instability and bias in the gradients computed without any entropy bonus; notice the erratic behavior of its ADI within the first $2000$ iterations.




\subsection{Very Large-Scale re. \S\ref{scale}}
\label{very_large_scale}
\begin{figure}[!ht]
    \centering
    \begin{subfigure}[b]{.49\textwidth}
    \centering
    \includegraphics[width=0.75\textwidth]{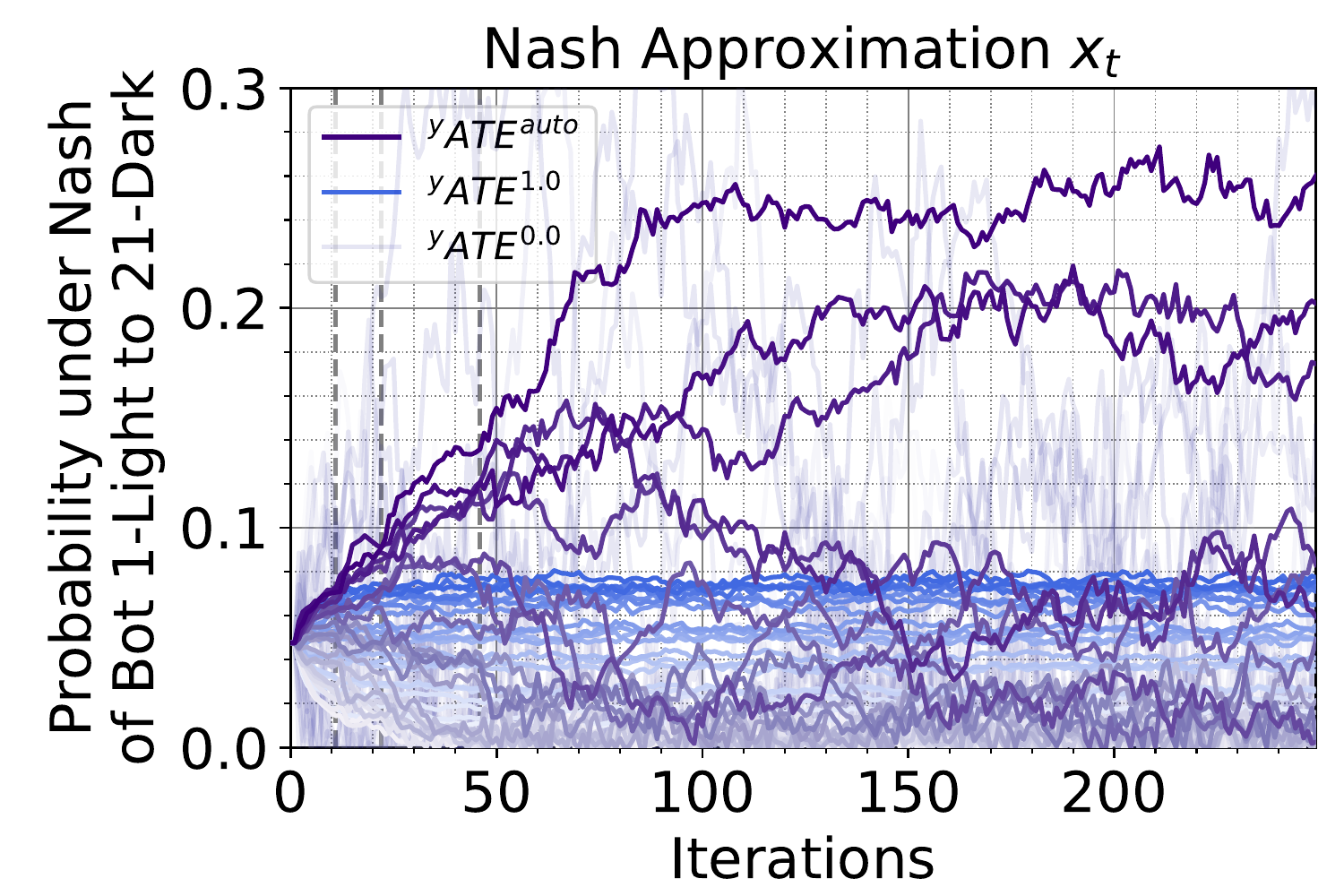}
    \caption{7-player, 21-action symmetric Nash ($x_t$) \label{fig:dipbignash_all_ate}}
    \end{subfigure}
    \begin{subfigure}[b]{.49\textwidth}
    \centering
    \includegraphics[width=0.75\textwidth]{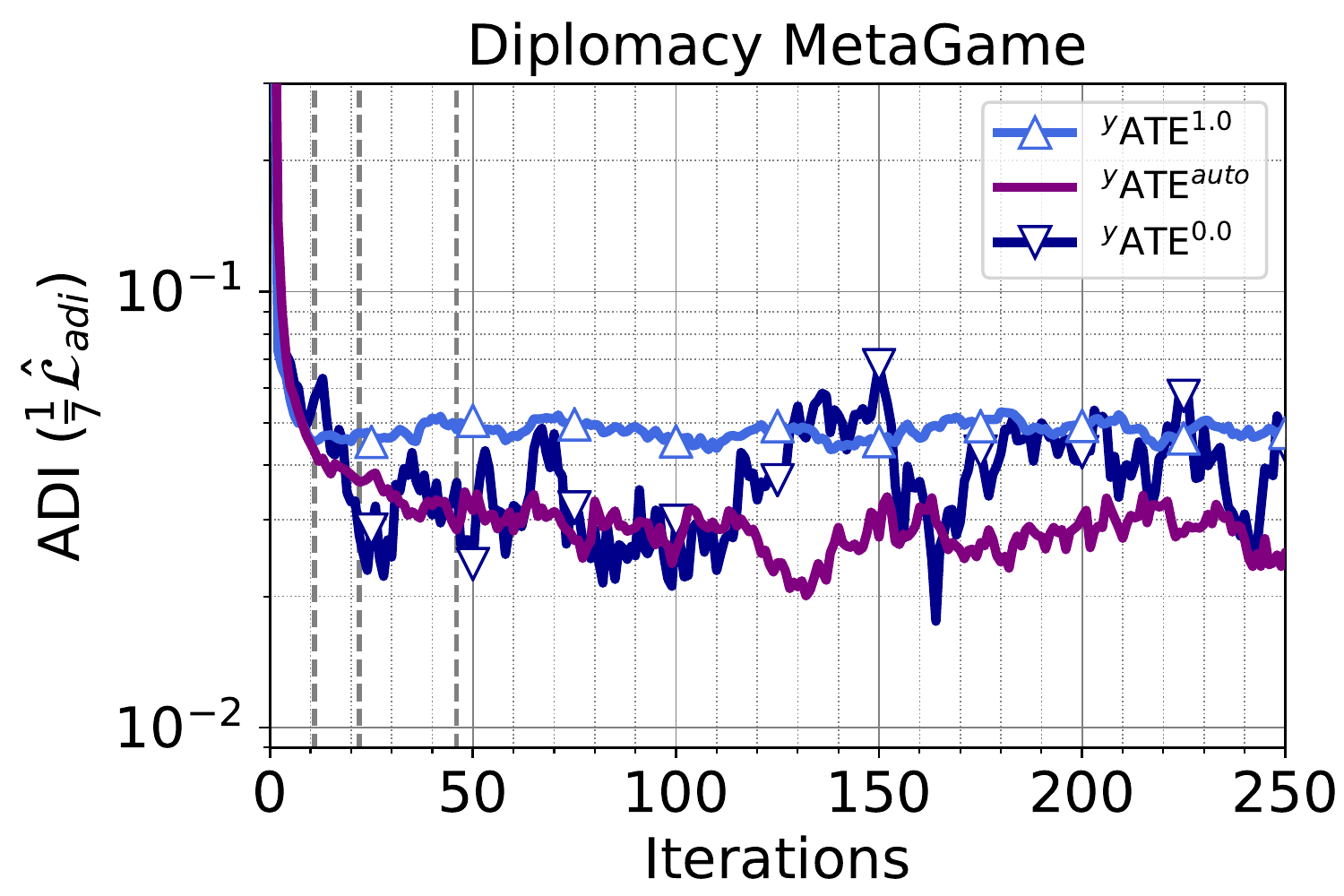}
    \caption{ADI Estimate \label{fig:dipbigexp_ate2}}
    \end{subfigure}
    \vspace{-5pt}
    \caption{(\subref{fig:dipbignash_all_ate}) Evolution of the symmetric Nash approximation returned by ADIDAS for the $7$-player, $21$-bot Diplomacy meta-game; (\subref{fig:dipbigexp_ate2}) ADI estimated from auxiliary variable $y_t$. Black vertical lines indicate the temperature $\tau$ was annealed.}
    \label{fig:diplomacy_big_ate}
\end{figure}

%
Finally, we repeat the above analysis with all $21$ bots.
To estimate winrate within $0.015$ (ADI within $0.03$) of the true estimate with probability $95\%$, a Chebyshev bound implies approximately $150$ samples are needed. The symmetric payoff tensor contains $888,030$ unique entries, requiring over $100$ million games in total. Note that ignoring the symmetry would require simulating $150 \times 21^7 \approx 270$ billion games and computing over a trillion payoffs ($\times7$ players). Simulating all games, as we show, is unnecessarily wasteful, and just storing the entire payoff tensor in memory, let alone computing with it would be prohibitive without special permutation-invariant data structures ($\approx 50$GB with \texttt{float32}). In Figure~\ref{fig:dipbignash_all_ate}, ADIDAS with $\eta_x = \eta_y = 0.1$ and $\epsilon = 0.001$ achieves a stable ADI below $0.03$ in less than $100$ iterations with $10$ samples of joint play per iteration and each game repeated $7$ times ($< 2.5\%$ of the games run by the naive alternative).
As expected, bots later in training (darker lines) have higher mass under the Nash distribution computed by $^y\texttt{ATE}^{auto}$. Runtime is discussed in Appx.~\ref{app:runtime}.

\paragraph{Importance of Entropy Bonus}

Figure~\ref{fig:dipbignash_all_ate} shows how the automated annealing mechanism ($^y\texttt{ATE}^{auto}$) seeks to maintain entropy regularization near a ``sweet spot" \textemdash too little entropy ($^y\texttt{ATE}^{0.0}$) results in an erratic evolution of the Nash approximation and too much entropy ($^y\texttt{ATE}^{1.0}$) prevents significant movement from the initial uniform distribution. Figure~\ref{fig:dipbigexp_ate2} shows that ADIDAS with the automated annealing mechanism meant to trace the QRE continuum achieves a lower ADI than its fixed temperature variants.



In the appendix, we perform additional ablation studies (e.g., no entropy, annealing), measure accuracy of $\hat{\mathcal{L}}^{\tau}_{adi}$, compare against more algorithms on other domains, and consider Tsallis entropy.


\section{Conclusion}

Existing algorithms either converge to Nash, but do not scale to large games or scale to large games, but do not converge to Nash. We proposed an algorithm to fill this void that queries necessary payoffs through sampling, obviating storing the full payoff tensor in memory. ADIDAS is principled and shown empirically to approximate Nash in large-normal form games.

\bibliographystyle{ACM-Reference-Format}
\bibliography{main}


\newpage

\onecolumn
\appendix
\appendixpage
\tableofcontents
\addtocontents{toc}{\protect\setcounter{tocdepth}{2}}

\newpage
\section{Runtime}
\label{app:runtime}

We briefly discussed runtime of ADIDAS in the main body within the context of the Colonel Blotto game. The focus of the paper is on the divide between algorithms that can solve for Nash in any reasonable amount of time (e.g., $\approx 3$ minutes) and those that cannot (e.g., GW with 53,000 hours). The modified Shapley's game and D7-Covariant game from GAMUT are both smaller than the Blotto game, so we omitted a runtime discussion for these.

The Diplomacy experiment required simulating Diplomacy games on a large shared compute cluster with simulated games taking anywhere from 3 minutes to 3 hours. Games were simulated at each iteration of ADIDAS asynchronously using a pool of 1000 workers (4 CPUs per worker, 1 worker per game); the Nash approximate $x_t$ was updated separately on a single CPU. The main computational bottleneck in this experiment was simulating the games themselves, rather than computing gradients from those games. Therefore, the number of games simulated (entries accessed in the payoff tensor) is a realistic metric of algorithmic efficiency.
\section{Two vs More Than Two Player Games}
\label{app:beyondtwo}

An $n$-player game for all $n \ge 3$ can be reduced in polynomial time to a $2$-player game such that the Nash equilibria of the $2$-player game can be efficiently used to compute approximate Nash equilibria of the $n$-player game~\citep{daskalakis2009complexity,chen2006settling,etessami2010complexity}.
\section{Symmetric Nash for Symmetric Games}
\label{app:sym}
Note that a symmetric Nash equilibrium is guaranteed to exist for a finite, normal-form game~\cite{fey2012symmetric}.

One of the reasons we enforce symmetry is that we had Nash-ranking in mind when designing the algorithm and experiments. In that case, for a symmetric meta-game, we desire a symmetric equilibrium so we have a single ranking to go by for evaluation. If each player, in for example the 7-player Diplomacy meta-game, returned a different distribution at Nash, then we'd have to figure out which player's side of the Nash to use for ranking.
\section{Convergence of ADIDAS}
\label{app:conv}

We first establish convergence of the simplified algorithm as described in the warm-up and then discuss convergence of the our more sophisticated, scalable algorithm ADIDAS.

\subsection{Convergence Warm-up: Full Access to In-Memory Payoff Tensor}
\label{app:conv:warmup}

The proof of this simple warm-up algorithm relies heavily on the detailed examination of the continuum of QREs proposed in~\cite{mckelvey1995quantal} and further analyzed in~\cite{turocy2005dynamic}. The Theorem presented below is essentially a succinct repetition of one of their results.

\begin{assumption}[No Principal Branching]
\label{nobranches}
The continuum of QREs from the uniform Nash to the limiting logit equilibrium is unique and contains no branching points.
\end{assumption}

\begin{assumption}[No Turning Points]
\label{noturning}
The continuum of QREs from the uniform Nash to the limiting logit equilibrium proceeds along a path with monotonically decreasing (increasing) $\tau$ ($\lambda$).
\end{assumption}

\begin{assumption}[Bounded sensitivity of QRE to temperature]
\label{sensitivity}
The shift in location of the QRE is upper bounded by an amount proportional to the increase in inverse temperature: $||x^*_{\lambda + \Delta \lambda} - x^*_{\lambda}|| \le \sigma \Delta \lambda.$
\end{assumption}

\begin{assumption}[Bound on BoA's of QRE's]
\label{boa}
Under gradient descent dynamics, the basin of attraction for any quantal response equilibrium, $x^*_{\lambda} = $ QRE$_{\tau=\lambda^{-1}}$, contains a ball of radius $r$. Formally, assuming $x_{t+1} \leftarrow x_t - \eta_t g_t$ with $g_t = \nabla_x \mathcal{L}^{\tau}_{adi}(x_t)$, $\eta_t$ a square-summable, not summable step size (e.g., $\propto t^{-1}$), and given $x_0 \in B(x^*_{\lambda}, r)$, there exists a $T$ such that $x_{t \ge T} \in B(x^*_{\lambda}, \epsilon)$ for any $\epsilon$.
\end{assumption}

\begin{reptheorem}{conv_basic}
Assume the QREs along the homotopy path have bounded sensitivity to $\lambda$ given by a parameter $\sigma$ (Assumption~\ref{sensitivity}), and basins of attraction with radii lower bounded by $r$ (Assumption~\ref{boa}). Let the step size $\Delta \lambda \le \sigma (r - \epsilon)$ with tolerance $\epsilon$. And let $T^*$ be the supremum over all $T$ such that Assumption~\ref{boa} is satisfied for any inverse temperature $\lambda \ge \Delta \lambda$. Then, assuming gradient descent for \texttt{OPT}, \Algref{alg_warmup} converges to the limiting logit equilibrium $x^*_{\lambda=\infty} = x^*_{\tau=0}$ in the limit as $T_{\lambda} \rightarrow \infty$.
\end{reptheorem}

\begin{proof}
Recall~\citet{mckelvey1995quantal} proved there exists a unique continuum of QREs tracing from infinite temperature ($\lambda=0$) to zero temperature ($\lambda=\infty$) for almost all games. Assumption~\ref{boa} effectively assumes the game in question is one from that class. \Algref{alg_warmup} initializes $\lambda=0$ and $x$ to the uniform distribution which is the exact QRE for that temperature. Next, in step~\ref{line:anneal}, the temperature is annealed by an amount that, by Lemma~\ref{sensitivity}, ensures $||x^*_{\lambda + \Delta \lambda} - x|| = ||x^*_{\lambda + \Delta \lambda} - x^*_{\lambda}|| \le r - \epsilon$, where $r$ is a minimal radius of the basin of attraction for any QRE. Then, in step~\ref{line:descend}, \texttt{OPT} returns an $\epsilon$-approximation, $x$, to the new QRE after $T^*$ steps, which implies $||x - x^*_{\lambda + \Delta \lambda}|| \le \epsilon$. The proof then continues by induction. The inverse temperature is increased by an amount ensuring then next QRE is within $r - \epsilon$ of the previous. The current approximation, $x$ is within $\epsilon$ of the previous, therefore, it is within $r - \epsilon + \epsilon = r$ of the next QRE, i.e., it is in its basin of attraction. The inverse temperature $\lambda$ is always increased by an amount such that the current approximation is always within the boundary of attraction for the next QRE. Therefore, in the limit of infinite annealing steps, $x$ converges to the QRE with zero temperature, known as the limiting logit equilibrium.
\end{proof}

\subsection{Convergence Sketch: Sampling the Payoff Tensor}
\label{app:conv:sto}

We do not rigorously prove any theoretical convergence result for the stochastic setting. A convergence proof is complicated by the fact that despite our efforts to reduce gradient bias, some bias will always remain. Although we make assumptions that ensure each iterate begins in the basin of attraction of the QRE of interest, even proving convergence of a hypothetically unbiased stochastic gradient descent to that specific local minimum could only be guaranteed with high probability (dependent on step size). Our goal was to outline a sensible argument that ADIDAS would converge to Nash asymptotically. Our claim of convergence stands on the shoulders of the work of~\citet{mckelvey1995quantal} who proved that there exists a unique path $P$ of Quantal Response Equilibria (QREs) parameterized by temperature $\tau$ which begins at the uniform distribution Nash ($\tau=\infty$) and ends at the limiting logit equilibium ($\tau=0$). \citet{turocy2005dynamic} solves for this path explicitly by solving the associated initial value problem (differential equation) where $t=\frac{1}{\tau}$ takes the place of the typical independent variable time. By numerically integrating this differential equation with infintessimally small steps $dt$,~\citet{turocy2005dynamic} can ensure the iterates progress along the path towards the limiting logit equilibrium (LLE). ADIDAS takes a conceptually similar approach. First, it initializes to the uniform equilibrium. Then it takes a small step $\Delta t$. In practice, the initial step we take increases $t$ from $0$ to $1$, which worked well enough, but one can imagine taking a smaller step, e.g., $0$ to $10^{-9}$. After such a small step, the QRE of the game with lower temperature will not have moved far from the initial uniform equilibrium. Therefore, we can minimize ADI to solve for the new QRE, thereby recovering to a point on the unique path $P$. The fact that we can only access the payoff tensor by samples means that we may need to sample many times ($s$ times) to obtain an accurate Monte Carlo estimate of the gradient of ADI. By repeating this process of decaying the temperature ($\tau_k > \tau_{k+1}$ $\Leftrightarrow$ $t_k < t_{k+1}$) and recovering the new QRE with gradient descent (possibly $n_k$ steps) on ADI ($\boldsymbol{x}_t = \boldsymbol{x}(\tau_k) \rightarrow \boldsymbol{x}_{t+n_k} = \boldsymbol{x}(\tau_{k+1})$), we too can follow $P$. In the limit as $s$, $n_k$, and $N=\sum_k n_k$ go to infinity and $\Delta t$ goes to zero, the issues identified in Figure~\ref{fig:manynasherror} are mitigated and we recover the LLE. Note, $n_k$ is effectively increased by reducing $\epsilon$ in~\Algref{alg_saped}. We claim ``ADIDAS is the first that can approximate Nash in large many-player, many-action normal-form games" because, in principle, it is technically sound according to the argument just presented but also efficient (does not require infinite samples in practice) as demonstrated empirically in our experiments. Note that because we only argue ADIDAS is asymptotically convergent (we provide no convergence rates), we do not contradict any Nash complexity results.

\section{Deviation Incentive Gradient}
\label{gen_pop_exp_grad}
We now provide the general form for the ADI gradient for normal form games.

\begin{align}
    \nabla_{x_i} \mathcal{L}_{adi}(\boldsymbol{x}) &= \textcolor{green}{\nabla_{x_i} [u^{\tau}_i(\br(x_{-i}), x_{-i}) - u^{\tau}_i(x_i,x_{-i})]} \nonumber
    \\ &+ \sum_{j \ne i} \textcolor{blue}{\nabla_{x_i} [u^{\tau}_j(\br(x_{-j}), x_{-j}) - u^{\tau}_j(x_j,x_{-j})]}.
\end{align}

\begin{align}
    \textcolor{green}{\nabla_{x_i} [u^{\tau}_i(\br(x_{-i}), x_{-i}) - u^{\tau}_i(x_i,x_{-i})]} &= \cancelto{0}{J_{x_i}(\br(x_{-i}))}^\top \big( \nabla_{z_i} u^{\tau}_i(z_i, x_{-i}) \vert_{\br_i, x_{-i}} \big) \label{self}
    \\ &+ \sum_{k \ne i} \cancelto{0}{J_{x_i}(x_{k})}^\top \big( \nabla_{z_{k}} u^{\tau}_i(\br(x_i), z_{-i}) \vert_{\br_i, x_{-i}} \big) \nonumber
    \\ &- \nabla_{x_i} u^{\tau}_i(x_i, x_{-i}) = \textcolor{green}{- \nabla_{x_i} u^{\tau}_i(x_i, x_{-i})}.\label{self_term_1}
\end{align}

\begin{align}
    \textcolor{blue}{\nabla_{x_i} [u^{\tau}_j(\br(x_{-j}), x_{-j}) - u^{\tau}_j(x_j,x_{-j})]} &= J_{x_i}(\br(x_{-j}))^\top \big( \nabla_{z_j} u^{\tau}_j(z_j, x_{-j}) \vert_{\br_j, x_{-j}} \big) \label{other}
    \\ &+ \sum_{k \ne j} J_{x_i}(x_{k})^\top \big( \nabla_{z_{k}} u^{\tau}_j(\br(x_{-j}), z_{-j}) \vert_{\br_j, x_{-j}} \big) \nonumber
    \\ &- \nabla_{x_i} u^{\tau}_j(x_j, x_{-j}) \nonumber
    \\ &= \textcolor{blue}{J_{x_i}(\br(x_{-j}))^\top \big( \nabla_{z_j} u^{\tau}_j(z_j, x_{-j}) \vert_{\br_j, x_{-j}} \big)} \label{other_term_1}
    \\ &\textcolor{blue}{+ \big( \nabla_{z_{i}} u^{\tau}_j(\br(x_{-j}), z_{-j}) \vert_{\br_j, x_{-j}} \big)} \label{other_term_2}
    \\ &\textcolor{blue}{- \nabla_{x_i} u^{\tau}_j(x_j, x_{-j})}. \label{other_term_3}
\end{align}

For entropy regularized utilities $u^\tau_i = u_i + S^\tau_i$, the policy gradient decomposes as
\begin{align}
    \nabla_{x_i} u^{\tau}_j(x_j, x_{-j}) &= \nabla_{x_i} u_j(x_j, x_{-j}) + \nabla_{x_i} S^{\tau}_j(x_j, x_{-j}). \label{util_reg_decomp}
\end{align}


\subsection{Gradient of Utility}
\label{appx:nfg_grad}
Before deriving the ADI gradient, we first show that the partial derivative of the utility can be defined in terms of expected utilities or payoffs. This eases presentation later. For example, in a 3-player game, player 1’s expected utility is defined as

\begin{align}
    u_1(a_1, a_2, a_3) &= \sum_{a_1 \in \mathcal{A}_1} \sum_{a_2 \in \mathcal{A}_2} \sum_{a_3 \in \mathcal{A}_3} u_1(a_1, a_2, a_3) x_{1 a_1} x_{2 a_2} x_{3 a_3}.
\end{align}

Taking the derivative with respect to player 1’s strategy $x_{1 a_1}$, i.e., the probability of player 1 specifically playing action $a_1$, we find

\begin{align}
    \frac{\partial u_1}{\partial x_{1 a_1}} &= \sum_{a_2 \in \mathcal{A}_2} \sum_{a_3 \in \mathcal{A}_3} u_1(a_1, a_2, a_3) x_{2 a_2} x_{3 a_3}
    \\ &= \mathbb{E}_{a_2 \sim x_2, a_3 \sim x_3} [ u_1(a_1, a_2, a_3) ].
\end{align}

From here, it should be clear that the full vector of partial derivatives, i.e., gradient, can be written as

\begin{align}
    \nabla_{x_1}(u_1) = \mathbb{E}_{a_2 \sim x_2, a_3 \sim x_3} [ u_1(a_1, a_2, a_3) ] \,\, \forall a_1 \in \mathcal{A}_1.
\end{align}

The Jacobian can be defined similarly (e.g, consider differentiating this result w.r.t. $x_{2 a_2}$ for all $a_2$.

\subsection{Tsallis-Entropy}
\label{app:tsallis_entropy_derivation}
First we derive gradients assuming utilities are carefully regularized using a Tsallis entropy bonus, $S^\tau_k$, parameterized by \emph{temperature} $\tau=p \in [0, 1]$:
\begin{align}
    S^{\tau}_k(x_k, x_{-k}) &= \frac{s_k}{p+1} ( 1 - \sum_m x_{km}^{p+1} ) = s_k \frac{p}{p+1} \Big[ \overbrace{\frac{1}{p} ( 1 - \sum_m x_{km}^{p+1} )}^{\text{Tsallis entropy}} \Big]
\end{align}
where $s_k = \Big( \sum_m (\nabla^k_{x_{km}})^{1/p} \Big)^p = ||\nabla^k_{x_k}||_{1/p}$. For Tsallis entropy, we assume payoffs in the game have been offset by a constant so that they are positive.


The coefficients in front of the Tsallis entropy term are chosen carefully such that a \emph{best response} for player $k$ can be efficiently computed:
\begin{align}
    \br(x_{-k}) &= \argmax_{z_k \in \Delta} z_k^\top \nabla^k_{x_k} + \frac{s_k}{p+1} ( 1 - \sum_m z_{km}^{p+1} ).
\end{align}
First note that the maximization problem above is strictly concave for $s_k > 0$ and $p \in (0, 1]$. If these assumptions are met, then any maximum is a unique global maximum. This is a constrained optimization problem, so in general the gradient need not be zero at the global optimum, but in this case it is. We will find a critical point by setting the gradient equal to zero and then prove that this point lies in the feasible set (the simplex) and satisfies second order conditions for optimality.
\begin{align}
    \nabla_{x_k} u_k^\tau(x_k, x_{-k}) &= \nabla^k_{x_k} - ||\nabla^k_{x_k}||_{1/p} x_{k}^{p} = 0
    \\ \implies \br(x_{-k}) &= \Big[ \frac{\nabla^k_{x_k}}{||\nabla^k_{x_k}||_{1/p}} \Big]^{\frac{1}{p}} = \Big[ \frac{\nabla^k_{x_k}}{s_k} \Big]^{\frac{1}{p}} = \frac{(\nabla^k_{x_k})^{\frac{1}{p}}}{||\nabla^k_{x_k}||_{1/p}^{1/p}} = \frac{(\nabla^k_{x_k})^{\frac{1}{p}}}{\sum_m (\nabla^k_{x_{km}})^{\frac{1}{p}}} \in \Delta.
\end{align}
The critical point is on the simplex as desired. Furthermore, the Hessian at the critical point is negative definite, $H(\br) = -p s_k\texttt{diag}(\br^{-1}) \prec 0$, so this point is a local maximum (and by strict concavity, a unique global maximum).

If the original assumptions are not met and $s_k = 0$, then this necessarily implies $u_k^\tau(x_k, x_{-k}) = 0$ for all $x_k$. As all actions achieve equal payoff, we define the best response in this case to be the uniform distribution. Likewise, if $p=0$, then the Tsallis entropy regularization term disappears ($1-\sum_m x_{km} = 0$) and the best response is the same as for the unregularized setting. Note in the unregularized setting, we define the best response to be a mixed strategy over all actions achieving the maximal possible utility.

\subsubsection{Gradients}
\label{grad_calculations}
We now derive the necessary derivatives for computing the deviation incentive gradient.


\paragraph{Entropy Gradients}

\begin{align}
    (A)\,\, \textcolor{green}{\nabla_{x_i} S^{\tau}_i (x_i, x_{-i})} &= -s_i x_i^p
    \\\nonumber\\ (B)\,\, \textcolor{blue}{\nabla_{x_i} s_j} &= p \Big( \sum_m (\nabla^j_{x_{jm}})^{1/p} \Big)^{p-1} \Big( \sum_m \frac{1}{p} \nonumber (\nabla^j_{x_{jm}})^{\frac{1}{p}-1} H^j_{j_m i} \Big)
    \\ &= \Big( \sum_m (\nabla^j_{x_{jm}})^{1/p} \Big)^{p-1} \Big( \sum_m (\nabla^j_{x_{jm}})^{\frac{1}{p}-1} H^j_{j_m i} \Big) \nonumber
    \\ &= \frac{1}{s_j^{\frac{1}{p}-1}} H^j_{ij} (\nabla^j_{x_j})^{\frac{1}{p}-1} = H^j_{ij} \br(x_{-j})^{1-p}
    \\ &\stackrel{p=1}{=} H^j_{ij} \mathbf{1} \nonumber
    \\ &\stackrel{p=\frac{1}{2}}{=} H^j_{ij} \frac{\nabla^j_{x_j}}{s_j} \nonumber
    \\\nonumber\\ (C)\,\, \textcolor{red}{\nabla_{x_i} S^{\tau}_j (x_j, x_{-j})} &= \frac{1}{s_j} S^{\tau}_j(x_j, x_{-j}) \textcolor{blue}{\underbrace{\nabla_{x_i} s_j}_{(B)}}
\end{align}

\paragraph{Best Response Gradients}

\begin{align}
    (D)\,\, \textcolor{orange}{J_{x_i} [(\nabla^j_{x_j})^{\frac{1}{p}}]} &= J_{x_i} [(H^j_{ji} x_i)^{\frac{1}{p}}] \nonumber
    \\ &= \frac{1}{p} (\nabla^j_{x_j})^{\frac{1}{p}-1} \odot H^j_{ji}
\end{align}
where $\odot$ denotes elementwise multiplication or, more generally, broadcast multiplication. In this case, $(\nabla^j_{x_j})^{\frac{1}{p}-1} \in \R^{d_j \times 1}$ is broadcast multiplied by $H^j_{ji} \in \R^{d_j \times d_i}$ to produce a Jacobian matrix in $\R^{d_j \times d_i}$.

\begin{align}
    (E)\,\, J_{x_i}(\br(x_{-j})) &= \frac{1}{\sum_m (\nabla^j_{x_{jm}})^{\frac{1}{p}}} J_{x_i} [(\nabla^j_{x_j})^{\frac{1}{p}}] - [(\nabla^j_{x_j})^{\frac{1}{p}}] [\frac{1}{\sum_m (\nabla^j_{x_{jm}})^{\frac{1}{p}}}]^2 \nabla_{x_i} [\sum_m (\nabla^j_{x_{jm}})^{\frac{1}{p}}]^\top \nonumber
    \\ &= \frac{1}{\sum_m (\nabla^j_{x_{jm}})^{\frac{1}{p}}} J_{x_i} [(\nabla^j_{x_j})^{\frac{1}{p}}] - [(\nabla^j_{x_j})^{\frac{1}{p}}] [\frac{1}{\sum_m (\nabla^j_{x_{jm}})^{\frac{1}{p}}}]^2 [\sum_m J_{x_i} [(\nabla^j_{x_{jm}})^{\frac{1}{p}}]]^\top \nonumber
    \\ &= \frac{1}{\sum_m (\nabla^j_{x_{jm}})^{\frac{1}{p}}} J_{x_i} [(\nabla^j_{x_j})^{\frac{1}{p}}] - [(\nabla^j_{x_j})^{\frac{1}{p}}] [\frac{1}{\sum_m (\nabla^j_{x_{jm}})^{\frac{1}{p}}}]^2 [\mathbf{1}^\top J_{x_i} [(\nabla^j_{x_{j}})^{\frac{1}{p}}]] \nonumber
    \\ &= [\frac{1}{\sum_m (\nabla^j_{x_{jm}})^{\frac{1}{p}}} \mathbf{I}_j - [(\nabla^j_{x_j})^{\frac{1}{p}}] [\frac{1}{\sum_m (\nabla^j_{x_{jm}})^{\frac{1}{p}}}]^2 \mathbf{1}^\top] J_{x_i} [(\nabla^j_{x_j})^{\frac{1}{p}}] \nonumber
    \\ &= \frac{1}{||\nabla^j_{x_j}||_{1/p}^{1/p}} [\mathbf{I}_j - \frac{(\nabla^j_{x_j})^{\frac{1}{p}}}{||\nabla^j_{x_j}||_{1/p}^{1/p}} \mathbf{1}^\top] J_{x_i} [(\nabla^j_{x_j})^{\frac{1}{p}}] \nonumber
    \\ &= \frac{1}{||\nabla^j_{x_j}||_{1/p}^{1/p}} [\mathbf{I}_j - \br(x_{-j}) \mathbf{1}^\top] \textcolor{orange}{\underbrace{J_{x_i} [(\nabla^j_{x_j})^{\frac{1}{p}}]}_{(D)}} \nonumber
    \\ &= \frac{1}{s_j^{1/p}} [\mathbf{I}_j - \br(x_{-j}) \mathbf{1}^\top] [\frac{1}{p} (\nabla^j_{x_j})^{\frac{1}{p}-1} \odot H^j_{ji}]
\end{align}

\paragraph{Deviation Incentive Gradient Terms}
Here, we derive each of the terms in the ADI gradient. The numbers left of the equations mark which terms we are computing in section~\ref{gen_pop_exp_grad}.
\begin{align}
    (\ref{self_term_1})\,\, \nabla_{x_i} [u^{\tau}_i(\br(x_{-i}), x_{-i}) - u^{\tau}_i(x_i,x_{-i})] &= -\nabla_{x_i} u^{\tau}_i(x_i,x_{-i}) \nonumber
    \\ &\stackrel{(\ref{util_reg_decomp})+(A)}{=} -(\nabla^i_{x_i} - s_i x_i^p).
\end{align}

\begin{align}
    (\ref{other_term_1})\,\, \nabla_{z_j} u^{\tau}_j(z_j, x_{-j}) \vert_{\br_j, x_{-j}} &= [\nabla_{z_j} u_j(z_j, x_{-j}) + \textcolor{green}{\underbrace{\nabla_{z_j} S^{\tau}_j(z_j, x_{-j})}_{(A)}}] \vert_{\br_j, x_{-j}} \nonumber
    \\ &= [ \nabla^j_{z_j} - s_j z_j^p ] \vert_{\br_j, x_{-j}} \nonumber
    \\ &= \nabla^j_{x_j} - s_j \br(x_{-j})^p \nonumber
    \\ &= \nabla^j_{x_j} - \nabla^j_{x_j} = 0.
\end{align}

\begin{align}
    (\ref{other_term_2})\,\, \nabla_{z_{i}} u^{\tau}_j(\br(x_{-j}), z_{-j}) \vert_{\br_j, x_{-j}} &= [\nabla_{z_i} u_j(\br(x_{-j}), z_{-j}) + \textcolor{red}{\underbrace{\nabla_{z_i} S^{\tau}_j(\br(x_{-j}), z_{-j})}_{(C)}}] \vert_{\br_j, x_{-j}} \nonumber
    \\ &= [H^j_{ij} \br(x_{-j}) + \frac{1}{s_j} S^{\tau}_j(\br(x_{-j}), x_{-j}) H^j_{ij} \br(x_{-j})^{1-p}] \nonumber
    \\ &= H^j_{ij} \br(x_{-j}) [1 + \frac{1}{s_j} S^{\tau}_j(\br(x_{-j}), x_{-j}) \br(x_{-j})^{-p}]
\end{align}

\begin{align}
    (\ref{other_term_3})\,\, \nabla_{z_{i}} u^{\tau}_j(x_j, z_{-j}) \vert_{x_j, x_{-j}} &= [\nabla_{z_i} u_j(x_j, z_{-j}) + \textcolor{red}{\underbrace{\nabla_{z_i} S^{\tau}_j(x_j, z_{-j})}_{(C)}}] \vert_{x_j, x_{-j}} \nonumber
    \\ &= [H^j_{ij} x_j + \frac{1}{s_j} S^{\tau}_j(x_j, x_{-j}) H^j_{ij} x_j^{1-p}] \nonumber
    \\ &= H^j_{ij} x_j [1 + \frac{1}{s_j} S^{\tau}_j(x_j, x_{-j}) x_j^{-p}]
\end{align}

\begin{align}
    (\ref{other})\,\, &\nabla_{x_i} [u^{\tau}_j(\br(x_{-j}), x_{-j}) - u^{\tau}_j(x_j,x_{-j})]
    \\ &\stackrel{(\ref{other_term_1})+(\ref{other_term_2})-(\ref{other_term_3})}{=} H^j_{ij} \br(x_{-j}) [1 + \frac{1}{s_j} S^{\tau}_j(\br(x_{-j}), x_{-j}) \br(x_{-j})^{-p}] - H^j_{ij} x_j [1 + \frac{1}{s_j} S^{\tau}_j(x_j, x_{-j}) x_j^{-p}] \nonumber
    \\ &= H^j_{ij} (\br(x_{-j}) - x_j) \nonumber
    \\ &+ \frac{1}{p+1} (1 - ||\br(x_{-j})||_{p+1}^{p+1}) H^j_{ij} \br(x_{-j})^{1-p} - \frac{1}{p+1} (1 - ||x_j||_{p+1}^{p+1}) H^j_{ij} x_j^{1-p} \nonumber
    \\ &= H^j_{ij} \Big[ (\br(x_{-j}) - x_j) + \frac{1}{p+1} \big( (1 - ||\br(x_{-j})||_{p+1}^{p+1}) \br(x_{-j})^{1-p} - (1 - ||x_j||_{p+1}^{p+1}) x_j^{1-p} \big) \Big]. \nonumber
\end{align}

\paragraph{Deviation Incentive Gradient (Tsallis Entropy)}
Finally, combining the derived terms gives:
\begin{align}
    &\nabla_{x_i} \mathcal{L}_{adi}(\boldsymbol{x}) = -(\nabla^i_{x_i} - x_i^p ||\nabla^i_{x_i}||_{1/p})
    \\ &+ \sum_{j \ne i} H^j_{ij} \Big[ (\br(x_{-j}) - x_j) + \frac{1}{p+1} \big( (1 - ||\br(x_{-j})||_{p+1}^{p+1}) \br(x_{-j})^{1-p} - (1 - ||x_j||_{p+1}^{p+1}) x_j^{1-p} \big) \Big]. \nonumber
\end{align}

Note that in the limit of zero temperature, the gradient approaches
\begin{align}
    \nabla_{x_i} \mathcal{L}_{adi}(\boldsymbol{x}) &\stackrel{p\rightarrow 0^+}{=} -\overbrace{(\nabla^i_{x_i} - \boldsymbol{1} ||\nabla^i_{x_i}||_{\infty})}^{\text{policy gradient}} + \sum_{j \ne i} H^j_{ij} (\br_j - x_j).
\end{align}

The second component of the policy gradient term is orthogonal to the tangent space of the simplex, i.e., it does not contribute to movement along the simplex so it can be ignored in the limit of $p \rightarrow 0^+$.

Also, a Taylor series expansion of the adaptive Tsallis entropy around $p=0$ shows $S^{\tau=p}_k = p s_k \mathcal{H}(x_k) + \mathcal{O}(p^2)$, so the Tsallis entropy converges to a multiplicative constant of the Shannon entropy in the limit of zero entropy. If a similar homotopy exists for Tsallis entropy, maybe its limit point is the same limiting logit equilibrium as with Shannon entropy. We leave this to future research.

\textit{Aside: } If you want to increase the entropy, just add a large constant to all payoffs which makes $\br = \frac{1}{d}$ in the limit; it can be shown that $\frac{1}{d}$ then becomes an equilibrium. Notice $\br$ is invariant to multiplicative scaling of the payoffs. Therefore, deviation incentive is linear with respect to multiplicative scaling. One idea to decrease entropy is to subtract a constant from the payoffs such that they are still positive but smaller. This can accomplish the desired effect, but will require more samples to estimate random variables with tiny values in their denominator. It seems like it won't be any more efficient than decreasing $p$.

\subsection{Shannon Entropy}
The Nash equilibrium of utilities regularized with Shannon entropy is well known as the Quantal Response Equilbrium or Logit Equilibrium. The best response is a scaled softmax over the payoffs. We present the relevant intermediate gradients below.

\begin{align}
    S_k^\tau(x_k, x_{-k}) &= -\tau \sum_i x_i \log(x_i)
    \\ \br(x_{-k}) &= \texttt{softmax}(\frac{\nabla^k_{x_k}}{\tau})
    \\ \nabla_{x_i} S^\tau_i(x_i, x_{-i}) &= -\tau (\log(x_i) + 1)
    \\ \nabla_{x_i} S^\tau_j(x_j, x_{-j}) &= 0
    \\ J_{x_i}(\br(x_{-j})) &= \frac{1}{\tau} (\texttt{diag}(\br_j) - \br_j \br_j^\top) H^j_{ji}
    \\ \nabla_{z_j} u^\tau_j(z_j, x_{-j})\vert_{\br_j, x_{-j}} &= \nabla^j_{x_j} - \tau (\log(\br_j) + 1)
    \\ \nabla_{x_i} [u^\tau_i(\br(x_{-i}), x_{-i}) - u^\tau_i(x_i,x_{-i})] &= -(\nabla^i_{x_i} - \tau (\log(\br_i) + 1))
    \\ \nabla_{z_i} u^\tau_j(\br(x_{-j}), z_{-j})\vert_{\br_j,x_{-j}} &= H^j_{ij} \br(x_{-j})
    \\ \nabla_{z_i} u^\tau_j(x_j, z_{-j})\vert_{x_j,x_{-j}} &= H^j_{ij} x_j
\end{align}

\begin{align}
    &\nabla_{x_i} [u^\tau_j(\br(x_{-j}), x_{-j}) - u^\tau_j(x_j, x_{-j})] \nonumber
    \\ &= [\frac{1}{\tau} (\texttt{diag}(\br_j) - \br_j \br_j^\top) H^j_{ji}]^\top (\nabla^j_{x_j} - \tau (\log(\br_j) + 1)) + H^j_{ij} \br(x_{-j}) - H^j_{ij} x_j
\end{align}

\paragraph{Deviation Incentive Gradient (Shannon Entropy)}
Combining the derived terms gives:
\begin{align}
    &\nabla_{x_i} \mathcal{L}_{adi}(\boldsymbol{x}) = -(\nabla^i_{x_i} - \tau (\log(x_i) + 1))
    \\ &+ \sum_{j \ne i} [\frac{1}{\tau} (\texttt{diag}(\br_j) - \br_j \br_j^\top) H^j_{ji}]^\top (\nabla^j_{x_j} - \tau (\log(\br_j) + 1)) + H^j_{ij} [\br(x_{-j}) - x_j]. \nonumber
\end{align}

\newpage
\section{Ablations}
\label{app:ablations}

We introduce some additional notation here. A superscript indicates the temperature of the entropy regularizer, e.g., \texttt{QRE}$^{0.1}$ uses $\tau=0.1$ and \texttt{QRE}$^{auto}$ anneals $\tau$ as before. \texttt{PED} minimizes $\mathcal{L}_{adi}$ without any entropy regularization or amortized estimates of payoff gradients (i.e., without the auxiliary variable $y$).


\subsection{Bias re. \S\ref{bias_intuition}+\S\ref{ped_grads}}
\label{appx:bias}
\begin{figure}[!ht]
    \centering
    \begin{subfigure}[b]{.49\textwidth}
    \includegraphics[width=\textwidth]{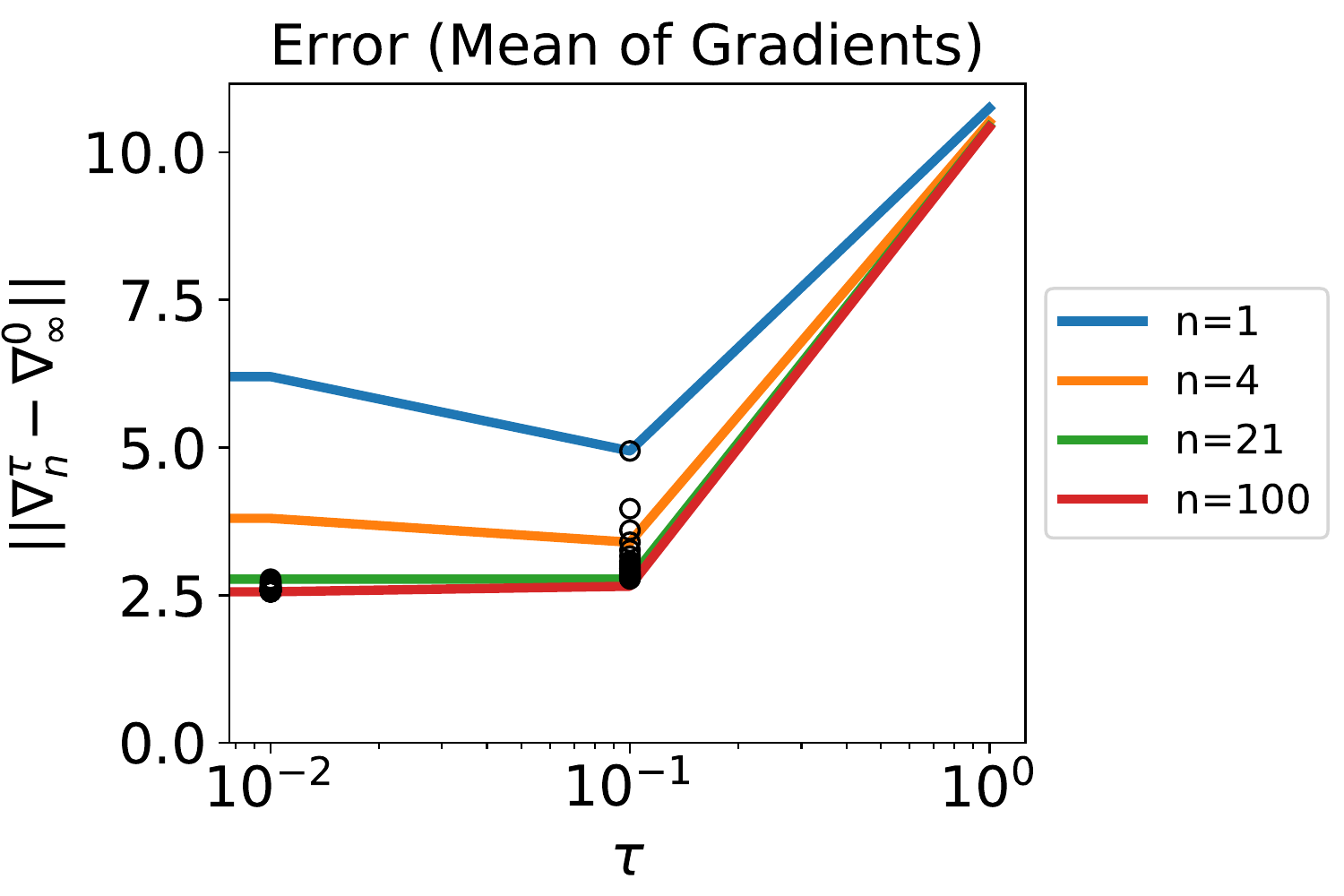}
    \caption{Bias \label{fig:bias_qre}}
    \end{subfigure}
    \begin{subfigure}[b]{.49\textwidth}
    \includegraphics[width=\textwidth]{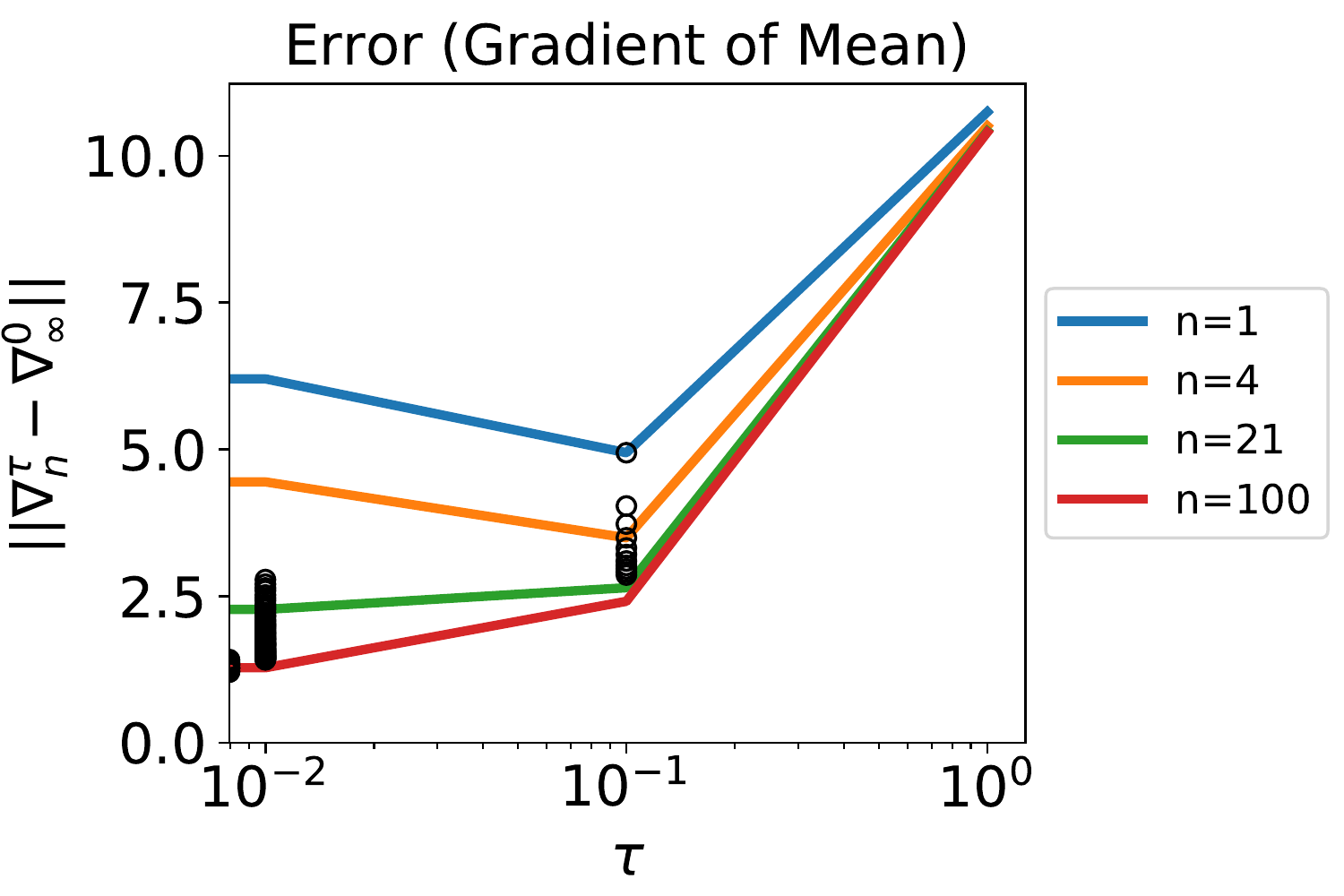}
    \caption{Concentration \label{fig:concentration_qre}}
    \end{subfigure}
    \vspace{-5pt}
    \caption{Bias-Bias Tradeoff on Blotto($10$ coins, $3$ fields, $4$ players). Curves are drawn for samples sizes of $n=\{1, 4, 21, 100\}$. Circles denote the minimum of each curve for all $n \in [1, 100]$. Zero entropy regularization results in high gradient bias, i.e., stochastic gradients, $\nabla^{\tau=0}_n$, do not align well with the expected gradient, $\nabla^{\tau=0}_{\infty}$, where $n$ is the number of samples. On the other hand, higher entropy regularization allows lower bias gradients but with respect to the entropy regularized utilities, not the unregularized utilities that we are interested in. The sweet spot lies somewhere in the middle. (\subref{fig:bias_qre}) SGD guarantees assume gradients are unbiased, i.e., the mean of sampled gradients is equal to the expected gradient in the limit of infinite samples $n$. Stochastic average deviation incentive gradients violate this assumption, the degree to which depends on the amount of entropy regularization $\tau$ and number of samples $n$; $\tau=10^{-2}$ appears to minimize the gradient bias for $n=100$ although with a nonzero asymptote around $2.5$. (\subref{fig:concentration_qre}) Computing a single stochastic gradient using more samples can reduce bias to zero in the limit. Note samples here refers to joint actions drawn from strategy profile $\boldsymbol{x}$, not gradients as in (\subref{fig:bias_qre}). Additional samples makes gradient computation more expensive, but as we show later, these sample estimates can be amortized over iterations by reusing historical play. Both of the effects seen in (\subref{fig:bias_qre}) and (\subref{fig:concentration_qre}) guide development of our proposed algorithm: (\subref{fig:bias_qre}) suggests using $\tau>0$ and (\subref{fig:concentration_qre}) suggests reusing recent historical play to compute gradients (with $\tau>0$).}
    \label{fig:gradient_bias_qre}
\end{figure}

Figure~\ref{fig:gradient_bias_qre} demonstrates there exists a sweet spot for the amount of entropy regularization\textemdash too little and gradients are biased, too much and we solve for the Nash of a game we are not interested in.

\subsection{Auxiliary $y$ re. \S\ref{amortized_estimates}} 
\begin{figure}[!ht]
    \centering
    \begin{subfigure}[b]{.49\textwidth}
    \includegraphics[width=\textwidth]{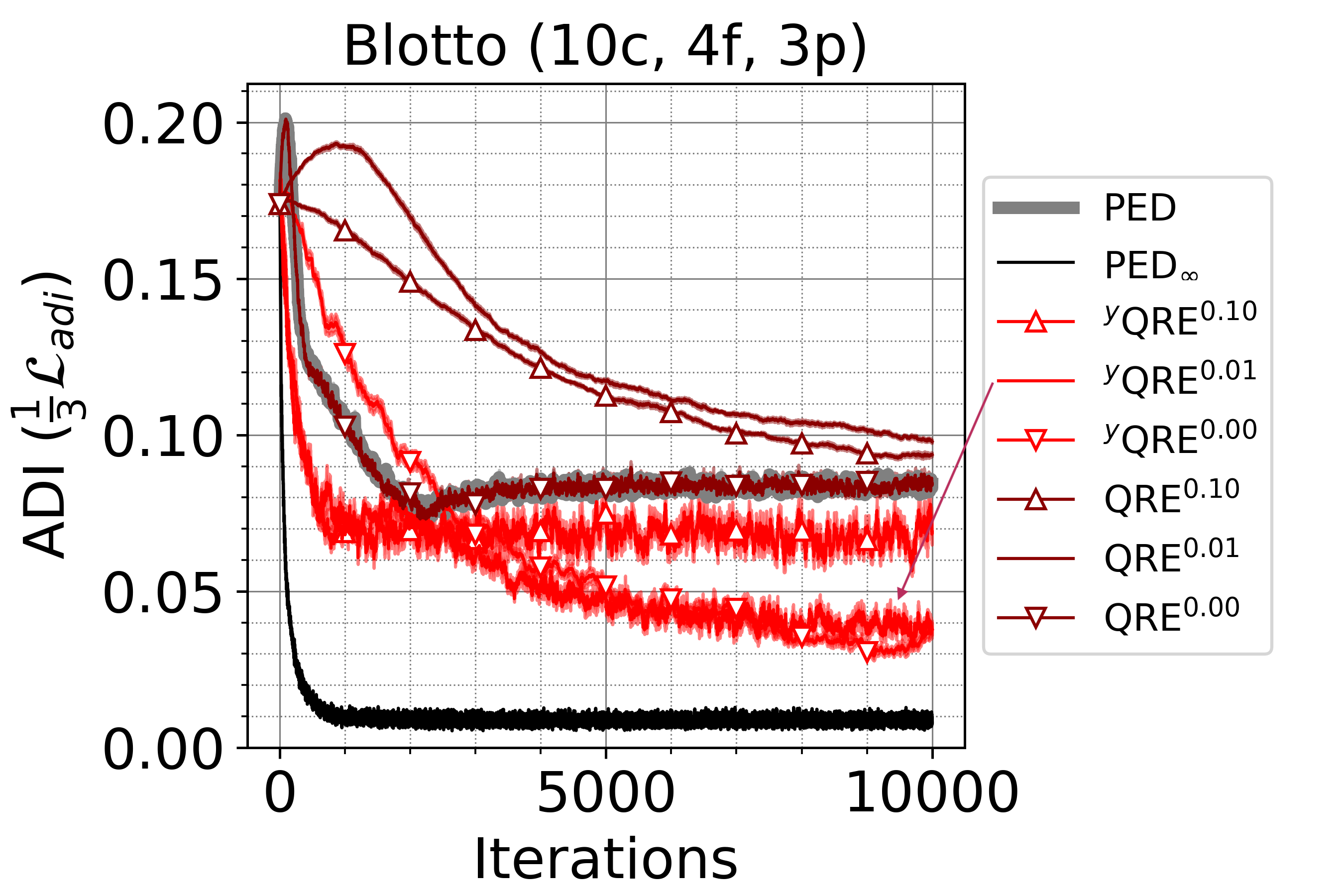}
    \caption{3-player Blotto \label{fig:app:blotto3track_qre}}
    \end{subfigure}
    \begin{subfigure}[b]{.49\textwidth}
    \includegraphics[width=\textwidth]{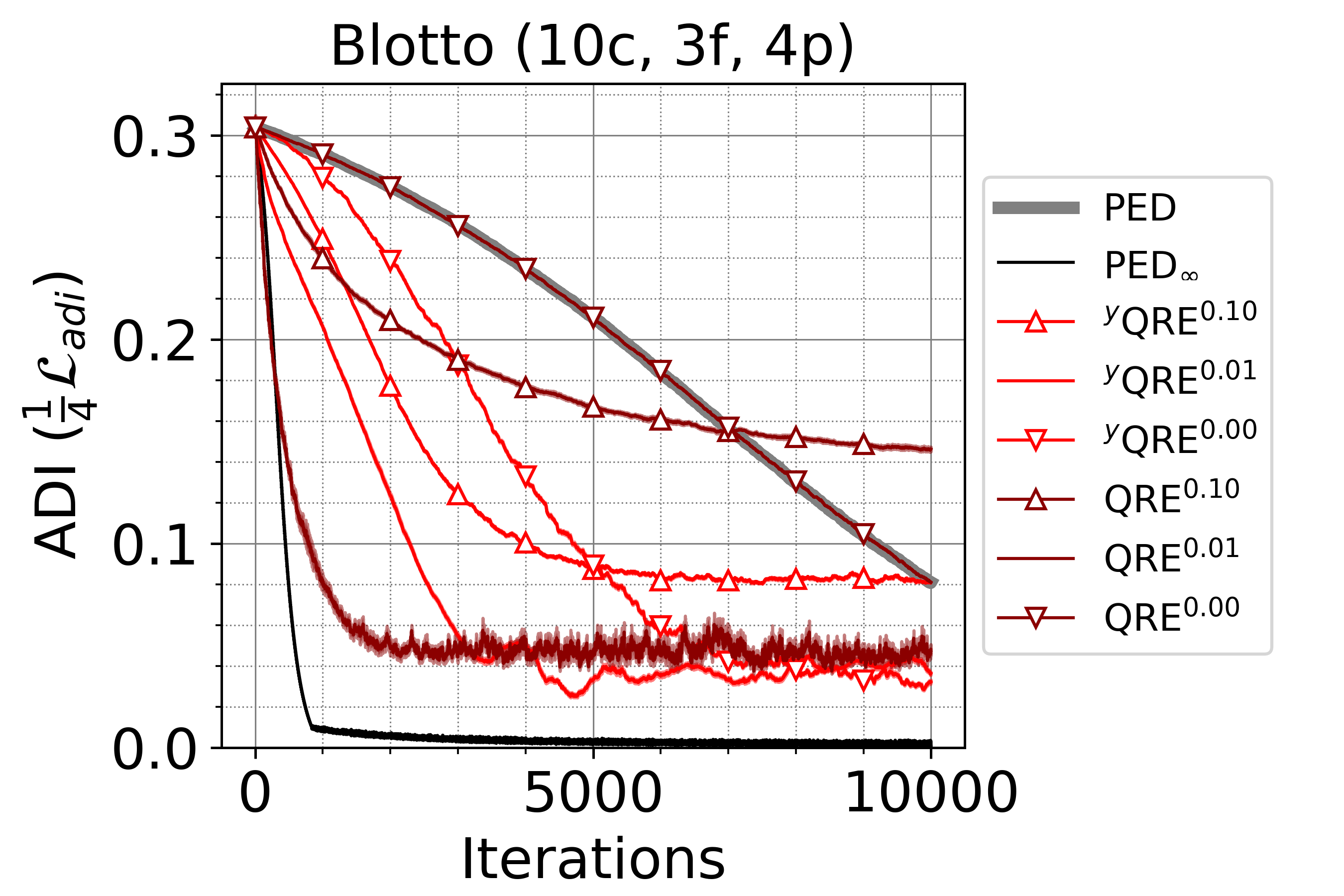}
    \caption{4-player Blotto \label{fig:app:blotto4track_qre}}
    \end{subfigure}
    \vspace{-5pt}
    \caption{Adding an appropriate level of entropy can accelerate convergence (compare PED to \texttt{QRE}$^{0.01}$ in (\subref{fig:blotto4track_qre})). And amortizing estimates of joint play using $y$ can reduce gradient bias, further improving performance (e.g., compare \texttt{QRE}$^{0.00}$ to $^y$\texttt{QRE}$^{0.00}$ in (\subref{fig:app:blotto3track_qre}) or (\subref{fig:app:blotto4track_qre})).}
    \label{fig:app:blotto_tracking_qre}
\end{figure}

The introduction of auxiliary variables $y_i$ are also supported by the results in Figure~\ref{fig:app:blotto_tracking_qre}\textemdash \texttt{QRE}$^{0.0}$ is equivalent to \texttt{PED} and $^y$\texttt{QRE}$^{0.0}$ is equivalent to \texttt{PED} augmented with $y$'s to estimate averages of payoff gradients.

\subsection{Annealing $\tau$ re. \S\ref{annealing}} 
\begin{figure}[!ht]
    \centering
    \begin{subfigure}[b]{.49\textwidth}
    \includegraphics[width=\textwidth]{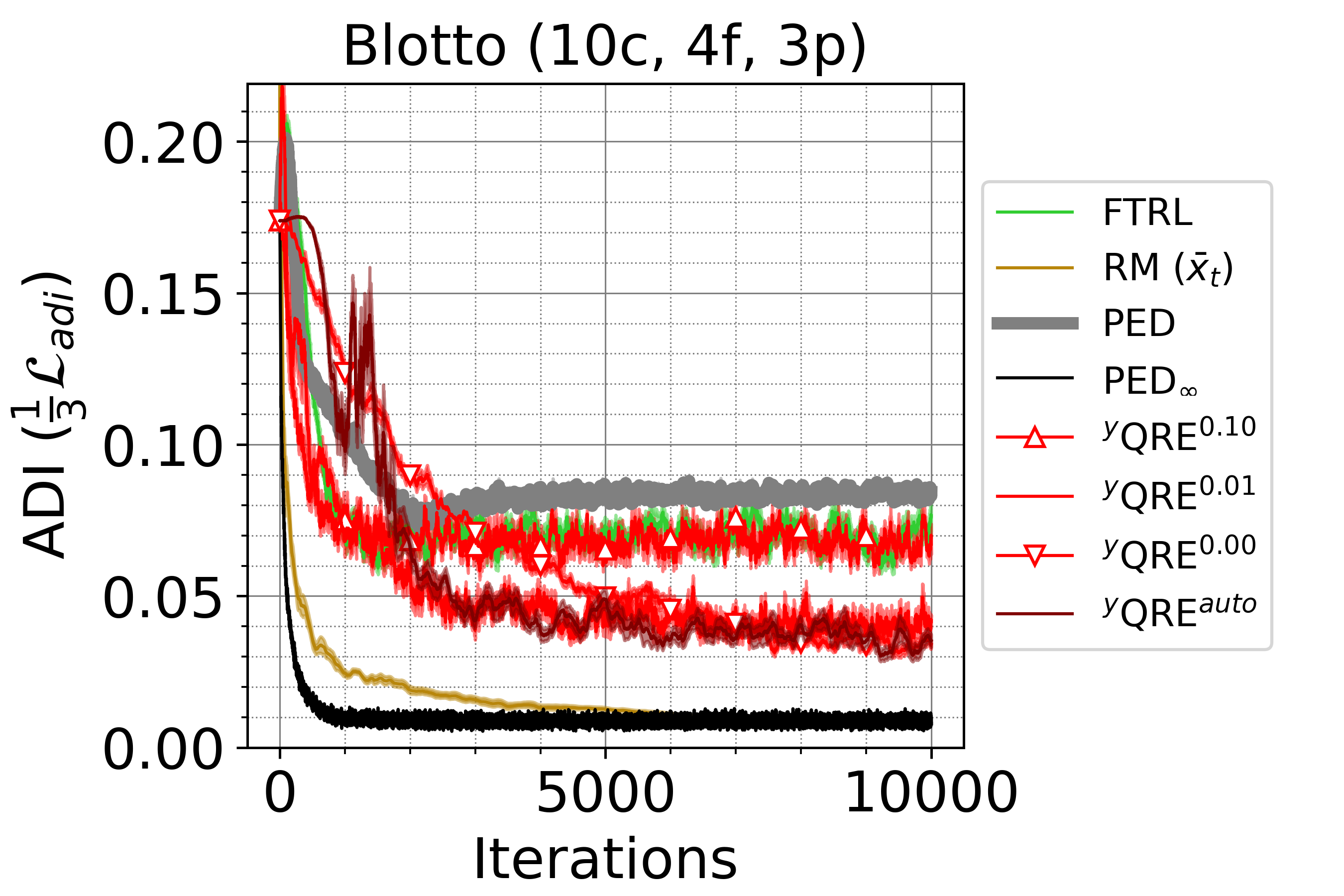}
    \caption{3-player Blotto \label{fig:app:blotto3_qre}}
    \end{subfigure}
    \begin{subfigure}[b]{.49\textwidth}
    \includegraphics[width=\textwidth]{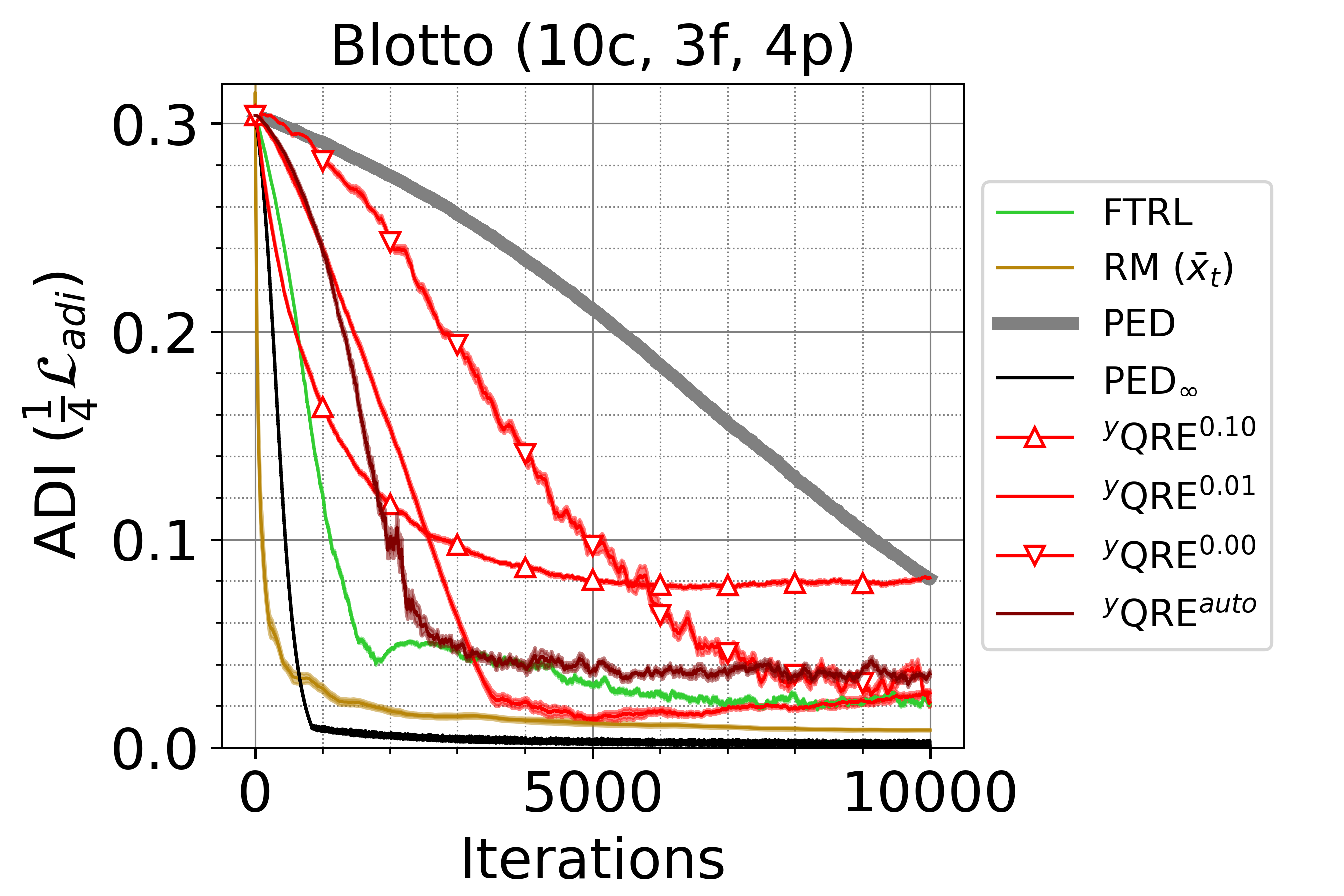}
    \caption{4-player Blotto \label{fig:app:blotto4_qre}}
    \end{subfigure}
    \vspace{-5pt}
    \caption{Average deviation incentive of the symmetric joint strategy $\boldsymbol{x}^{(t)}$ is plotted against algorithm iteration $t$. Despite \texttt{FTRL}'s lack of convergence guarantees, it converges quickly in these games.}
    \label{fig:app:blotto_qre}
\end{figure}

ADIDAS includes temperature annealing, replacing the need to preset $\tau$ with instead an ADI threshold $\epsilon$. Figure~\ref{fig:app:blotto_qre} compares this approach against other variants of the algorithm and shows this automated annealing mechanism reaches comparable final levels of ADI.

\subsection{Convergence re. \S\ref{convergence}} 
\begin{figure}[!h]
    \centering
    \begin{subfigure}[b]{.49\textwidth}
    \includegraphics[width=\textwidth]{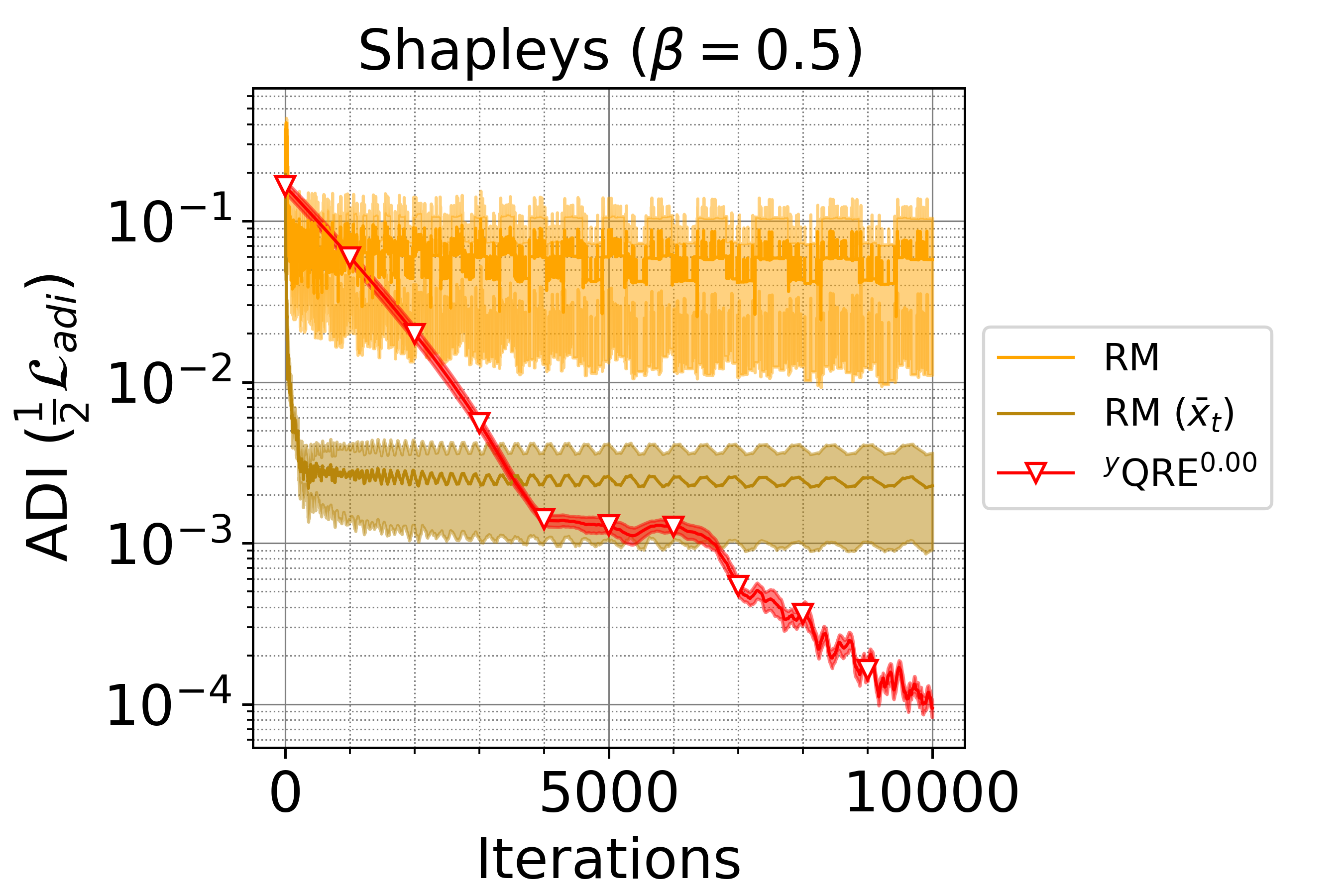}
    \caption{Modified-Shapley's \label{fig:app:modshapley_qre}}
    \end{subfigure}
    \begin{subfigure}[b]{.49\textwidth}
    \includegraphics[width=\textwidth]{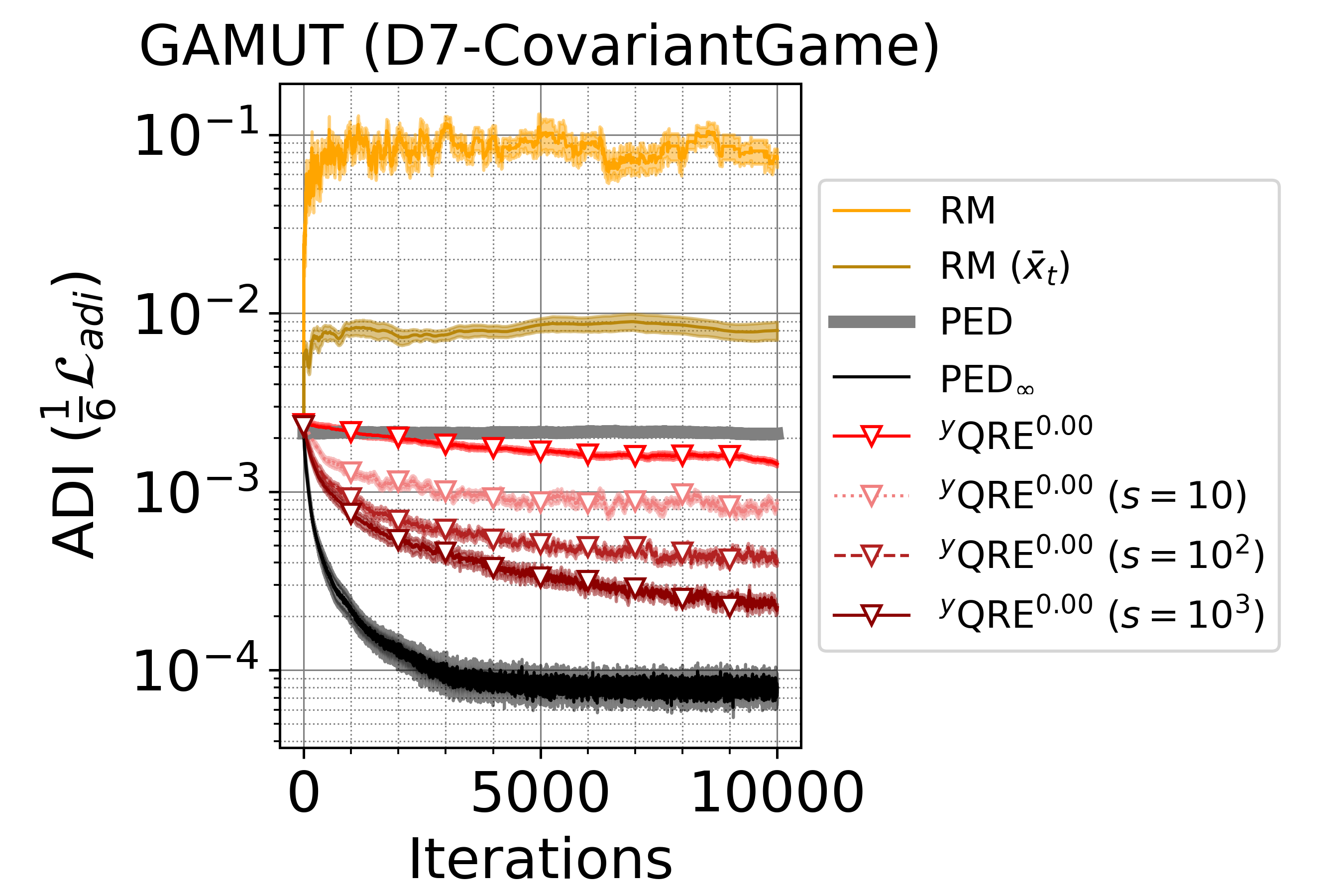}
    \caption{GAMUT-D7 \label{fig:app:gamutd7_qre}}
    \end{subfigure}
    \vspace{-5pt}
    \caption{ADIDAS reduces $\mathcal{L}_{adi}$ in both games. In game~(\subref{fig:app:modshapley_qre}), created by~\protect\citet{ostrovski2013payoff}, to better test performance, $\boldsymbol{x}$ is initialized randomly rather than with the uniform distribution because the Nash is at uniform. In~(\subref{fig:app:gamutd7_qre}), computing gradients using full expectations (in black) results in very low ADI. Computing gradients using only single samples plus historical play allows a small reduction in ADI. More samples (e.g., $n=10^3$) allows further reduction.}
    \label{fig:app:noregfail}
\end{figure}

In Figure~\ref{fig:app:blotto_qre}, \texttt{FTRL} and \texttt{RM} achieve low ADI quickly in some cases. \texttt{FTRL} has recently been proven not to converge to Nash, and this is suggested to be true of no-regret algorithms in general~\citep{flokas2020no,mertikopoulos2018cycles}. Before proceeding, we demonstrate empirically in Figure~\ref{fig:app:noregfail} that \texttt{FTRL} and \texttt{RM} fail on games where minimizing $\mathcal{L}^{\tau}_{adi}$ still makes progress, even without an annealing schedule.


\subsection{ADI stochastic estimate} 
\begin{figure}[h!]
    \centering
    \begin{subfigure}[b]{.49\textwidth}
    \includegraphics[width=\textwidth]{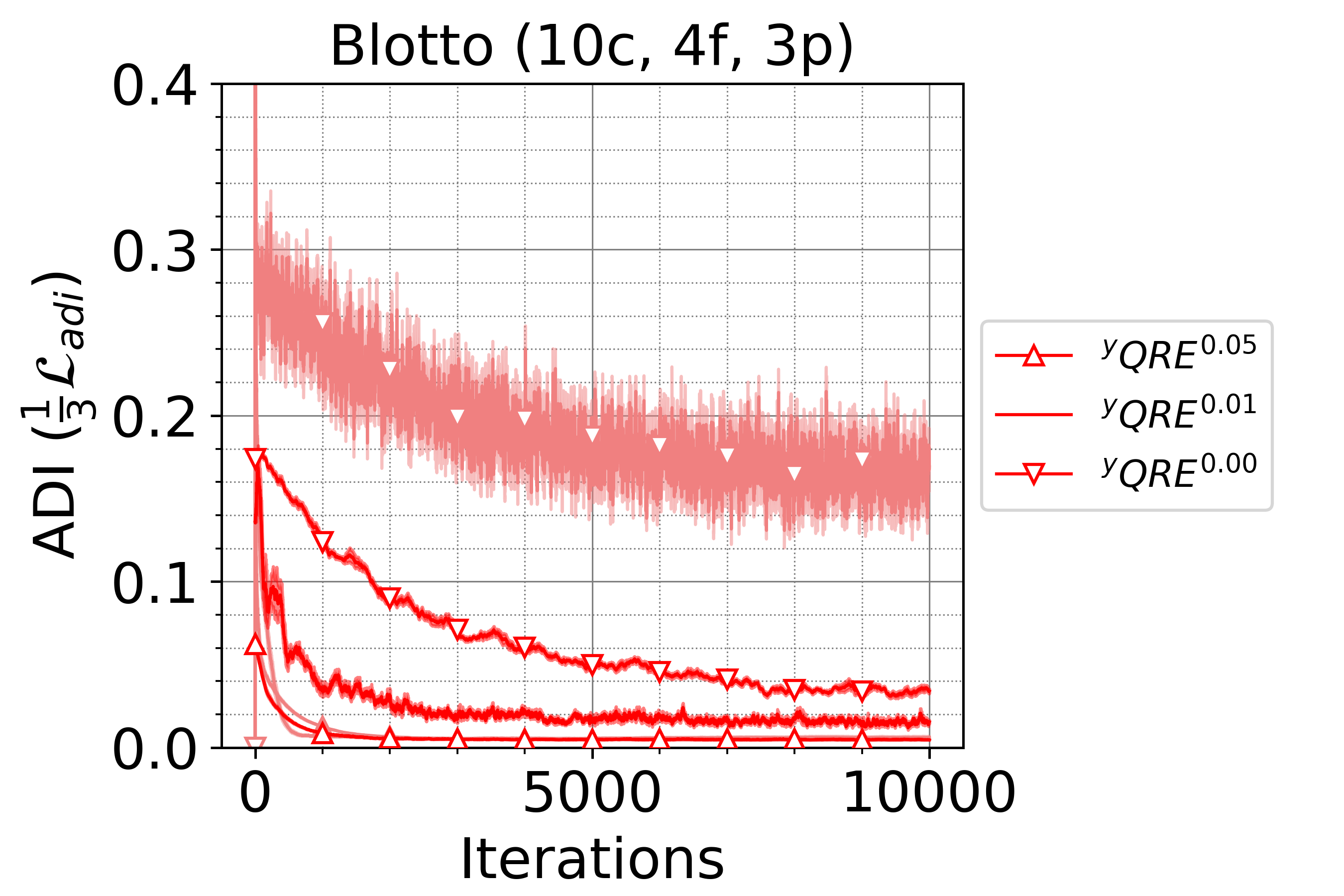}
    \caption{Blotto-3 \label{fig:exp_bias_blotto3_qre}}
    \end{subfigure}
    \begin{subfigure}[b]{.49\textwidth}
    \includegraphics[width=\textwidth]{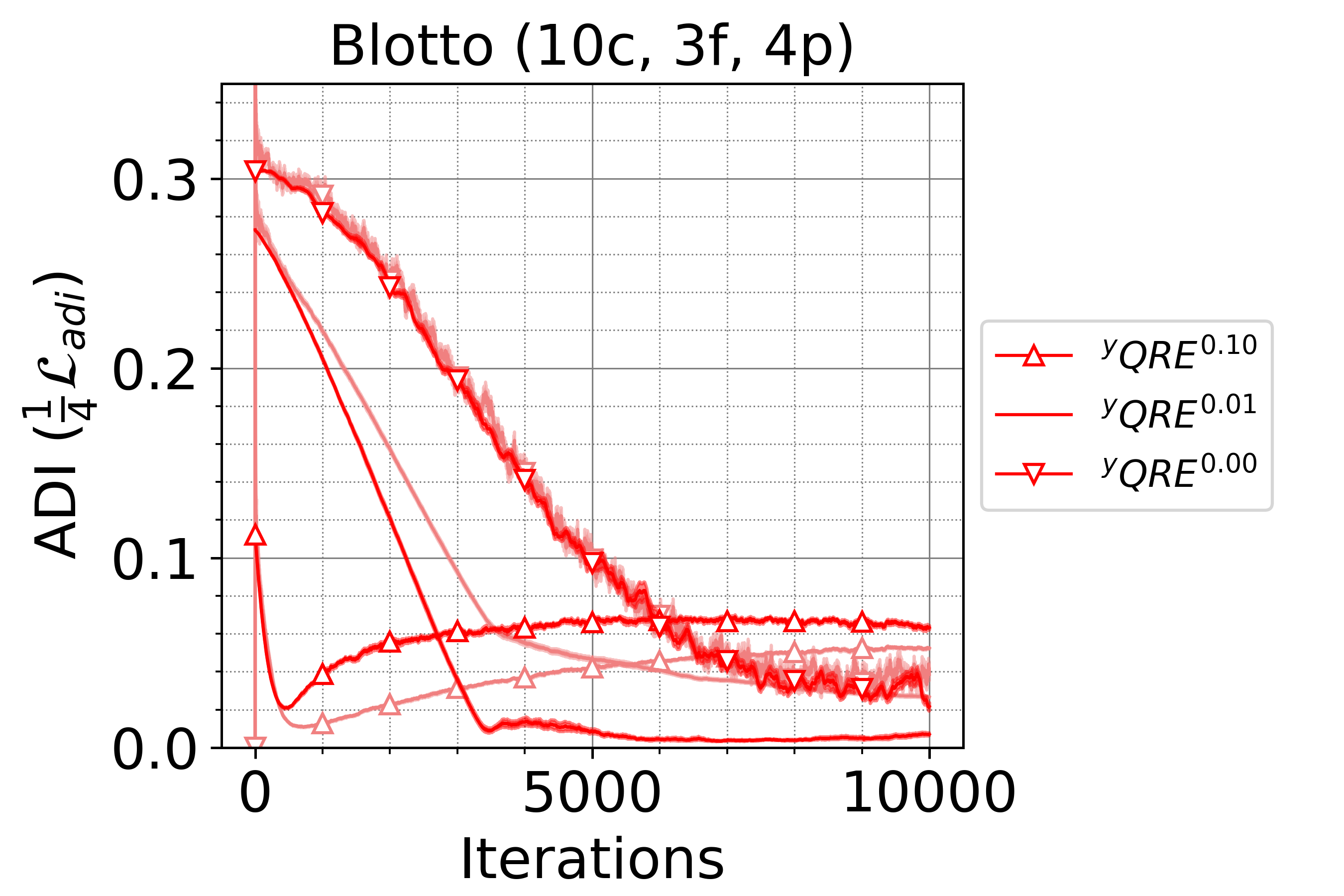}
    \caption{Blotto-4 \label{fig:exp_bias_blotto4_qre}}
    \end{subfigure}
    \vspace{-5pt}
    \caption{Accuracy of running estimate of $\mathcal{L}^{\tau}_{adi}$ computed from $y^{(t)}$ (in light coral) versus true value (in red).}
    \label{fig:exp_est_qre}
\end{figure}

Computing ADI exactly requires the full payoff tensor, so in very large games, we must estimate ADI. Figure~\ref{fig:exp_est_qre} shows how estimates of $\mathcal{L}_{adi}$ computed from historical play track their true expected value throughout training.

\section{Experiments Repeated with \texttt{ATE}}
\label{ate_exp}

\begin{figure}[!ht]
    \centering
    \begin{subfigure}[b]{.49\textwidth}
    \includegraphics[width=\textwidth]{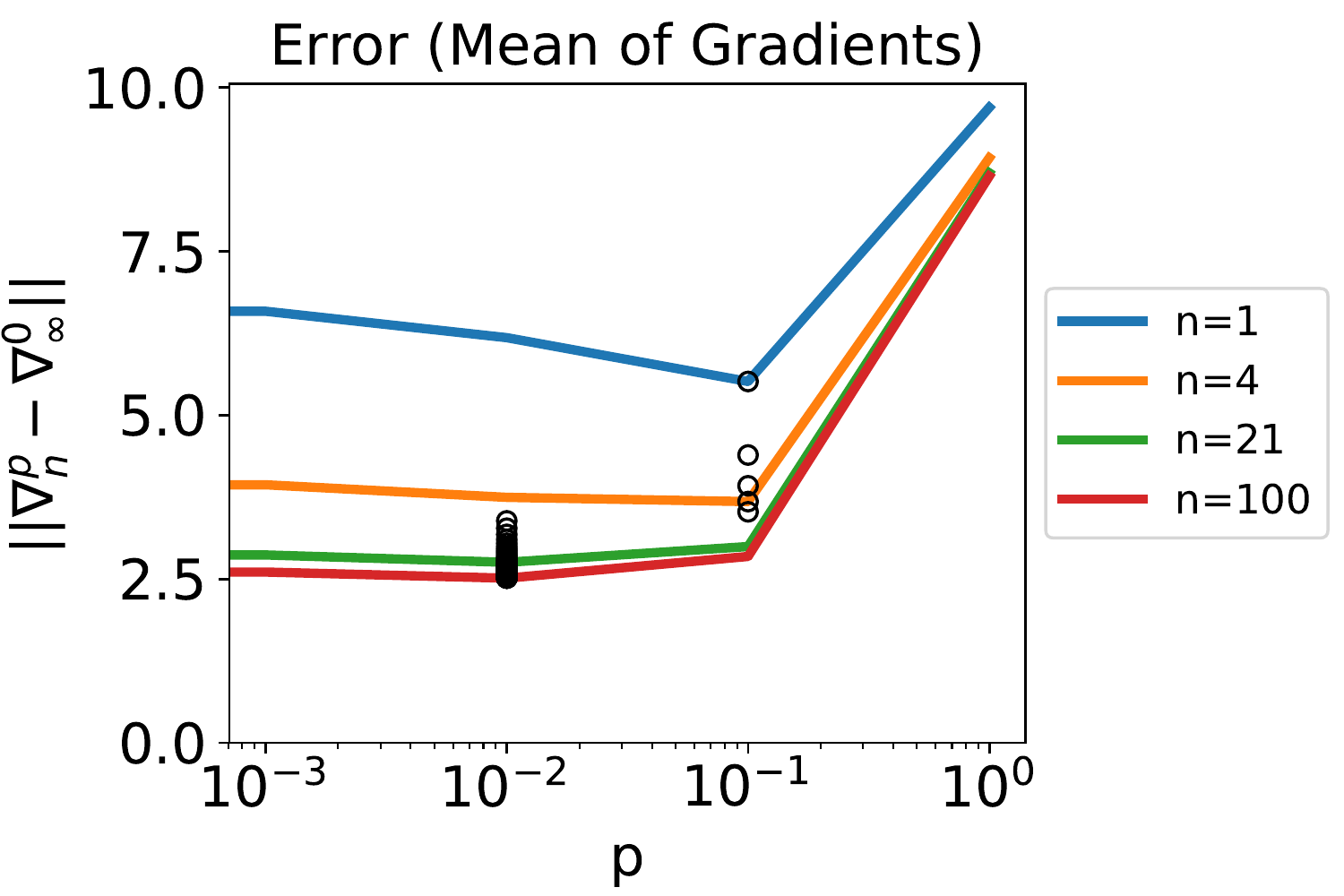}
    \caption{Bias \label{fig:bias_ate}}
    \end{subfigure}
    \begin{subfigure}[b]{.49\textwidth}
    \includegraphics[width=\textwidth]{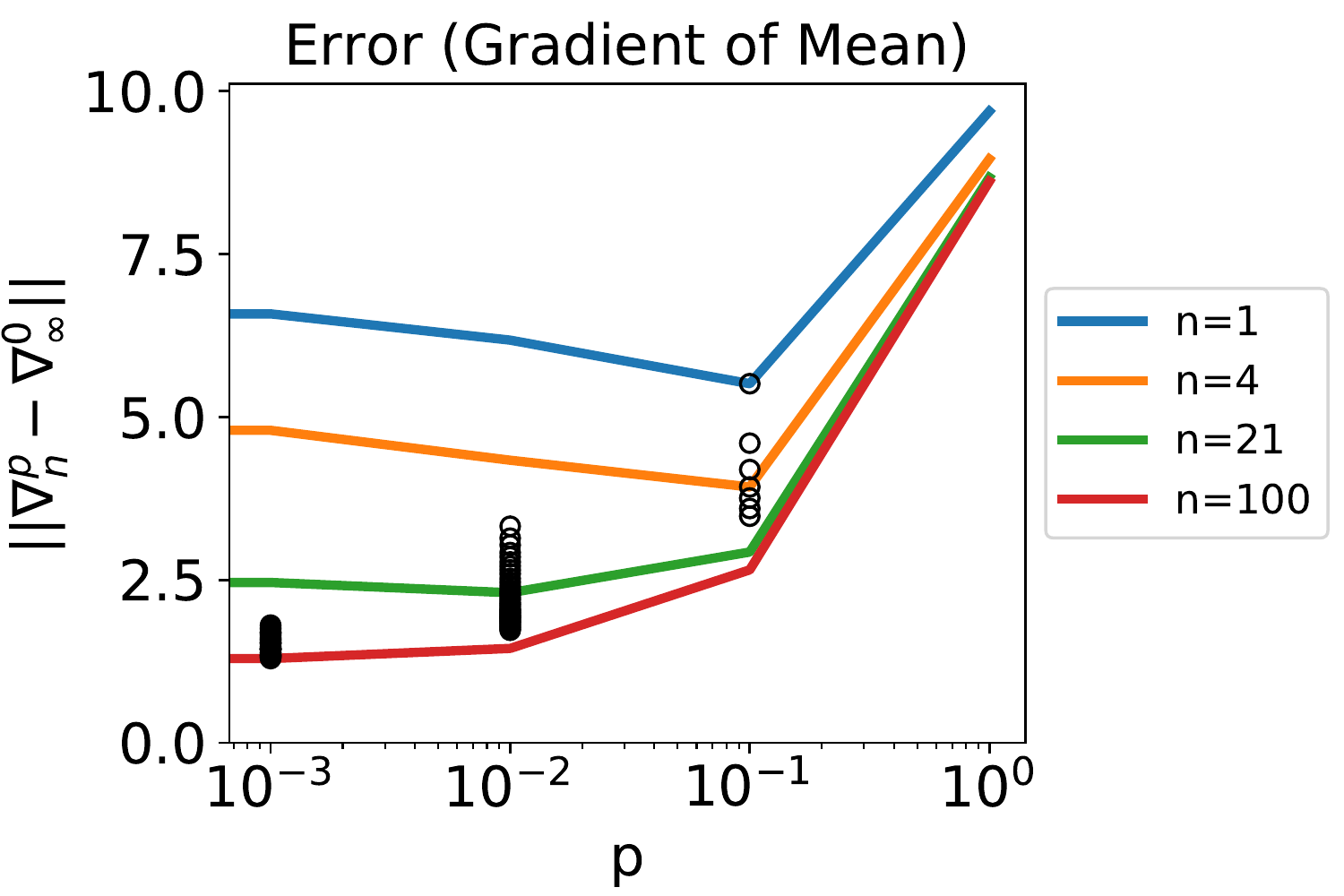}
    \caption{Concentration \label{fig:concentration_ate}}
    \end{subfigure}
    \vspace{-5pt}
    \caption{Bias-Bias Tradeoff on Blotto($10$ coins, $3$ fields, $4$ players). Curves are drawn for samples sizes of $n=\{1, 4, 21, 100\}$. Circles denote the minimum of each curve for all $n \in [1, 100]$. Zero entropy regularization results in high gradient bias, i.e., stochastic gradients, $\nabla^{\tau=0}_n$, do not align well with the expected gradient, $\nabla^{\tau=0}_{\infty}$, where $n$ is the number of samples. On the other hand, higher entropy regularization allows lower bias gradients but with respect to the entropy regularized utilities, not the unregularized utilities that we are interested in. The sweet spot lies somewhere in the middle. (\subref{fig:bias_ate}) SGD guarantees assume gradients are unbiased, i.e., the mean of sampled gradients is equal to the expected gradient in the limit of infinite samples $n$. Stochastic average deviation incentive gradients violate this assumption, the degree to which depends on the amount of entropy regularization $\tau$ and number of samples $n$; $p=10^{-2}$ appears to minimize the gradient bias for $n=100$ although with a nonzero asymptote around $2.5$. (\subref{fig:concentration_ate}) Computing a single stochastic gradient using more samples can reduce bias to zero in the limit. Note samples here refers to joint actions from strategy profile $\boldsymbol{x}$, not gradients as in (\subref{fig:bias_ate}). Additional samples makes gradient computation more expensive, but as we show later, these sample estimates can be amortized over iterations by reusing historical play. Both the effects seen in (\subref{fig:bias_ate}) and (\subref{fig:concentration_ate}) guide development of our proposed algorithm: (\subref{fig:bias_ate}) suggests using $\tau>0$ and (\subref{fig:concentration_ate}) suggests reusing recent historical play to compute gradients (with $\tau>0$).}
    \label{fig:gradient_bias_ate}
\end{figure}

\paragraph{Bias re. \S\ref{bias_intuition}+\S\ref{ped_grads}} We first empirically verify that adding an entropy regularizer to the player utilities introduces a trade-off: set entropy regularization too low and the best-response operator will have high bias; set entropy regularization too high and risk solving for the Nash of a game we are not interested in. Figure~\ref{fig:gradient_bias_ate} shows there exists a sweet spot in the middle for moderate amounts of regularization (temperatures).

\begin{figure}[!ht]
    \centering
    \begin{subfigure}[b]{.49\textwidth}  
    \includegraphics[width=\textwidth]{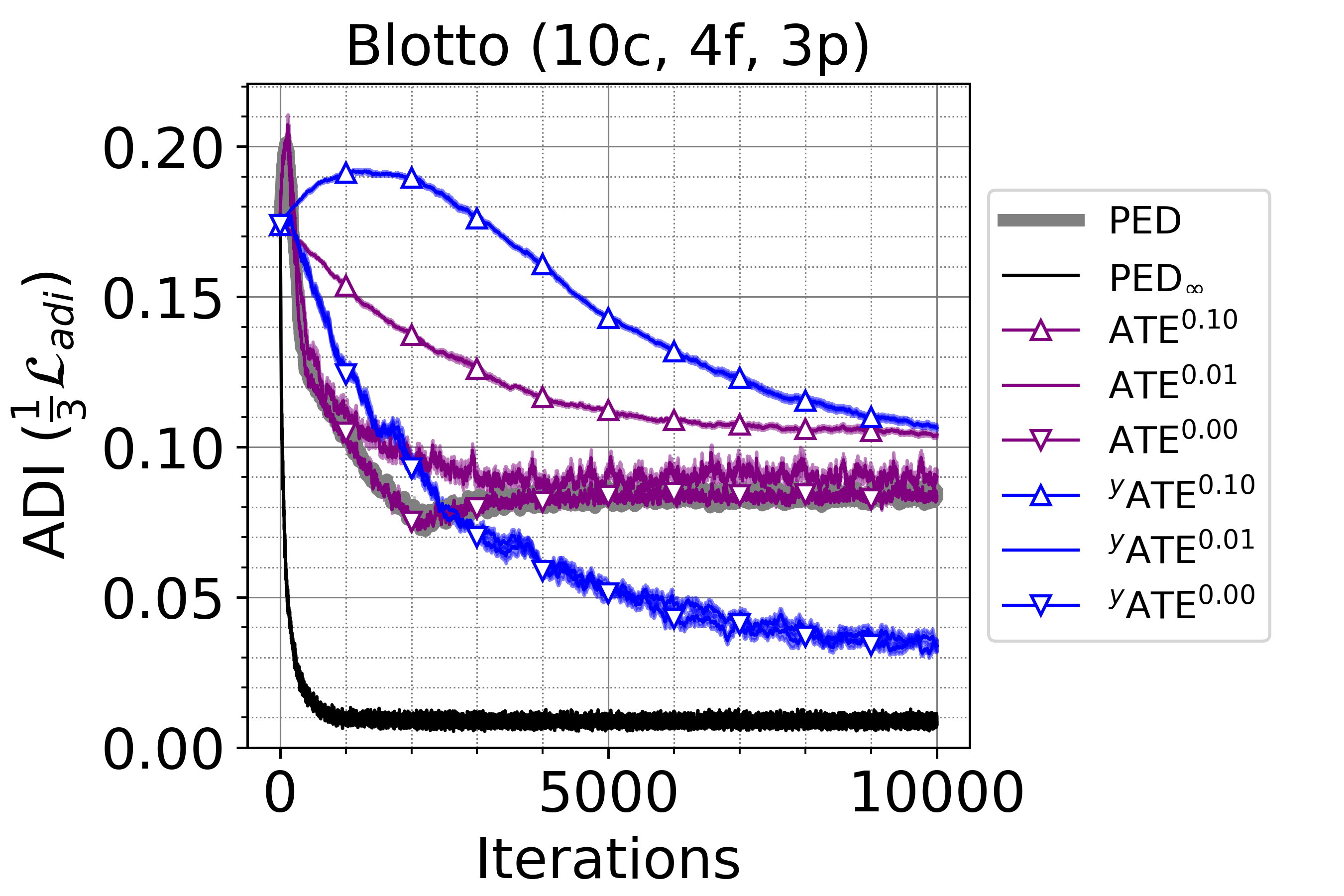}
    \caption{Blotto-3 \label{fig:blotto3track_ate}}
    \end{subfigure}
    \begin{subfigure}[b]{.49\textwidth}
    \includegraphics[width=\textwidth]{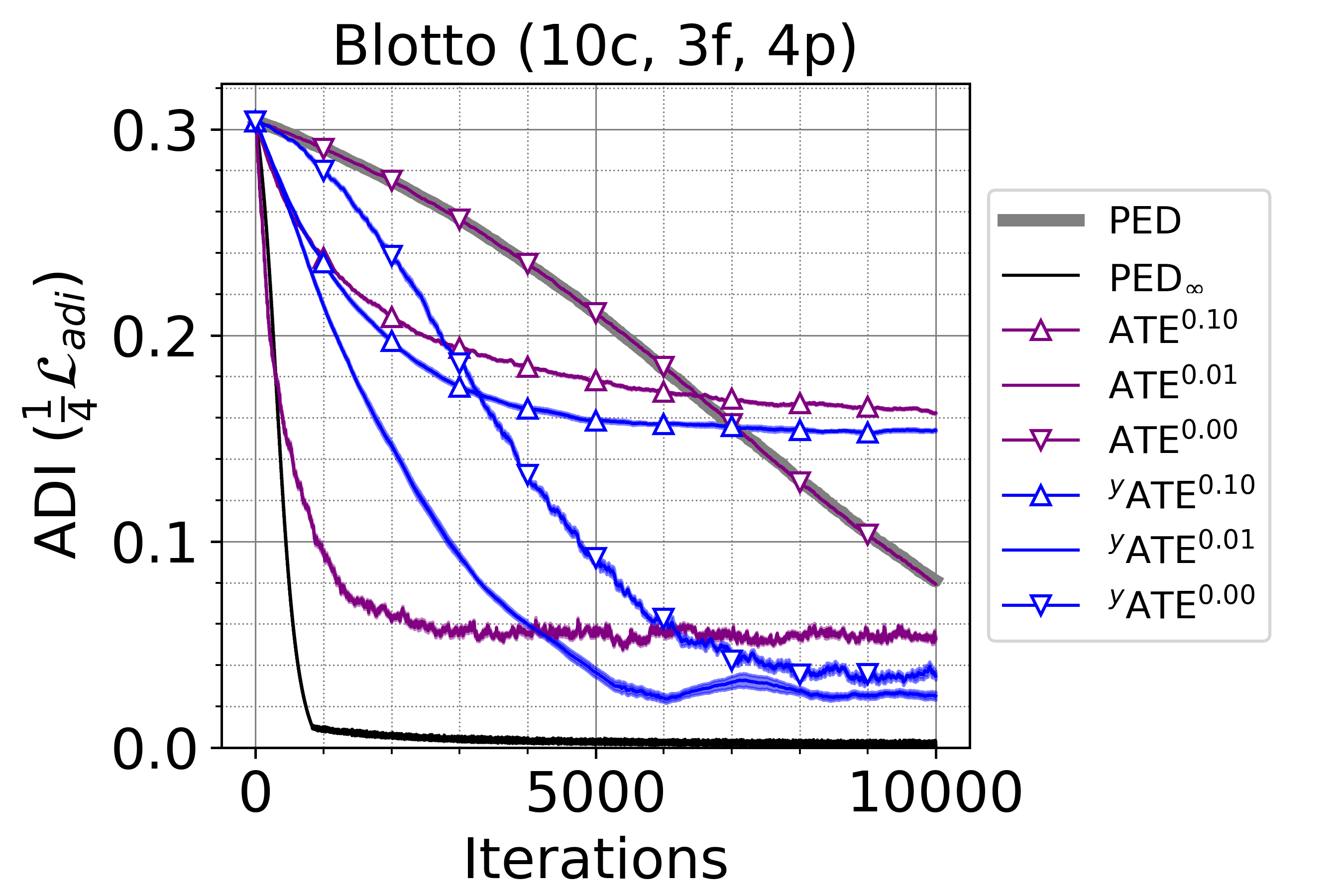}
    \caption{Blotto-4 \label{fig:blotto4track_ate}}
    \end{subfigure}
    \vspace{-5pt}
    \caption{(\subref{fig:blotto3track_ate}) $3$-player Blotto game; (\subref{fig:blotto4track_ate}) $4$-player Blotto game. Adding an appropriate level of entropy (e.g., $\tau=0.01$) can accelerate convergence (compare PED to \texttt{ATE}$^{0.01}$ in (\subref{fig:blotto4track_ate})). And amortizing estimates of joint play can reduce gradient bias, further improving performance (e.g., compare \texttt{ATE}$^{0.01}$ to $^y$\texttt{ATE}$^{0.01}$ in (\subref{fig:blotto3track_ate}) or (\subref{fig:blotto4track_ate})).}
    \label{fig:blotto_tracking_ate}
\end{figure}

\paragraph{Auxiliary $y$ re. \S\ref{amortized_estimates}} The introduction of auxiliary variables $y_i$ are supported by the results in Figure~\ref{fig:blotto_tracking_ate}\textemdash \texttt{ATE}$^{0.0}$ is equivalent to \texttt{PED} and $^y$\texttt{ATE}$^{0.0}$ is equivalent to \texttt{PED} augmented with $y$'s to estimate averages of payoff gradients.

In Figure~\ref{fig:blotto_tracking_ate}, we also see a more general relationship between temperature and convergence rate. Higher temperatures appear to result in faster initial convergence ($\mathcal{L}_{adi}$ spikes initially in Figure~\ref{fig:blotto3track_ate} for $\tau<0.1$) and lower variance but higher asymptotes, while the opposite holds for lower temperatures. These results suggest annealing the temperature over time to achieve fast initial convergence and lower asymptotes. Lower variance should also be possible by carefully annealing the learning rate to allow $y$ to accurately perform tracking. Fixed learning rates were used here; we leave investigating learning rate schedules to future work.

Figure~\ref{fig:blotto4track_ate} shows how higher temperatures (through a reduction in gradient bias) can result in accelerated convergence.

\begin{figure}[!ht]
    \centering
    \begin{subfigure}[b]{.49\textwidth}
    \includegraphics[width=\textwidth]{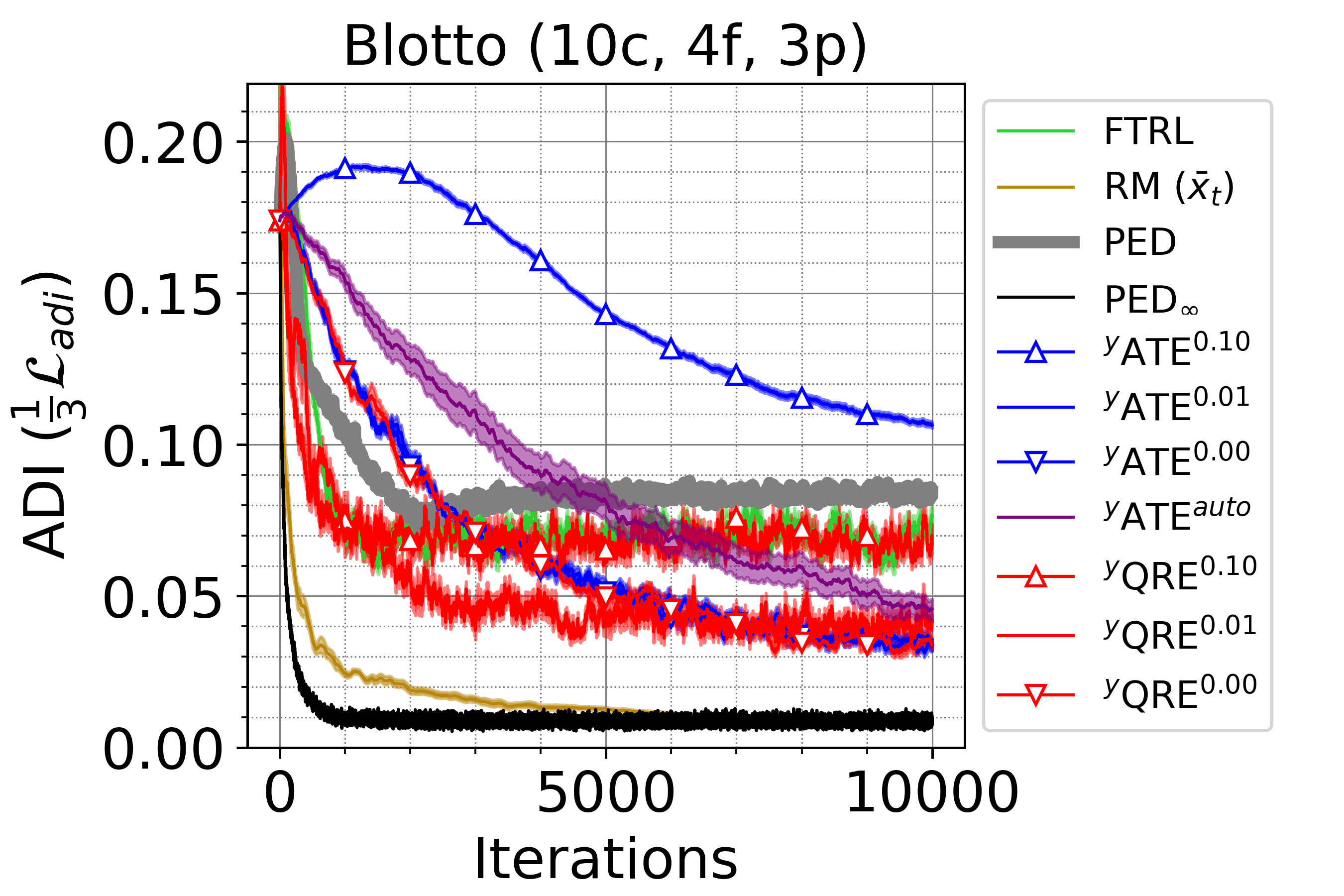}
    \caption{Blotto-3 \label{fig:blotto3_ate}}
    \end{subfigure}
    \begin{subfigure}[b]{.49\textwidth}
    \includegraphics[width=\textwidth]{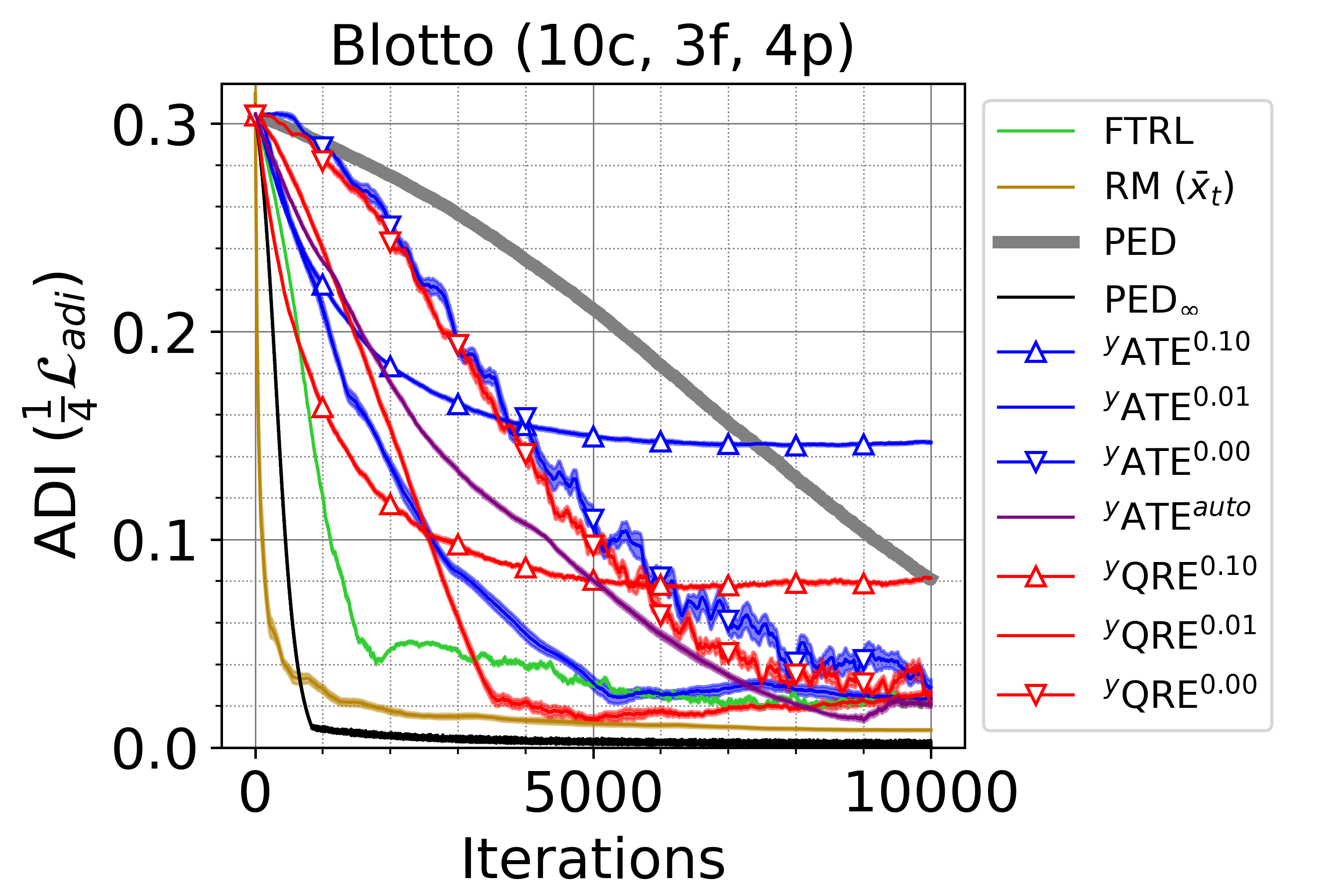}
    \caption{Blotto-4 \label{fig:blotto4_ate}}
    \end{subfigure}
    \vspace{-5pt}
    \caption{(\subref{fig:blotto3_ate}) $3$-player Blotto game; (\subref{fig:blotto4_ate}) $4$-player Blotto game. The maximum a single agent can exploit the symmetric joint strategy $\boldsymbol{x}^{(t)}$ is plotted against algorithm iteration $t$. Despite \texttt{FTRL}'s lack of convergence guarantees, it converges quickly in these Blotto games.}
    \label{fig:blotto_ate}
\end{figure}

\paragraph{Annealing $\tau$ re. \S\ref{annealing}} ADIDAS includes temperature annealing replacing the need for setting the hyperparameter $\tau$ with instead an ADI threshold $\epsilon$. Figure~\ref{fig:blotto_ate} compares this approach against several other variants of the algorithm and shows this automated annealing mechanism reaches comparable final levels of ADI.

\begin{figure}[!ht]
    \centering
    \begin{subfigure}[b]{.49\textwidth}
    \includegraphics[width=\textwidth]{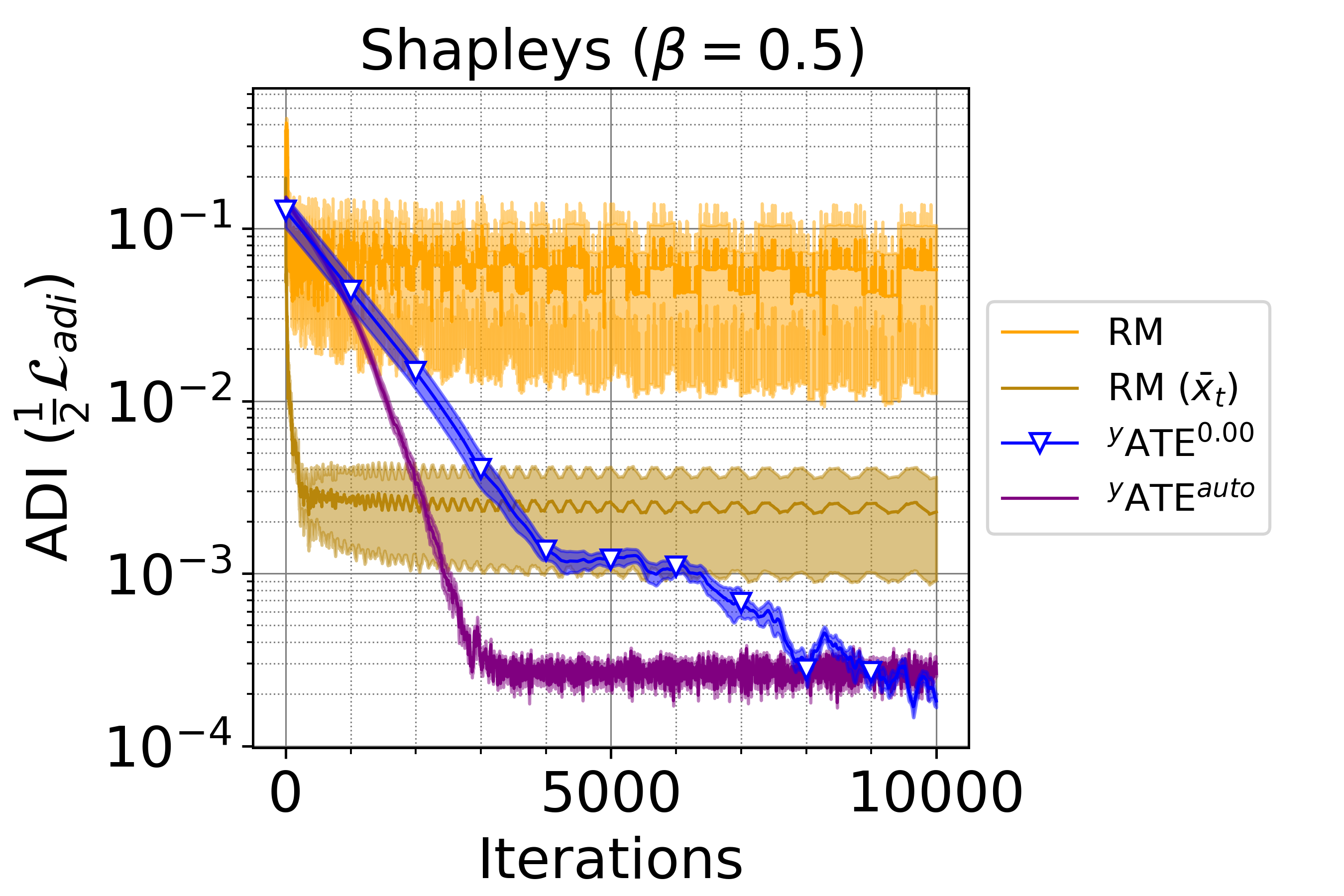}
    \caption{Modified-Shapley's \label{fig:modshapley_ate}}
    \end{subfigure}
    \begin{subfigure}[b]{.49\textwidth}
    \includegraphics[width=\textwidth]{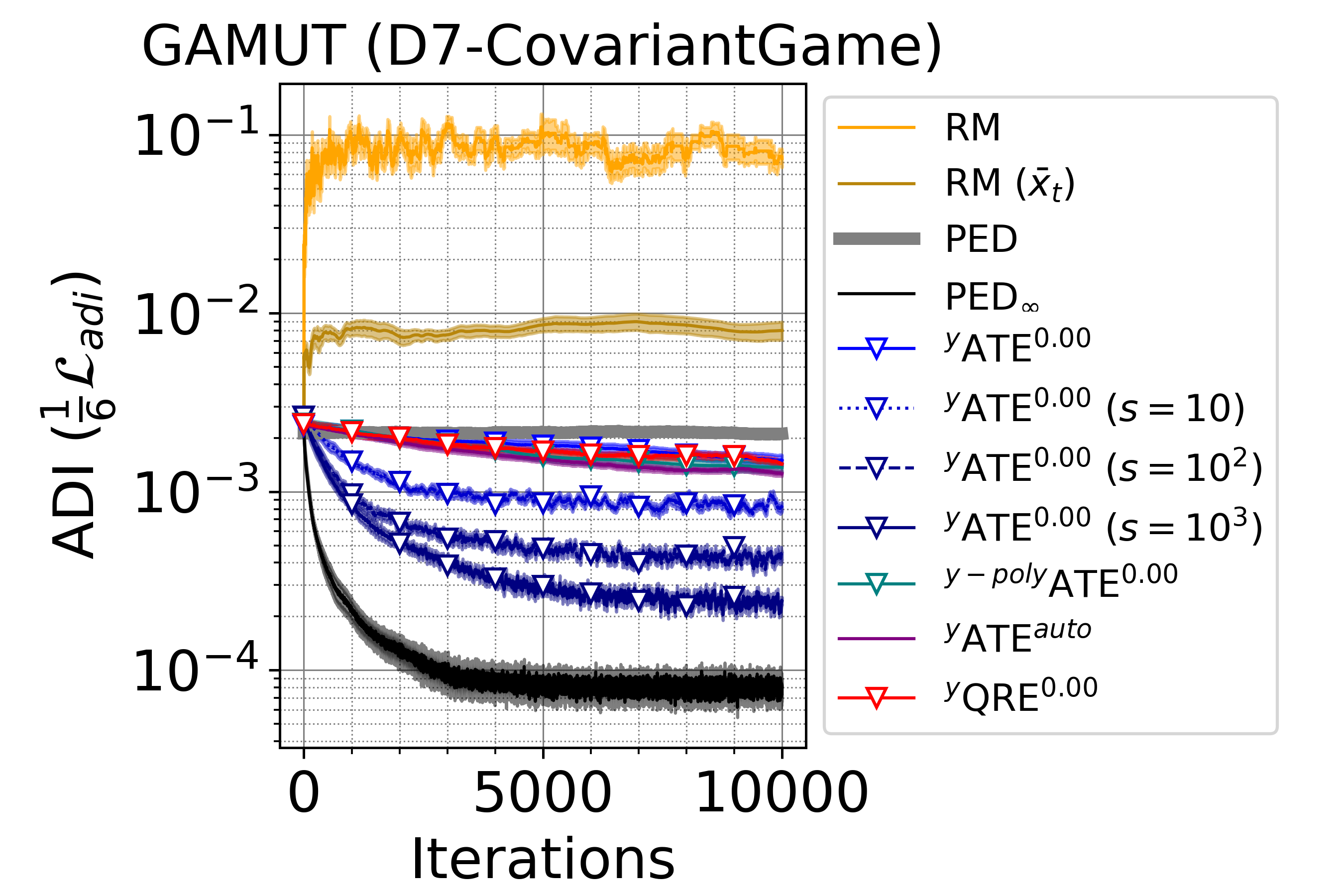}
    \caption{GAMUT-D7 \label{fig:gamutd7_ate}}
    \end{subfigure}
    \vspace{-5pt}
    \caption{(\subref{fig:modshapley_ate}) Modified-Shapley's; (\subref{fig:gamutd7_ate}) GAMUT-D7. Deviation incentive descent reduces $\mathcal{L}_{adi}$ in both games. In~(\subref{fig:modshapley_ate}), to better test the performance of the algorithms, $x$ is initialized randomly rather than with the uniform distribution because the Nash is at uniform. In~(\subref{fig:gamutd7_ate}), computing ADI gradients using full expectations (in black) results in very low levels of ADI. Computing estimates using only single samples plus historical play allows a small reduction in ADI. More samples (e.g., $n=10^3$) allows further reduction.}
    \label{fig:noregfail_ate}
\end{figure}

\paragraph{Convergence re. \S\ref{convergence}} In Figure~\ref{fig:blotto_ate}, \texttt{FTRL} and \texttt{RM} achieve low levels of ADI quickly in some cases. \texttt{FTRL} has recently been proven not to converge to Nash, and this is suggested to be true of no-regret algorithms such as \texttt{RM} in general~\citep{flokas2020no,mertikopoulos2018cycles}. Before proceeding, we demonstrate empirically in Figure~\ref{fig:noregfail_ate} that \texttt{FTRL} and \texttt{RM} fail on some games where ADIDAS still makes progress.

\begin{figure}[h!]
    \centering
    \begin{subfigure}[b]{.49\textwidth}
    \includegraphics[width=\textwidth]{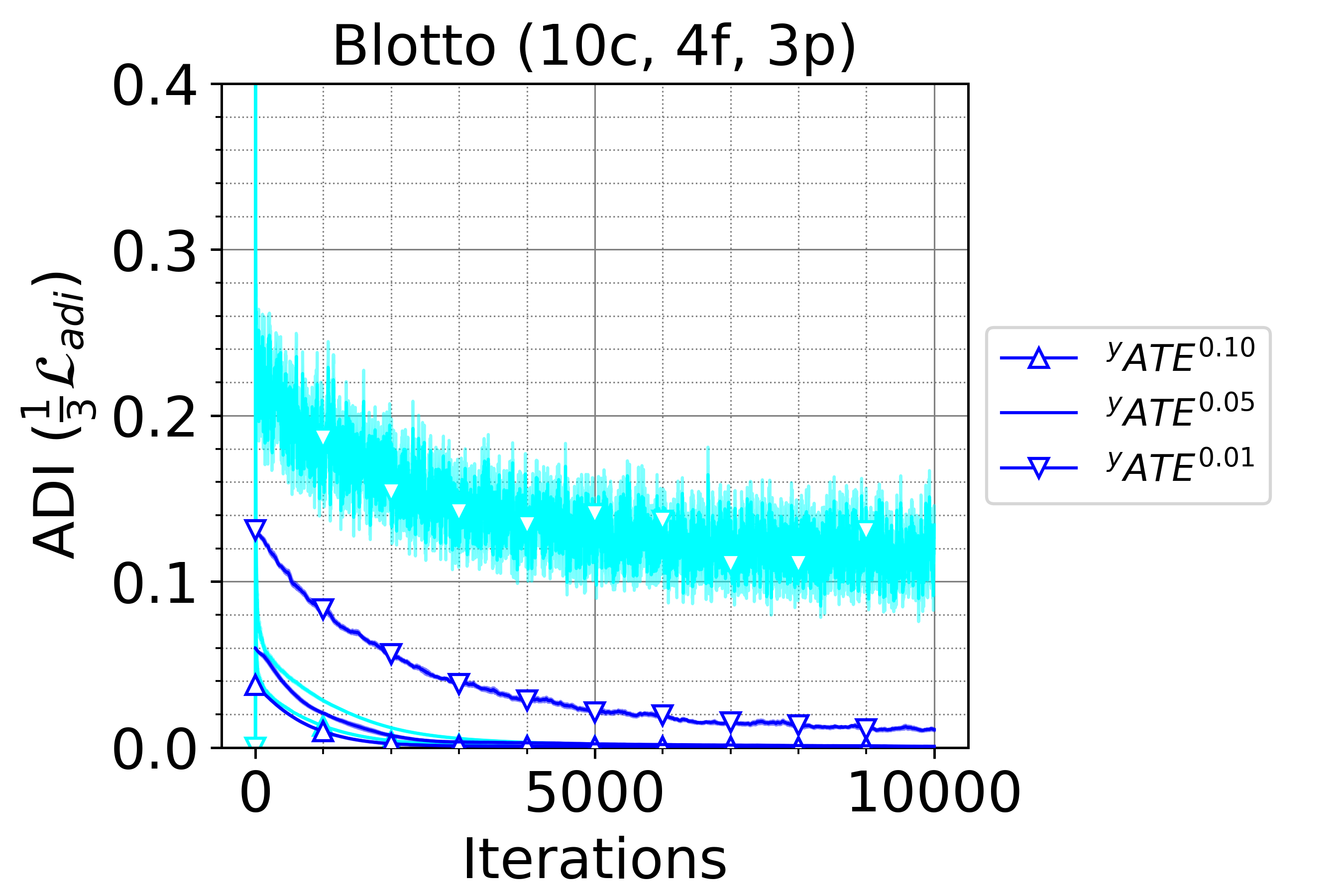}
    \caption{Blotto-3 \label{fig:exp_bias_blotto3_ate}}
    \end{subfigure}
    \begin{subfigure}[b]{.49\textwidth}
    \includegraphics[width=\textwidth]{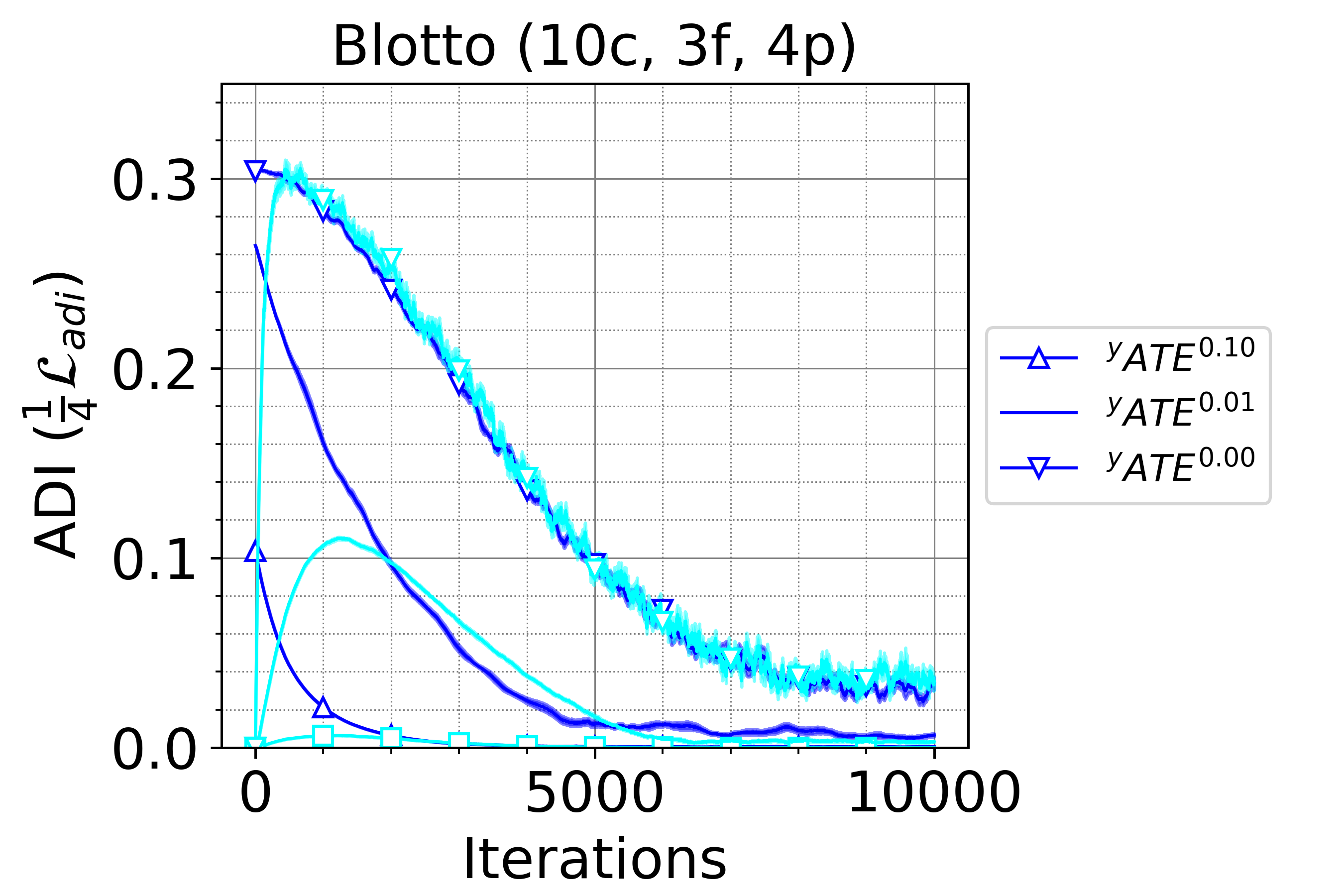}
    \caption{Blotto-4 \label{fig:exp_bias_blotto4_ate}}
    \end{subfigure}
    \vspace{-5pt}
    \caption{Accuracy of running estimate of $\mathcal{L}^{\tau}_{adi}$ computed from $y^{(t)}$ (in light coral) versus true value (in blue).}
    \label{fig:exp_est_ate}
\end{figure}

\paragraph{Large-Scale re \S\ref{scale}} Computing ADI exactly requires the full payoff tensor, so in very large games, we must estimate the ADI. Figure~\ref{fig:exp_est_ate} shows how estimates of $\mathcal{L}_{adi}$ computed from historical play track their true expected value throughout training.

\newpage
\section{Comparison Against Additional Algorithms}
\label{app:morealgs}

\subsection{ED and FP Fail}
\label{app:xtra_comparisons}

We chose not to include Exploitability Descent (ED) or Fictitious Play (FP) in the main body as we considered them to be ``straw men". ED is only expected to converge in 2-player, zero-sum games. FP is non-convergent in some 2-player games as well~\citep{goldberg2013approximation}.
\begin{figure}
    \centering
    \includegraphics[scale=0.49]{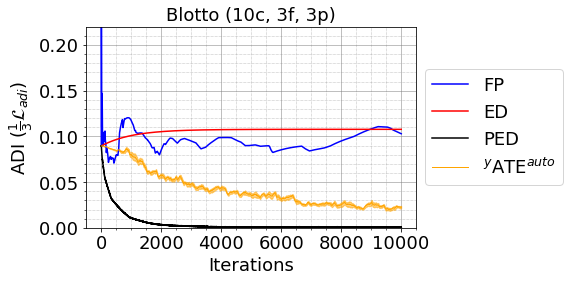}
    \caption{FP, ED, PED access the full tensor. $^y$ATE$^{auto}$ samples.}
    \vspace{-0.4cm}
    \label{fig:fp_ed}
\end{figure}
We run ED and FP with true expected gradients \& best responses ($s$$=$$\infty$) on the $3$ player game in Figure~\ref{fig:fp_ed} to convince the reader that failure to converge is not due to stochasticity.

\subsection{Gambit Solvers}
\label{app:gambit_solvers}

We ran all applicable gambit solvers  on the 4-player, 10-coin, 3-field Blotto game (comand listed below). All solvers fail to return a Nash equilibrium except \texttt{gambit-enumpoly} which returns all $36$ permutations of the following pure, non-symmetric Nash equilibrium:
\begin{align}
    x^* &= [(10,0,0), (10,0,0), (0,10,0), (0,0,10)]
\end{align}
where each of the four players places 10 coins on one of the three fields.

\begin{itemize}
    \item \texttt{gambit-enumpoly}
    \item \texttt{gambit-gnm}
    \item \texttt{gambit-ipa}
    \item \texttt{gambit-liap}
    \item \texttt{gambit-simpdiv}
    \item \texttt{gambit-logit}
\end{itemize}

Command:
\begin{lstlisting}
timeout 3600s gambit-enumpoly -H < blotto_10_3_4.nfg >> enumpoly.txt; timeout 3600s gambit-gnm < blotto_10_3_4.nfg >> gnm.txt; timeout 3600s gambit-ipa < blotto_10_3_4.nfg >> ipa.txt; timeout 3600s gambit-liap < blotto_10_3_4.nfg >> liap.txt; timeout 3600s gambit-simpdiv < blotto_10_3_4.nfg >> simpdiv.txt; timeout 3600s gambit-logit -m 1.0 -e < blotto_10_3_4.nfg >> logit.txt
\end{lstlisting}

\section{Additional Game Domains}
\label{app:xtra_games}

\subsection{Diplomacy Experiments - Subsampled Games}
\begin{figure}[h!]
    \centering
    \includegraphics[width=.49\textwidth]{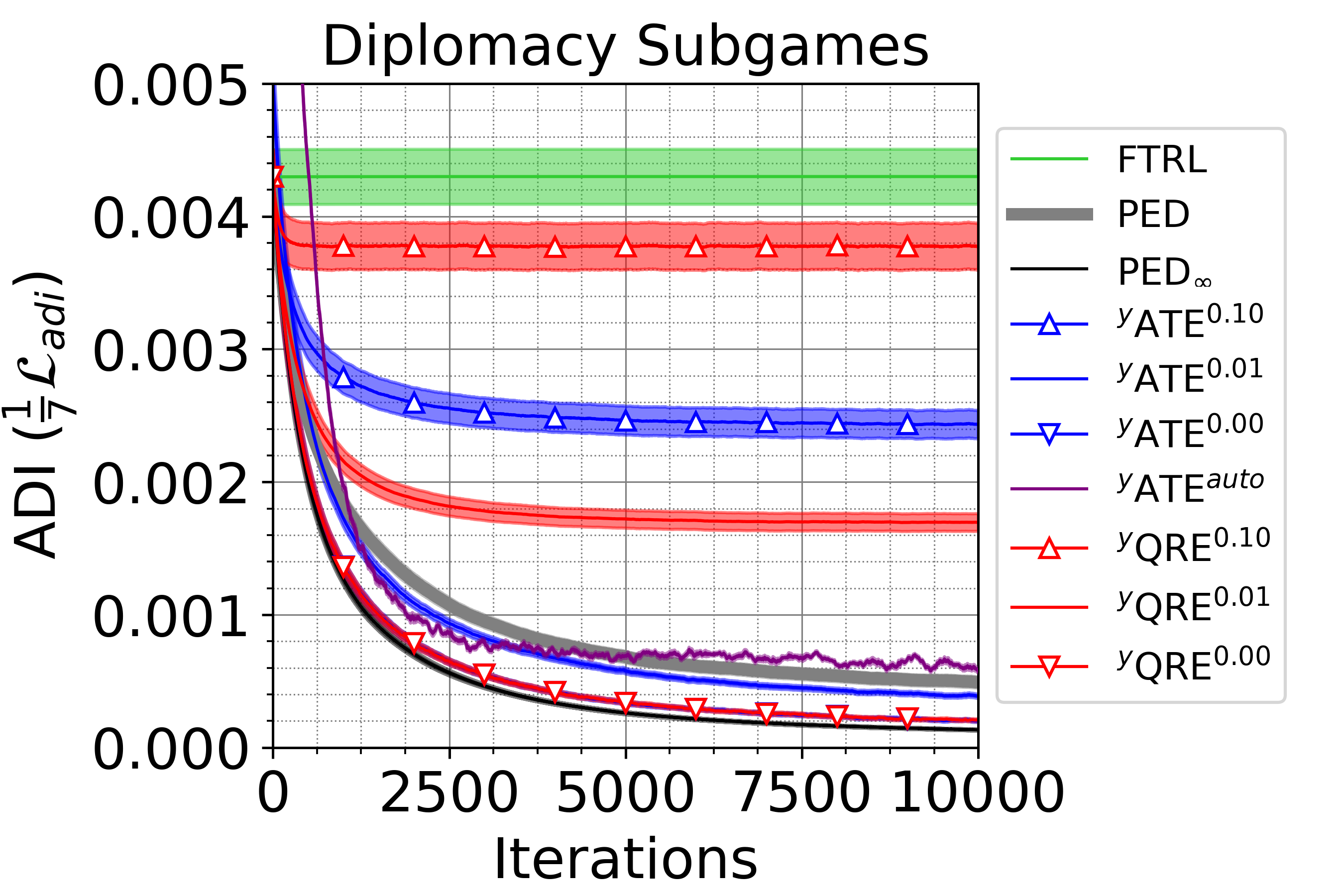}
    \caption{Subsampled games.}
    \label{fig:dipsub}
\end{figure}

Figure~\ref{fig:dipsub} runs a comparison on $40$ subsampled tensors ($7$-players, $4$-actions each) taken from the $40$ turns of a single Diplomacy match. The four actions selected for each player are sampled from the corresponding player's trained policy.

\begin{figure}[h!]
    \centering
    \begin{subfigure}[b]{.49\textwidth}
    \includegraphics[width=\textwidth]{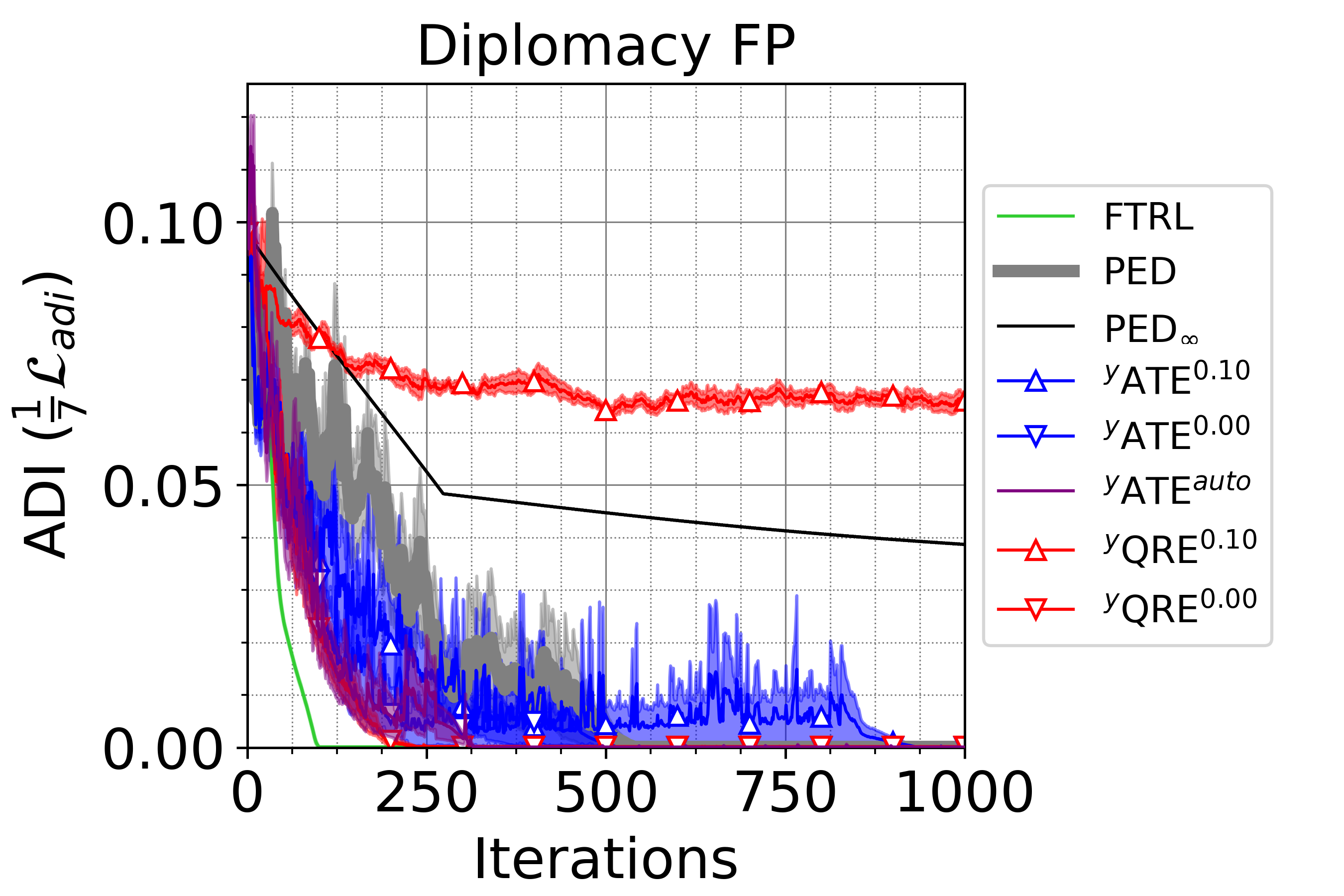}
    \caption{Diplomacy FP \label{fig:dipfp}}
    \end{subfigure}
    \begin{subfigure}[b]{.49\textwidth}
    \includegraphics[width=\textwidth]{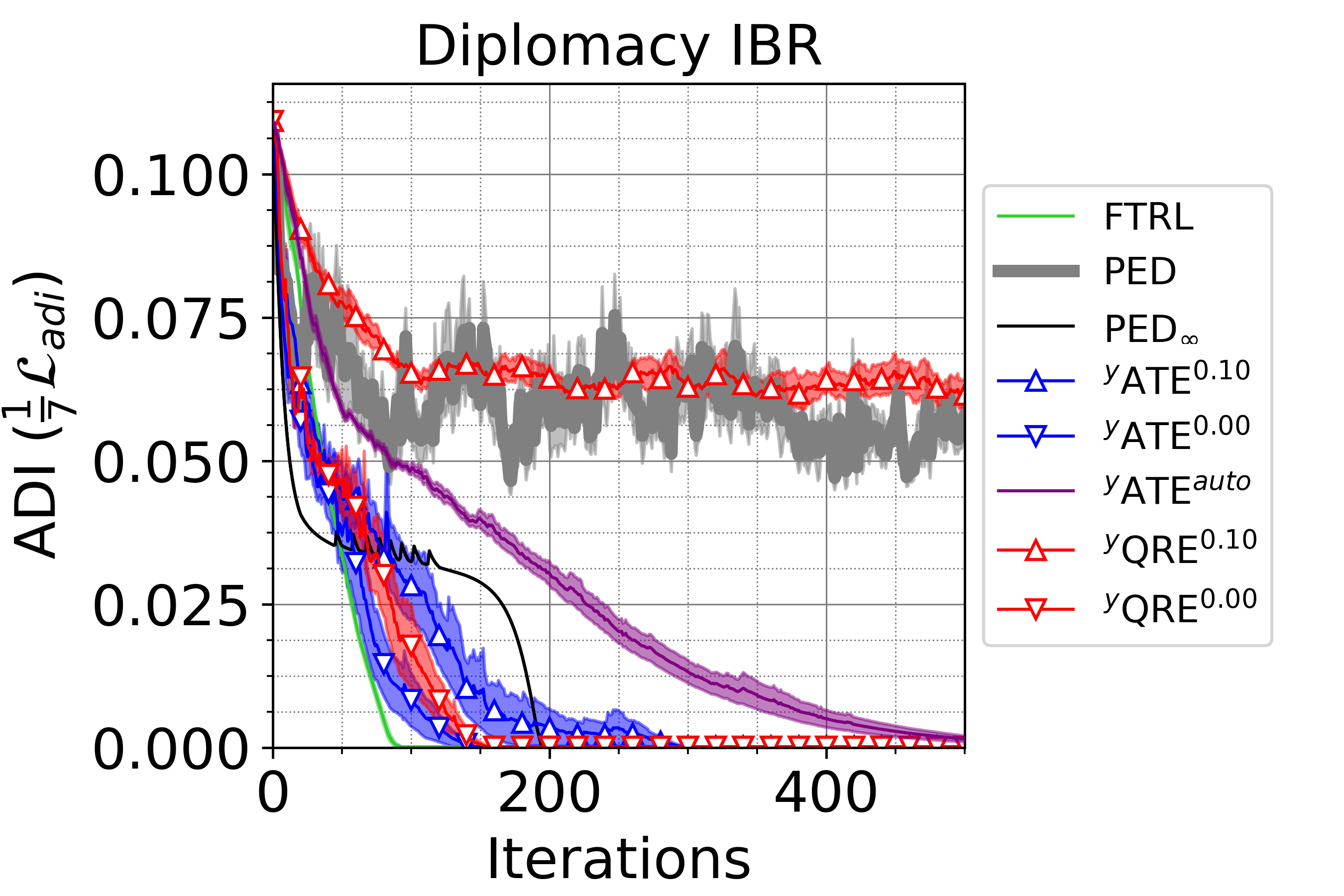}
    \caption{Diplomacy IBR \label{fig:dipibr}}
    \end{subfigure}
    \vspace{-5pt}
    \caption{(\subref{fig:dipfp}) FP; (\subref{fig:dipibr}) IBR. The maximum a single agent can exploit the symmetric joint strategy $\boldsymbol{x}^{(t)}$ is plotted against algorithm iteration $t$. Many of the algorithms quickly achieve near zero $\mathcal{L}_{adi}$, so unlike in the other experiments, hyperparameters are selected according according to the earliest point at which exploitablity falls below $0.01$ with ties split according to the final value.}
    \label{fig:diplomacy}
\end{figure}

Figure~\ref{fig:diplomacy} runs a comparison on two Diplomacy meta-games, one with 5 bots trained using Fictious Play and the other with bots trained using Iterated Best Response (IBR) \textemdash these are the same meta-games analyzed in Figure 3 of~\citep{anthony2020learning}.

Figure~\ref{fig:diplomacy_med_ate} demonstrates an empirical game theoretic analysis~\citep{wellman2006methods,jordan2007empirical,wah2016empirical} of a large symmetric $7$-player Diplomacy meta-game where each player elects $1$ of $5$ trained bots to play on their behalf. In this case, the expected value of each entry in the payoff tensor represents a winrate. Each entry can only be estimated by simulating game play, and the result of each game is a Bernoulli random variable. To obtain a winrate estimate within $0.01$ of the true estimate with probability $95\%$, a Chebyshev bound implies more than $223$ samples are needed. The symmetric payoff tensor contains $330$ unique entries, requiring over $74$ thousand games in total. In the experiment below, ADIDAS achieves negligible ADI in less than $7$ thousand iterations with $50$ samples of joint play per iteration ($\approx 5 \times$ the size of the tensor).
%
%
%

\subsection{Diplomacy Experiments - Empirical Game Theoretic Analysis}
\label{app:more_dip}

\begin{figure}[!ht]
    \centering
    \begin{subfigure}[b]{.49\textwidth}
    \includegraphics[width=\textwidth]{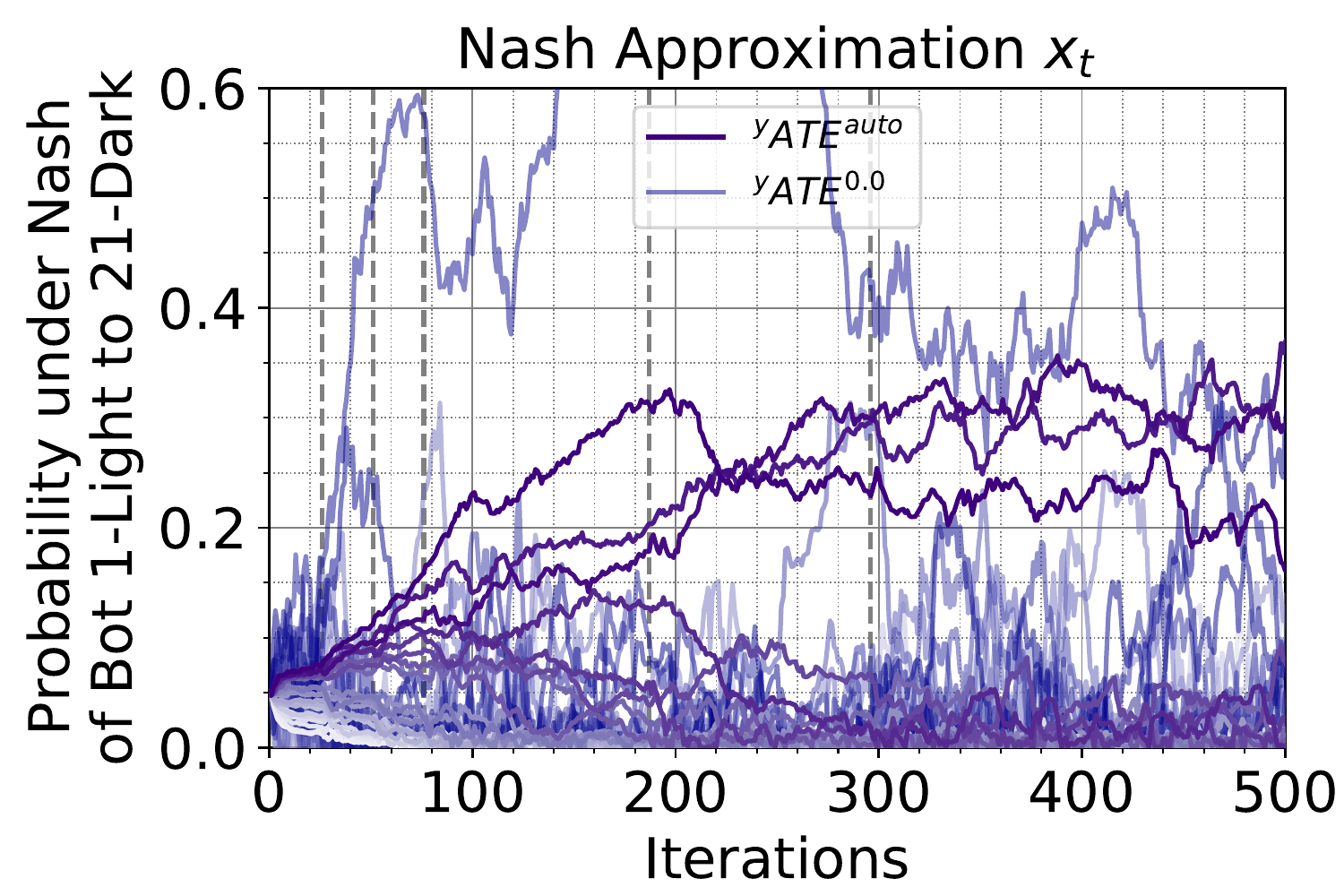}
    \caption{7-player, 21-action symmetric Nash ($x_t$) \label{fig:dipbignash_all_ate_25}}
    \end{subfigure}
    \begin{subfigure}[b]{.49\textwidth}
    \includegraphics[width=\textwidth]{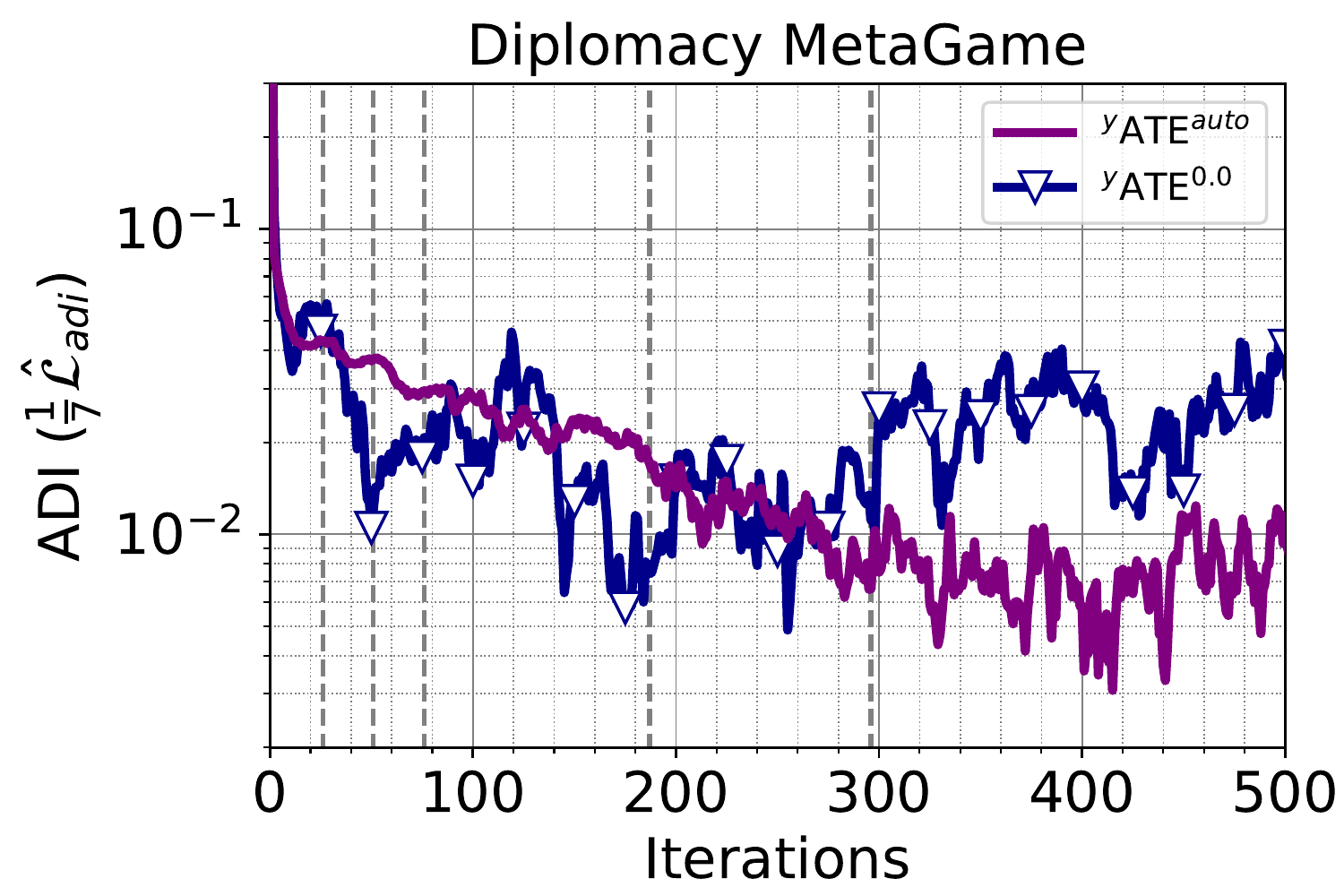}
    \caption{ADI Estimate \label{fig:dipbigexp_ate2_25}}
    \end{subfigure}
    \vspace{-5pt}
    \caption{(\subref{fig:dipbignash_all_ate}) Evolution of the symmetric Nash approximation; (\subref{fig:dipbigexp_ate2}) ADI estimated from auxiliary variable $y_t$. Black vertical lines indicate the temperature $\tau$ was annealed. Auxiliary learning rate $\eta_y = 1/25$. In addition to the change in $\eta_y$ from $1/10$, also note the change in axes limits versus Figure~\ref{fig:diplomacy_big_ate}.}
    \label{fig:diplomacy_big_all_ate_25}
\end{figure}

Figure~\ref{fig:diplomacy_big_all_ate_25} repeats the computation of Figure~\ref{fig:diplomacy_big_ate} with a smaller auxiliary learning rate $\eta_y$ and achieves better results.


\subsection{El Farol Bar Stage Game}
\label{app:elfarol}

\begin{figure}[!ht]
    \centering
    \includegraphics[width=0.5\textwidth]{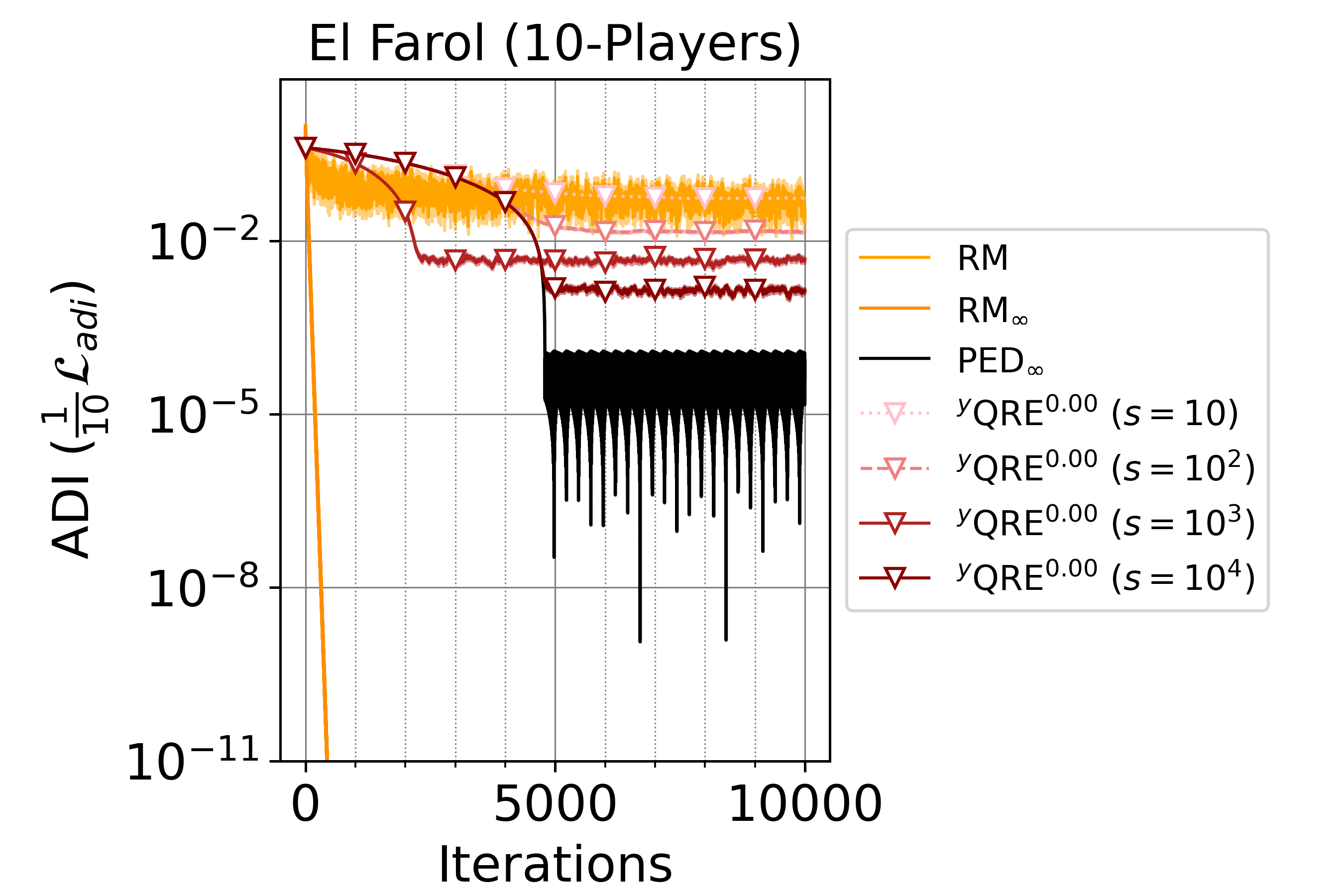}
    \caption{(10-player, 2-action El Farol stage game) ADIDAS and regret matching both benefit from additional samples. Both find the same unique mixed Nash equilibrium. In this case, regret matching finds it much more quickly.}
    \label{fig:elfarol}
\end{figure}

We compare ADIDAS variants and regret matching in Figure~\ref{fig:elfarol} on the 10-player symmetric El Farol Bar stage game with hyperparameters $n = 10$, $c = 0.7$, $C = nc$, $B = 0$, $S = 1$, $G = 2$ (see Section 3.1, The El Farol stage game in~\citep{whitehead2008farol}). Recall that the homotopy that ADIDAS attempts to trace is displayed in Figure~\ref{fig:el_farol_homotopy} of the main body.

\section{Description of Domains}
\label{app:descrip_games}

\subsection{Modified Shapley's}
\label{mod_shap_def}

The modified Shapley's game mentioned in the main body (Figure~\ref{fig:modshapley_qre}) is defined in Table~\ref{tab:mod_shap_def}~\citep{ostrovski2013payoff}.
\begin{table}[ht!]
    \centering
    \begin{subfigure}[b]{.49\textwidth}
    \centering
    \begin{tabular}{|c|c|c|}
        \hline
        $1$ & $0$ & $\beta$ \\ \hline
        $\beta$ & $1$ & $0$ \\ \hline
        $0$ & $\beta$ & $1$ \\ \hline
    \end{tabular}
    \caption{Player A's Payoff Matrix \label{fig:mod_shap_player_A}}
    \end{subfigure}
    \begin{subfigure}[b]{.49\textwidth}
    \centering
    \begin{tabular}{|c|c|c|}
        \hline
        $-\beta$ & $1$ & $0$ \\ \hline
        $0$ & $-\beta$ & $1$ \\ \hline
        $1$ & $0$ & $-\beta$ \\ \hline
    \end{tabular}
    \caption{Player B's Payoff Matrix \label{fig:mod_shap_player_B}}
    \end{subfigure}
    \caption{(\subref{fig:mod_shap_player_A}) Player A; (\subref{fig:mod_shap_player_B}) Player B. We set $\beta=0.5$ in experiments and subtract $-\beta$ from each payoff matrix to ensure payoffs are non-negative; \texttt{ATE} requires non-negative payoffs.}
    \label{tab:mod_shap_def}
\end{table}

\subsection{Colonel Blotto}
\label{app:blotto}

Despite its apparently simplicity, Colonel Blotto is a complex challenge domain and one under intense research~\citep{behnezhad2017faster,behnezhad2018battlefields,behnezhad2019optimal,ahmadinejad2019duels,boix2020multiplayer}.
\section{Connections to Other Algorithms}
\label{app:alg_connections}

\subsection{Consensus Algorithm}
\label{app:consensus_connection}
ADIDAS with Tsallis entropy and temperature fixed to $\tau=p=1$ recovers the regularizer proposed for the Consensus algorithm~\citep{mescheder2017numerics} plus entropy regularization. To see this, recall from Appx.~\ref{app:tsallis_entropy_derivation}, the Tsallis entropy regularizer:
\begin{align}
    S^{\tau=p=1}_k(x_k, x_{-k}) &= \frac{s_k}{2} ( 1 - \sum_m x_{km}^{2} )
\end{align}
where $s_k = \Big( \sum_m (\nabla^k_{x_{km}})^{1/1} \Big)^1 = ||\nabla^k_{x_k}||_{1/1}$ is treated as a constant w.r.t. $\boldsymbol{x}$.

In the case with $\tau=p=1$, $\br_k = \br(x_{-k}) = \frac{1}{s_k} \nabla^k_{x_{k}}$ where we have assumed the game has been offset by a constant so that it contains only positive payoffs. Plugging these into the definition of $\mathcal{L}^{\tau}_{adi}$, we find
\begin{align}
    \mathcal{L}^{\tau}_{adi}(\boldsymbol{x}) &= \sum_k u^{\tau}_k(\br_k, x_{-k}) - u^{\tau}_k(x_k,x_{-k})
    \\ &\approx \sum_k u_k(\frac{1}{s_k}\nabla^k_{x_k}, x_{-k}) - u_k(x_k, x_{-k})
    \\ &= \sum_k \frac{1}{s_k} \underbrace{||\nabla^k_{x_k}||^2}_{\text{consensus regularizer}} - x_k^\top \nabla^k_{x_k}.
\end{align}

Note the Consensus regularizer can also be arrived at by replacing the best response with a 1-step gradient ascent lookahead, i.e., $\br_k = x_k + \eta \nabla^k_{x_k}$:
\begin{align}
    \mathcal{L}^{\tau}_{adi}(\boldsymbol{x}) &= \sum_k u^{\tau}_k(\br_k, x_{-k}) - u^{\tau}_k(x_k,x_{-k})
    \\ &\approx \sum_k u_k(x_k + \eta \nabla^k_{x_k}, x_{-k}) - u_k(x_k, x_{-k})
    \\ &= \sum_k (x_k^\top \nabla^k_{x_k}) + \eta ||\nabla^k_{x_k}||^2 - (x_k^\top \nabla^k_{x_k})
    \\ &= \sum_k \eta ||\nabla^k_{x_k}||^2.
\end{align}

\subsection{Exploitability Descent as Extragradient}
\label{app:ed_connection}

In normal-form games, Exploitability Descent (ED)~\citep{Lockhart19ED} is equivalent to Extragradient~\citep{korpelevich1976extragradient} (or Mirror Prox~\citep{juditsky2011solving}) with an infinite intermediate step size. Recall $\br_k = \argmax_{x \in \Delta^{m_k - 1}} u_k(x_k, x_{-k}) = \argmax_{x \in \Delta^{m_k - 1}} x^\top \nabla^k_{x_k}$. Using the convention that ties between actions result in vectors that distribute a $1$ uniformly over the maximizers, the best response can be rewritten as $\br_k = \lim_{\hat{\eta} \rightarrow \infty} \Pi[x_k + \hat{\eta} \nabla^k_{x_k}]$ where $\Pi$ is the Euclidean projection onto the simplex. Define $F(\boldsymbol{x})$ such that its $k$th component $F(\boldsymbol{x})_k = -\nabla^k_{x_k} = -\nabla^k_{x_k}(x_{-k})$ where we have simply introduced $x_{-k}$ in parentheses to emphasize that player $k$'s gradient is a function of $x_{-k}$ only, and not $x_k$. Equations without subscripts imply they are applied in parallel over the players.

ED executes the following update in parallel for all players $k$:
\begin{align}
    x_{k+1} &\leftarrow \Pi[x_k + \eta \nabla_{x_k} \{ u_k(x_k, x_{-k}) \} \vert_{x_{-k} = \br_{-k}}].
\end{align}

Define $\hat{x}_k = x_k - \hat{\eta} F(\boldsymbol{x})_k$. And as an abuse of notation, let $\hat{x}_{-k} = x_{-k} - \hat{\eta} F(\boldsymbol{x})_{-k}$. Extragradient executes the same update in parallel for all players $k$:
\begin{align}
    x_{k+1} &\leftarrow \Pi[x_k - \eta F(\Pi[\boldsymbol{x} - \hat{\eta} F(\boldsymbol{x})])_k]
    \\ &= \Pi[x_k - \eta F(\Pi[\boldsymbol{x} + \hat{\eta} \nabla_{\boldsymbol{x}}])_k]
    \\ &= \Pi[x_k - \eta F(\br)_k]
    \\ &= \Pi[x_k + \eta \nabla_{x_k} \{ u_k(x_k, x_{-k}) \} \vert_{x_{-k} = \br_{-k}}].
\end{align}

Extragradient is known to converge in two-player zero-sum normal form games given an appropriate step size scheme. The main property that Extragradient relies on is monotonicity of the vector function $F$. All two-player zero-sum games induce a monotone $F$, however, this is not true in general of two-player general-sum or games with more players. ED is only proven to converge for two-player zero-sum, but this additional connection provides an additional reason why we do not expect ED to solve many-player general-sum normal-form games, which are the focus of this work. Please see Appx.~\ref{app:xtra_comparisons} for an experimental demonstration.
\section{Python Code}
\label{app:code}

For the sake of reproducibility we have included code in python+numpy.

\begin{lstlisting}[language=Python, caption=Header.]
"""
Copyright 2020 ADIDAS Authors. 


Licensed under the Apache License, Version 2.0 (the "License");
you may not use this file except in compliance with the License.
You may obtain a copy of the License at

https://www.apache.org/licenses/LICENSE-2.0

Unless required by applicable law or agreed to in writing, software
distributed under the License is distributed on an "AS IS" BASIS,
WITHOUT WARRANTIES OR CONDITIONS OF ANY KIND, either express or implied.
See the License for the specific language governing permissions and
limitations under the License.
"""
import numpy as np
from scipy import special

def simplex_project_grad(g):
  """Project gradient onto tangent space of simplex."""
  return g - g.sum() / g.size
\end{lstlisting}
\begin{lstlisting}[language=Python, caption=ADIDAS Gradient.]
def gradients_qre_nonsym(dist, y, anneal_steps, payoff_matrices,
                         num_players, temp=0., proj_grad=True,
                         exp_thresh=1e-3, lrs=(1e-2, 1e-2),
                         logit_clip=-1e5):
    """Computes exploitablity gradient and aux variable gradients.

    Args:
      dist: list of 1-d np.arrays, current estimate of nash
      y: list of 1-d np.arrays, current est. of payoff gradient
      anneal_steps: int, elapsed num steps since last anneal
      payoff_matrices: dict with keys as tuples of agents (i, j) and
        values of (2 x A x A) arrays, payoffs for each joint action.
        keys are sorted and arrays are indexed in the same order.
      num_players: int, number of players
      temp: non-negative float, default 0.
      proj_grad: bool, if True, projects dist gradient onto simplex
      exp_thresh: ADI threshold at which temp is annealed
      lrs: tuple of learning rates (lr_x, lr_y)
      logit_clip: float, minimum allowable logit
    Returns:
      gradient of ADI w.r.t. (dist, y, anneal_steps)
      temperature (possibly annealed, i.e., reduced)
      unregularized ADI (stochastic estimate)
      shannon regularized ADI (stochastic estimate)
    """
    # first compute policy gradients and player effects (fx)
    policy_gradient = []
    other_player_fx = []
    grad_y = []
    unreg_exp = []
    reg_exp = []
    for i in range(num_players):

      nabla_i = np.zeros_like(dist[i])
      for j in range(num_players):
        if j == i:
          continue
        if i < j:
          hess_i_ij = payoff_matrices[(i, j)][0]
        else:
          hess_i_ij = payoff_matrices[(j, i)][1].T

        nabla_ij = hess_i_ij.dot(dist[j])
        nabla_i += nabla_ij / float(num_players - 1)

      grad_y.append(y[i] - nabla_i)

      if temp >= 1e-3:  # numerical under/overflow for temp < 1e-3
        br_i = special.softmax(y[i] / temp)
        br_i_mat = (np.diag(br_i) - np.outer(br_i, br_i)) / temp
        log_br_i_safe = np.clip(np.log(br_i), logit_clip, 0)
        br_i_policy_gradient = nabla_i - temp * (log_br_i_safe + 1)
      else:
        power = np.inf
        s_i = np.linalg.norm(y[i], ord=power)
        br_i = np.zeros_like(dist[i])
        maxima_i = (y[i] == s_i)
        br_i[maxima_i] = 1. / maxima_i.sum()
        br_i_mat = np.zeros((br_i.size, br_i.size))
        br_i_policy_gradient = np.zeros_like(br_i)

      policy_gradient_i = np.array(nabla_i)
      if temp > 0:
        log_dist_i_safe = np.clip(np.log(dist[i]), logit_clip, 0)
        policy_gradient_i -= temp * (log_dist_i_safe + 1)
      policy_gradient.append(policy_gradient_i)

      unreg_exp_i = np.max(y[i]) - y[i].dot(dist[i])
      unreg_exp.append(unreg_exp_i)

      entr_br_i = temp * special.entr(br_i).sum()
      entr_dist_i = temp * special.entr(dist[i]).sum()

      reg_exp_i = y[i].dot(br_i - dist[i]) + entr_br_i - entr_dist_i
      reg_exp.append(reg_exp_i)

      other_player_fx_i = (br_i - dist[i])
      other_player_fx_i += br_i_mat.dot(br_i_policy_gradient)
      other_player_fx.append(other_player_fx_i)

    # then construct ADI gradient
    grad_dist = []
    for i in range(num_players):

      grad_dist_i = -policy_gradient[i]
      for j in range(num_players):
        if j == i:
          continue
        if i < j:
          hess_j_ij = payoff_matrices[(i, j)][1]
        else:
          hess_j_ij = payoff_matrices[(j, i)][0].T

        grad_dist_i += hess_j_ij.dot(other_player_fx[j])

      if proj_grad:
        grad_dist_i = simplex_project_grad(grad_dist_i)

      grad_dist.append(grad_dist_i)

    unreg_exp_mean = np.mean(unreg_exp)
    reg_exp_mean = np.mean(reg_exp)

    _, lr_y = lrs
    if (reg_exp_mean < exp_thresh) and (anneal_steps >= 1 / lr_y):
      temp = np.clip(temp / 2., 0., 1.)
      if temp < 1e-3:  # consistent with numerical issue above
        temp = 0.
      grad_anneal_steps = -anneal_steps
    else:
      grad_anneal_steps = 1

    return ((grad_dist, grad_y, grad_anneal_steps), temp,
      unreg_exp_mean, reg_exp_mean)
\end{lstlisting}
\begin{lstlisting}[language=Python, caption=ADIDAS Gradient (assuming symmetric Nash and with Tsallis entropy).]
def gradients_ate_sym(dist, y, anneal_steps, payoff_matrices,
                      num_players, p=1, proj_grad=True,
                      exp_thresh=1e-3, lrs=(1e-2, 1e-2)):
    """Computes ADI gradient and aux variable gradients.

    Args:
      dist: list of 1-d np.arrays, current estimate of nash
      y: list of 1-d np.arrays, current est. of payoff gradient
      anneal_steps: int, elapsed num steps since last anneal
      payoff_matrices: dict with keys as tuples of agents (i, j) and
        values of (2 x A x A) arrays, payoffs for each joint action.
        keys are sorted and arrays are indexed in the same order.
      num_players: int, number of players
      p: float in [0, 1], Tsallis entropy-regularization
      proj_grad: bool, if True, projects dist gradient onto simplex
      exp_thresh: ADI threshold at which p is annealed
      lrs: tuple of learning rates (lr_x, lr_y)
    Returns:
      gradient of ADI w.r.t. (dist, y, anneal_steps)
      temperature, p (possibly annealed, i.e., reduced)
      unregularized ADI (stochastic estimate)
      tsallis regularized ADI (stochastic estimate)
    """
    nabla = payoff_matrices[0].dot(dist)
    if p >= 1e-2:  # numerical under/overflow when power > 100.
      power = 1. / float(p)
      s = np.linalg.norm(y, ord=power)
      if s == 0:
        br = np.ones_like(y) / float(y.size)  # uniform dist
      else:
        br = (y / s)**power
    else:
      power = np.inf
      s = np.linalg.norm(y, ord=power)
      br = np.zeros_like(dist)
      maxima = (y == s)
      br[maxima] = 1. / maxima.sum()

    unreg_exp = np.max(y) - y.dot(dist)
    br_inv_sparse = 1 - np.sum(br**(p + 1))
    dist_inv_sparse = 1 - np.sum(dist**(p + 1))
    entr_br = s / (p + 1) * br_inv_sparse
    entr_dist = s / (p + 1) * dist_inv_sparse
    reg_exp = y.dot(br - dist) + entr_br - entr_dist

    entr_br_vec = br_inv_sparse * br**(1 - p)
    entr_dist_vec = dist_inv_sparse * dist**(1 - p)

    policy_gradient = nabla - s * dist**p
    other_player_fx = (br - dist)
    other_player_fx += 1 / (p + 1) * (entr_br_vec - entr_dist_vec)

    other_player_fx_translated = payoff_matrices[1].dot(
      other_player_fx)
    grad_dist = -policy_gradient
    grad_dist += (num_players - 1) * other_player_fx_translated
    if proj_grad:
      grad_dist = simplex_project_grad(grad_dist)
    grad_y = y - nabla

    _, lr_y = lrs
    if (reg_exp < exp_thresh) and (anneal_steps >= 1 / lr_y):
      p = np.clip(p / 2., 0., 1.)
      if p < 1e-2:  # consistent with numerical issue above
        p = 0.
      grad_anneal_steps = -anneal_steps
    else:
      grad_anneal_steps = 1

    return ((grad_dist, grad_y, grad_anneal_steps), p, unreg_exp,
      reg_exp)
\end{lstlisting}

\end{document}